\documentclass{article}

\usepackage[final]{neurips_2025}

\usepackage[utf8]{inputenc} % allow utf-8 input
\usepackage[T1]{fontenc}    % use 8-bit T1 fonts
\usepackage{hyperref}       % hyperlinks
\usepackage{url}            % simple URL typesetting
\usepackage{graphicx}
\usepackage{floatrow}
\usepackage{subcaption}
\usepackage{booktabs}       % professional-quality tables
\usepackage{amsfonts}       % blackboard math symbols
\usepackage{nicefrac}       % compact symbols for 1/2, etc.
\usepackage{microtype}      % microtypography
\usepackage{xcolor}         % colors
\usepackage{bm}
\usepackage{wrapfig}
\usepackage{amsmath, amsfonts}

\definecolor{mplblue}{rgb}{0.0, 0.447, 0.741}
\definecolor{mplgreen}{rgb}{0.0, 0.621, 0.451}
\definecolor{mplorange}{rgb}{0.850, 0.325, 0.098}

\newcommand\blfootnote[1]{%
  \begingroup
  \renewcommand\thefootnote{}\footnote{#1}%
  \addtocounter{footnote}{-1}%
  \endgroup
}

\title{Bridging Critical Gaps in Convergent Learning: How Representational Alignment Evolves Across Layers, Training, and Distribution Shifts}

\author{%
  Chaitanya Kapoor,\; Sudhanshu Srivastava,\; Meenakshi Khosla \\
  Department of Cognitive Science\\
  University of California, San Diego\\
  La Jolla, CA 92093 \\
  \texttt{\{chkapoor, sus021, mkhosla\}@ucsd.edu} \\
}

\begin{document}

\maketitle

\begin{abstract}
    Understanding convergent learning---the degree to which independently trained neural systems---whether multiple artificial networks or brains and models---arrive at similar internal representations---is crucial for both neuroscience and AI. Yet, the literature remains narrow in scope---typically examining just a handful of models with one dataset, relying on one alignment metric, and evaluating networks at a single post-training checkpoint. We present a large-scale audit of convergent learning, spanning dozens of vision models and thousands of layer-pair comparisons, to close these long-standing gaps.  First, we pit three alignment families against one another---linear regression (affine-invariant), orthogonal Procrustes (rotation-/reflection-invariant), and permutation/soft-matching (unit-order-invariant). We find that orthogonal transformations align representations nearly as effectively as more flexible linear ones, and although permutation scores are lower, they significantly exceed chance, indicating a privileged representational basis. Tracking convergence throughout training further shows that nearly all eventual alignment crystallizes within the first epoch---well before accuracy plateaus---indicating it is largely driven by shared input statistics and architectural biases, not by the final task solution. Finally, when models are challenged with a battery of out-of-distribution images, early layers remain tightly aligned, whereas deeper layers diverge in proportion to the distribution shift. These findings fill critical gaps in our understanding of representational convergence, with implications for neuroscience and AI.\blfootnote{All code is publicly available at: \href{https://github.com/NeuroML-Lab/representation-alignment}{https://github.com/NeuroML-Lab/representation-alignment}}
\end{abstract}

\section{Introduction}
Deep Neural Networks (DNNs) are becoming increasingly popular in neuroscience for predicting neural responses~\citep{yamins2014performance, yamins2016using, khaligh2014deep}, or as models for reverse-engineering algorithms of neural computation~\citep{schrimpf2018brain, schrimpf2020integrative, cichy2016comparison}. This congruence invokes the necessity to gain a deep understanding of how DNNs learn to represent information. A core question in this domain is whether independently trained networks converge on similar internal representations---and if so, under what conditions and along which dimensions this convergence unfolds? Comparative analysis of model representations helps reverse engineering neural networks by linking architectural components, training objectives, and data inputs to learned representations and, in turn, model behavior.
%Deep Neural Networks (DNNs) are becoming increasingly popular in neuroscience for predicting neural responses~\citep{yamins2014performance, yamins2016using, khaligh2014deep}, or as models for reverse-engineering algorithms of neural computation~\citep{schrimpf2018brain, schrimpf2020integrative, cichy2016comparison}. This congruence invokes the necessity to gain a deep understanding of how DNNs learn to represent information. A core question in this domain is whether independently trained networks converge on similar internal representations---and if so, under what conditions and along which dimensions this convergence unfolds? Comparative analysis of model representations helps reverse engineering neural networks by linking architectural components, training objectives, and data inputs to learned representations and, in turn, model behavior. For neuroscientific modeling using DNNs, a foundational question lies in ascertaining which aspects of their representations vary across different architectural choices, and which aspects---if any---are universal across such choices. Thus, studying this representational convergence has far-reaching implications. 

Over the past decade, there has been growing recognition that similar representations emerge across diverse models, despite differing in architecture, training procedures, or data modalities. Early work demonstrated that independent training runs of the same architecture develop a core set of features that align well across networks. For example, early layers in convolutional networks learn Gabor-like filters, across a range of architectures and tasks~\citep{yosinski2014transferable, krizhevsky2012imagenet}. Efforts to quantify these similarities have employed techniques such as canonical correlation analysis (CCA) and its variants~\citep{morcos2018insights, raghu2017svcca}, centered kernel alignment (CKA)~\citep{kornblith2019similarity}, representational similarity analysis (RSA)~\citep{mehrer2020individual} and model stitching~\citep{bansal2021revisiting}. These studies have underscored the high degree of alignment in early layers and noted convergence even in later stages of the network. These findings provide empirical grounding for theories of representational convergence, hinting at the existence of universal principles governing learning, which may also shed light on how biological neural circuits process information.

%Over the past decade, there has been growing recognition that similar representations can emerge across diverse models, despite models differing in architecture, training procedures, or data modalities. Early work demonstrated that independent training runs of the same architecture develop a core set of features that align well across networks. For example, early layers in convolutional networks learn Gabor-like filters, across a range of architectures and tasks~\citep{yosinski2014transferable, krizhevsky2012imagenet}. Efforts to quantify these similarities have employed techniques such as canonical correlation analysis (CCA) and its variants~\citep{morcos2018insights, raghu2017svcca}, centered kernel alignment (CKA)~\citep{kornblith2019similarity}, representational similarity analysis (RSA)~\citep{mehrer2020individual} and model stitching~\citep{bansal2021revisiting}. These studies have underscored the high degree of alignment in early layers and noted convergence even in later stages of the network. These findings provide empirical grounding for theories of representational convergence, hinting at the existence of universal principles governing learning, which may also shed light on how biological neural circuits process information.

Recent studies have extended this line of inquiry, demonstrating that increasing model capability---through increased scale, multitask training, or cross-modal learning---drives representational convergence not just within, but also \emph{across} modalities~\citep{moschella2023relative}. The Platonic Representation Hypothesis further argues that as models scale and solve more tasks, they are driven to discover a universal, modality-agnostic representation of reality~\citep{huh2024platonic}. Yet, identifying precise conditions of network convergence on similar representations and its implications remain open questions.

%Recent studies have extended this line of inquiry, demonstrating that increasing model capability---through increased scale, multitask training, or cross-modal learning---their representations converge not only within a single modality, but also across modalities ~\citep{moschella2023relative}. For instance, the Platonic Representation Hypothesis argues that as models solve a larger number of tasks and scale up in capacity, they are pressured into discovering a universal, modality-agnostic representation of the underlying reality~\citep{huh2024platonic}. Yet, many open questions remain about the conditions under which different networks converge on similar representations and the implications of such convergence.
To address these gaps, in our work, we examine representational alignment along three key axes:\\
\newline
\textbf{Across layers:} %One important axis of inquiry is representational alignment across layers. Prior work has shown that early layers tend to extract general, low-level features---such as edge detectors in vision networks---while deeper layers develop more task-specific representations. Studies by~\citep{kornblith2019similarity},~\citep{mehrer2020individual}, and~\citep{li2015convergent} have quantified these changes using methods like canonical correlation analysis (CCA) and CKA. However, these approaches often rely on single metrics that obscure the minimal transformations needed to align representations. Understanding the precise nature of these transformations is critical for dissecting how representations in different networks relate to each other (\emph{e.g.}, are they similar in information content, representational geometry, or even at the level of single-neuron tuning?). It is also unknown whether the hierarchical (layer-wise correspondence) results hold true for other metrics with more restricted invariances than affine transformations.
Prior work has shown that early layers tend to extract general, low-level features---such as edge detectors in vision networks---while deeper layers develop task-specific representations. Studies~\citep{kornblith2019similarity, mehrer2020individual, li2015convergent} have quantified these changes using methods like CCA and CKA. However, these approaches often rely on single metrics that obscure minimal transformations needed to align representations. Understanding the precise nature of these transformations is critical for dissecting how representations in different networks relate to each other (\emph{e.g.}, are they similar in information content, representational geometry, or even at the level of single-neuron tuning?). It is also unknown whether hierarchical (layer-wise) correspondence holds for other metrics with more restricted invariances than affine transformations.\\
\newline
\textbf{Across training:} %A second key question concerns the evolution of representational convergence over the course of training. Most studies compare networks at their converged state. Yet, a deeper understanding requires examining not only the final representations, but also the learning trajectories that lead there. Conventionally, it is assumed that as different networks optimize on a task, their internal representations become more similar, driven by the final task solution. This assumption is the basis for the \emph{contravariance principle}~\citep{cao2021explanatory}, which posits that when a network is pushed to achieve a challenging task, there is less room for variation in the final solution, forcing representations to converge. However, the question of when representational convergence occurs \emph{during} training remains underexplored. Understanding this dynamic can illuminate the roles of initialization, early data statistics, architectural biases, learning dynamics, and the final task solution in shaping alignment. Previous studies have shown a \emph{``lower layers learn first''}~\citep{raghu2017svcca} behavior by comparing layers across time on CIFAR-10 using SVCCA, but little else is known about the dynamics of convergence, especially for more complex vision networks and using other metrics. Recent theoretical studies such as~\citep{domine2024lazy},~\citep{saxe2013exact} present an analytical framework for the temporal dynamics of learning in deep \emph{linear} networks, distinguishing an early \emph{``lazy''} phase, characterized by kernel regression dynamics~\citep{jacot2018neural}---from a subsequent \emph{``rich''} phase characterized by non-linear feature learning. While these models provide valuable insights into training dynamics and convergence in linear regimes, they fall short of fully explaining representational alignment in complex, non-linear architectures. Though deep linear models are instructive, further work is needed to elucidate how representational alignment develops in and generalizes to the rich, nonlinear regimes typical of contemporary neural networks.
A second key question is \emph{when} convergence emerges. Most studies snapshot networks only after training has finished, but revealing the mechanisms behind alignment demands tracking how representations co-evolve during training, not just examining the final state. Conventionally, it is assumed that as different networks optimize on a task, their internal representations become more similar, driven by the final task solution. This assumption is the basis for the \emph{contravariance principle}~\citep{cao2021explanatory}, which posits that when a network is pushed to achieve a difficult task, there is less room for variation in the final solution, forcing representations to converge. However, the question of when representational convergence occurs \emph{during} training remains underexplored. Understanding this dynamic can illuminate the roles of initialization, early data statistics, architectural biases, learning dynamics, and the final task solution in shaping alignment. Previous studies have shown a \emph{``lower layers learn first''}~\citep{raghu2017svcca} behavior by comparing layers over time on CIFAR-10 using SVCCA, but little is known about the dynamics of convergence, especially for complex vision networks and using other metrics. More recently, Atanasov et al.~\citep{atanasov2021neural} described a \emph{``silent alignment effect''} in which a network's output aligns with the target function early in training well before the loss falls. Theoretical studies such as~\citep{braun2025not, domine2024lazy, saxe2013exact} present an analytical framework for the temporal dynamics of learning in deep \emph{linear} networks, distinguishing an early \emph{``lazy''} phase, characterized by kernel regression dynamics~\citep{jacot2018neural}---from a subsequent \emph{``rich''} phase characterized by non-linear feature learning. While these models provide valuable insights into training dynamics and convergence in linear regimes, they fall short of fully explaining representational alignment in non-linear architectures. Though deep linear models provide valuable theoretical traction, further work is needed to elucidate how representational alignment develops and generalizes to the rich, non-linear regimes typical of contemporary DNNs.\\
\newline
\textbf{Across distribution shifts:} Although many DNNs exhibit highly human-like responses to in-distribution stimuli, there is mounting evidence that their responses can diverge dramatically under out-of-distribution (OOD) conditions~\citep{prasad2022exploring, geirhos2021partial, geirhos2018generalisation}. Despite this, the effect of OOD inputs on model-to-model representational convergence remains poorly understood. Exploring this axis is crucial for rigorously testing the universality of learned representations.\\
\newline

\textbf{Key contributions.}  In this work, we address these critical gaps by performing a large-scale systematic audit of representational convergence along these three axes. First, we employ three alignment metrics---linear regression~\citep{yamins2014performance, schrimpf2018brain, zhuang2021unsupervised, kubilius2019brain}, Procrustes analysis~\citep{sucholutsky2023getting, williams2021generalized}, and permutation-based methods~\citep{li2015convergent, khosla2024soft, khosla2024privileged}---each with different restrictions on the freedom of the mapping function, to identify the minimal set of transformations needed to align representations across networks reasonably well for each layer. This approach allows us to dissect how representations relate to each other, whether in terms of information content, representational geometry, or single-neuron tuning. Second, we examine the temporal dynamics of convergence during training, revealing that nearly all alignment occurs within the first epoch, challenging the assumption that convergence is tied to task-specific learning. Finally, we explore how changes in input statistics affect representational alignment across layers, demonstrating how OOD inputs differentially affect later versus early layers.

\section{Problem Statement}
We consider two representations $
\bm{X}_i\in \mathbb{R}^{M\times N_x} \quad \text{and} \quad \bm{X}_j\in \mathbb{R}^{M\times N_y}$
obtained from different models over \(M\) unique stimuli, where \(N_x\) and \(N_y\) denote the number of neurons (or units) in each representation, respectively. To systematically identify the minimal transformations needed for alignment, we use three metrics that quantify similarities between networks while ignoring nuisance transformations. Among these are \textbf{(a)} the \textbf{linear regression score} which discounts affine transformations, \textbf{(b)} the \textbf{Procrustes score} which discounts rotations and reflections, treating them as nuisance factors and \textbf{(c)} the \textbf{permutation score} which considers the order of units in the representations as arbitrary. These metrics are ordered to reflect progressively less permissive mapping functions (from flexible to strict). When representations have an equal number of neurons $(N_x=N_y=N)$, each of these similarity metrics seeks to minimize their Euclidean distance by optimizing over a set of $N\times N$-dimensional mapping matrices $\bm{M}$, $\underset{\bm{M}}{\text{min}} \Vert \bm{X}_i - \bm{MX}_j\Vert_2^2$. Linear regression score imposes no constraint on $\bm{M}$, Procrustes score enforces the matrices to be orthogonal (i.e. {$\bm{M} \in \mathbf{O}(N)$}) and permutation score requires the matrices to be permutation matrices {$\left(\bm{W} \in \mathcal{P}(N)\right)$}). When \(N_x \neq N_y\), we use a generalized version of permutation-based alignment called the soft-matching score~\citep{khosla2024soft}. Here, the mapping matrix \(\bm{M} \in \mathbb{R}^{N_x \times N_y}\) is constrained such that its entries are nonnegative and satisfy $\sum_{j=1}^{N_y} \bm{M}_{ij} = \frac{1}{N_x} \quad \forall i = 1, \dots, N_x$, $
    \sum_{i=1}^{N_x} \bm{M}_{ij} = \frac{1}{N_y} \quad \forall j = 1, \dots, N_y.$ These constraints place \(\bm{M}\) in the transportation polytope \(\mathcal{T}(N_x, N_y)\)~\cite{de2013combinatorics}.

Once the optimal mapping matrix \(\bm{M}\) is computed for each metric, we report the alignment using the pairwise correlation $
\texttt{Alignment} = \texttt{corr}\left(\bm{X}_j,\ \bm{M}\bm{X}_i\right).$ For asymmetric metrics, we report the average alignment score computed in both directions: \texttt{corr}$\left(\bm{X}_j,\  \bm{M}_1\bm{X}_{i}\right)$ and \texttt{corr}$\left(\bm{X}_i,\ \bm{M}_2\bm{X}_{j}\right)$, where $\bm{M}_1$ and $\bm{M}_2$ are the respective transformations. We also report our results using Spearman's rank correlation coefficient in Appendix~\ref{sec: spearman-results} to address the possibility of being susceptible to high-variance activation dimensions.

\noindent
By leveraging these metrics---which progressively relax mapping constraints from strict (soft-matching/permutation) to flexible (linear regression)---we can dissect the nature of representational alignment across networks, distinguishing between similarity in representational form (captured by the soft-matching score, which reduces to the permutation score when \(N_x = N_y\)), geometric shape (captured by the Procrustes score), and information content (captured by the linear regression score). 

\section{Method}
Below, we outline our framework for evaluating alignment between different deep convolutional vision models (ResNet$18$, ResNet$50$~\citep{he2016deep}, VGG$16$, VGG$19$~\citep{simonyan2014very}), with their training procedures described in Appendix~\ref{sec: network-training}. We demonstrate the robustness of all these results on CIFAR100 in Appendix~\ref{sec: all-cifar100-results}.
%Below, we outline our framework for evaluating alignment between different deep convolutional vision models (ResNet18, ResNet50~\citep{he2016deep}, VGG16, VGG19~\citep{simonyan2014very}), with their training procedures described in Appendix~\ref{sec: network-training}. We also demonstrate the robustness of these results on self-supervised vision networks (Appendix~\ref{sec: self-sup-alignment}), vision transformers (Appendix~\ref{sec: vit-alignment}) and language models (Appendix~\ref{sec: lang-alignment}).
%\subsection{Network Training}
%Consider a model type denoted by $M$. We train a pair of models, $\{M_1, M_2\}$ initialized with two different random seeds. We initialize the models using a uniform Xavier distribution~\citep{glorot2010understanding}. This setup ensures that the two models are identical in architecture and achieve comparable task performance, allowing us to isolate the effects of stochastic variations in the SGD process (such as initialization differences and input order). By comparing the representations from these models, we can quantify the minimal set of transformations required to align them. All models are trained from scratch on CIFAR100 and ImageNet for $100$ and $80$ epochs respectively. We save model weights at every epoch and additionally store the best-performing weights based on test-set performance for each dataset.
%

\subsection{Comparing Convolutional Layers}
For a given convolutional layer, let the activations be represented by $\bm{X} \in \mathbb{R}^{m \times h \times w \times c}$, where \(m\) is the number of stimuli, \(h\) and \(w\) denote the spatial height and width, and \(c\) is the number of channels (i.e., convolutional filters). Each convolutional layer produces a feature map whose spatial dimensions are equivariant to translations. That is, a circular shift along the spatial dimensions yields an equivalent representation (up to a shift). As a result, one could compare the full spatial activation patterns between networks by considering an equivalence relation that allows for spatial shifts. However, evaluating an alignment that optimizes over all possible shifts together with another alignment (\emph{e.g.}, Procrustes) is computationally costly.

Previous work has shown that optimal spatial shift in convolutional layers tend to be close to zero~\citep{williams2021generalized}. This motivates our simpler approach: rather than collapsing the spatial dimensions by flattening the entire feature map (which would yield $\bm{X}\in\mathbb{R}^{m \times (h \cdot w \cdot c)}$), we extract a single representative value from each channel. In our experiments, we choose the value at the center pixel of each channel, reducing the activation tensor \(\bm{X}\) to a two-dimensional matrix $\bm{X}^{\prime} \in \mathbb{R}^{m \times c}$, where each row corresponds to a stimulus and each column to a channel. This drastically reduces the computational complexity of computing the optimal mapping. For instance, aligning full, spatially-flattened representations would incur a runtime of \(\mathcal{O}(m h^2w^2c + h^3w^3c^3)\), whereas our center-pixel approach reduces the problem to aligning an \(m \times c\) matrix, resulting in a more tractable complexity.
%Previous work has demonstrated that in many convolutional layers the optimal spatial shift parameters tend to be close to zero~\citep{williams2021generalized}. This motivates our simpler approach: rather than collapsing the spatial dimensions by flattening the entire feature map (which would yield an activation matrix of size \(m \times (h \cdot w \cdot c)\)), we instead extract a single representative value from each channel. In our experiments, we choose the value at the center pixel of each channel. This reduces the activation tensor \(\bm{X}\) to a two-dimensional matrix $\bm{X}^{\prime} \in \mathbb{R}^{m \times c}$, where each row corresponds to a stimulus and each column to a channel. This strategy drastically reduces the computational complexity of computing the optimal mapping. For instance, aligning the full spatially flattened representations would incur a runtime of \(\mathcal{O}(m h^2w^2c + h^3w^3c^3)\), whereas our center-pixel approach reduces the problem to aligning an \(m \times c\) matrix, resulting in a more tractable complexity.

\subsection{Computing Alignment}

Alignment was quantified in four regimes:  
(1) \textbf{Within–architecture:} each layer aligned with its counterpart in an independently initialized instance of the same architecture (trained on CIFAR100 or ImageNet);  
(2) \textbf{Across architectures:} all layer–layer alignments between different architectures;  
(3) \textbf{Across training:} analysis repeated after each of the first ten epochs;  
(4) \textbf{Under distribution shift:} within–architecture alignments tested on ImageNet-trained models using 17 OOD datasets from~\citep{geirhos2021partial}.  
All scores are averaged over five-fold cross-validation $(k=5)$.

\section{Results}
\label{sec: dcnn-results}

\subsection{Evolution of Convergence Across the Network Hierarchy}
\label{sec: convergence-evolution}

\begin{figure*}[htbp!]
    \centering
    % cifar100
    %\includegraphics[width=.25\textwidth]{figures/network-hierarchy/r18_cifar100.png}\hfill
    %\includegraphics[width=.25\textwidth]{figures/network-hierarchy/r50_cifar100.png}\hfill
    %\includegraphics[width=.25\textwidth]{figures/network-hierarchy/vgg16_cifar100.png}\hfill
    %\includegraphics[width=.25\textwidth]{figures/network-hierarchy/vgg19_cifar100.png}\\
    % imagenet
    \includegraphics[width=.25\textwidth]{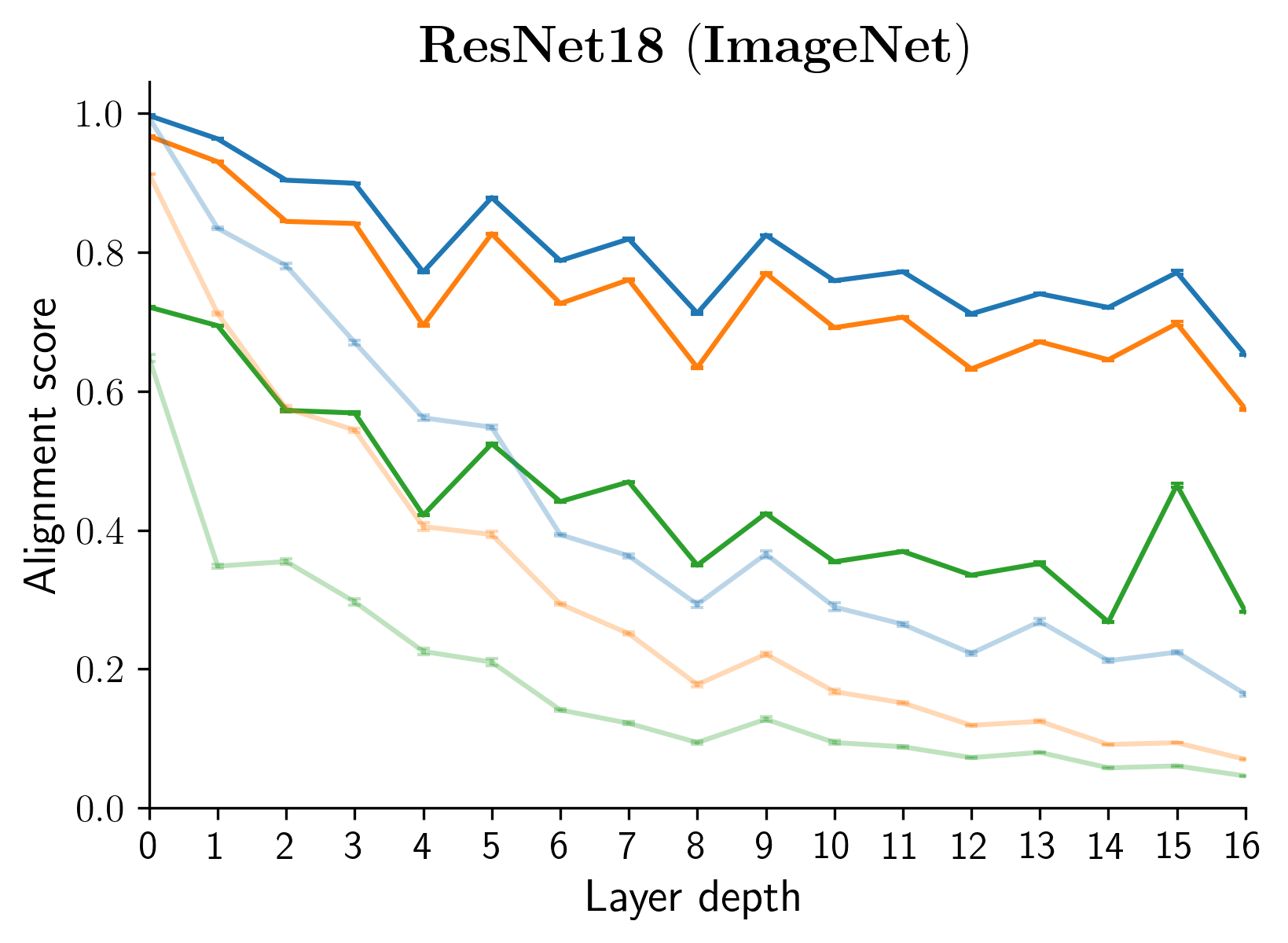}\hfill
    \includegraphics[width=.25\textwidth]{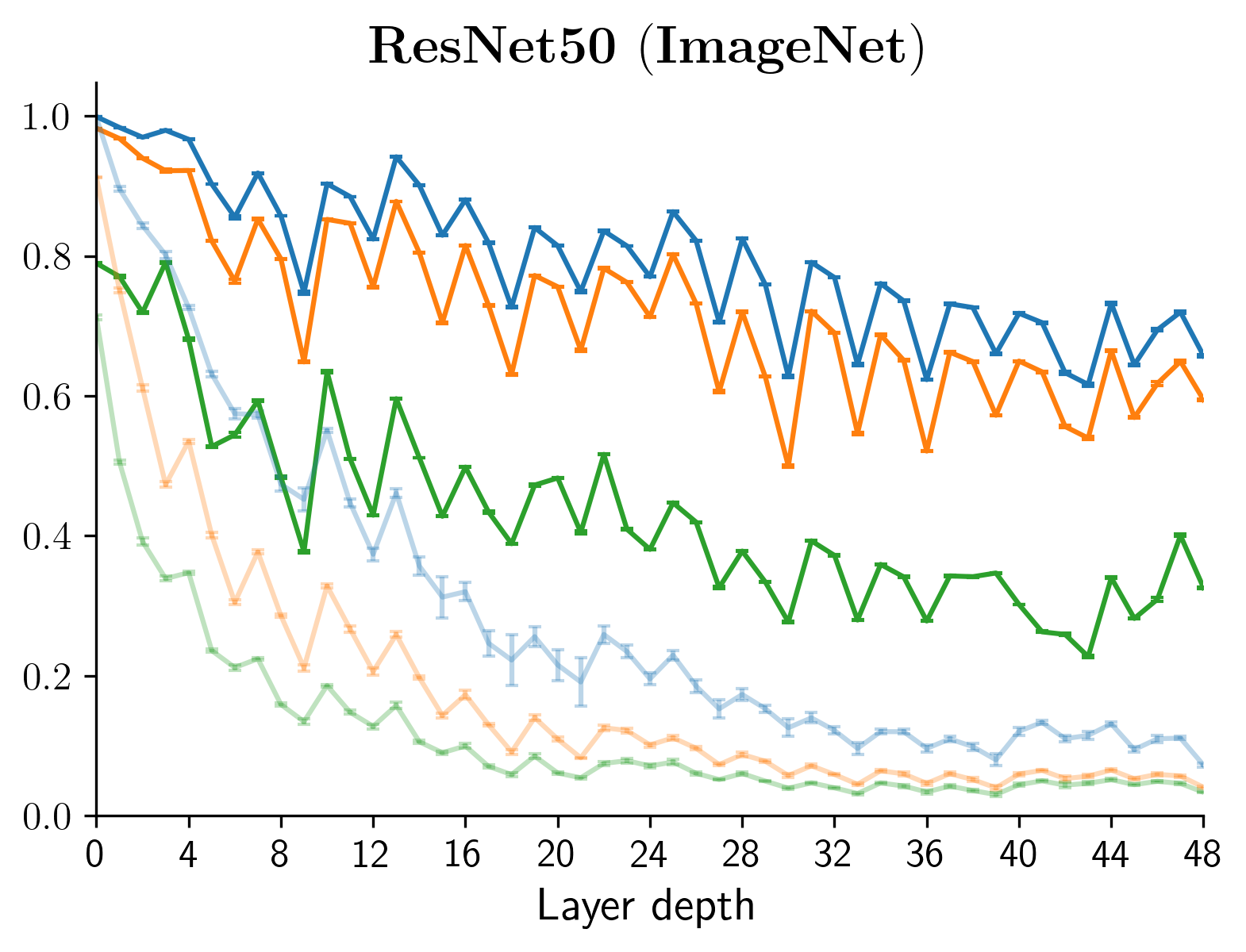}\hfill
    \includegraphics[width=.25\textwidth]{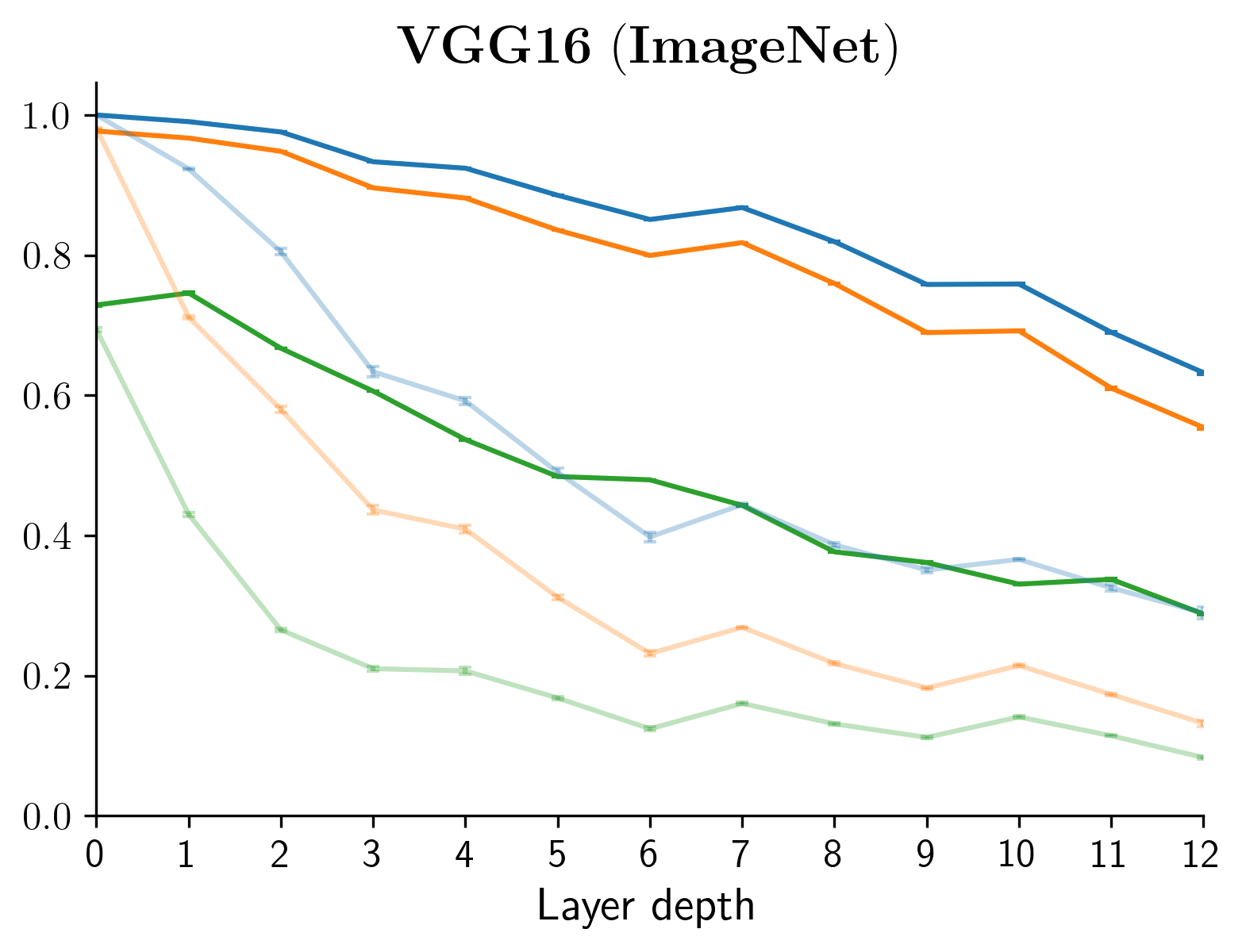}\hfill
    \includegraphics[width=.25\textwidth]{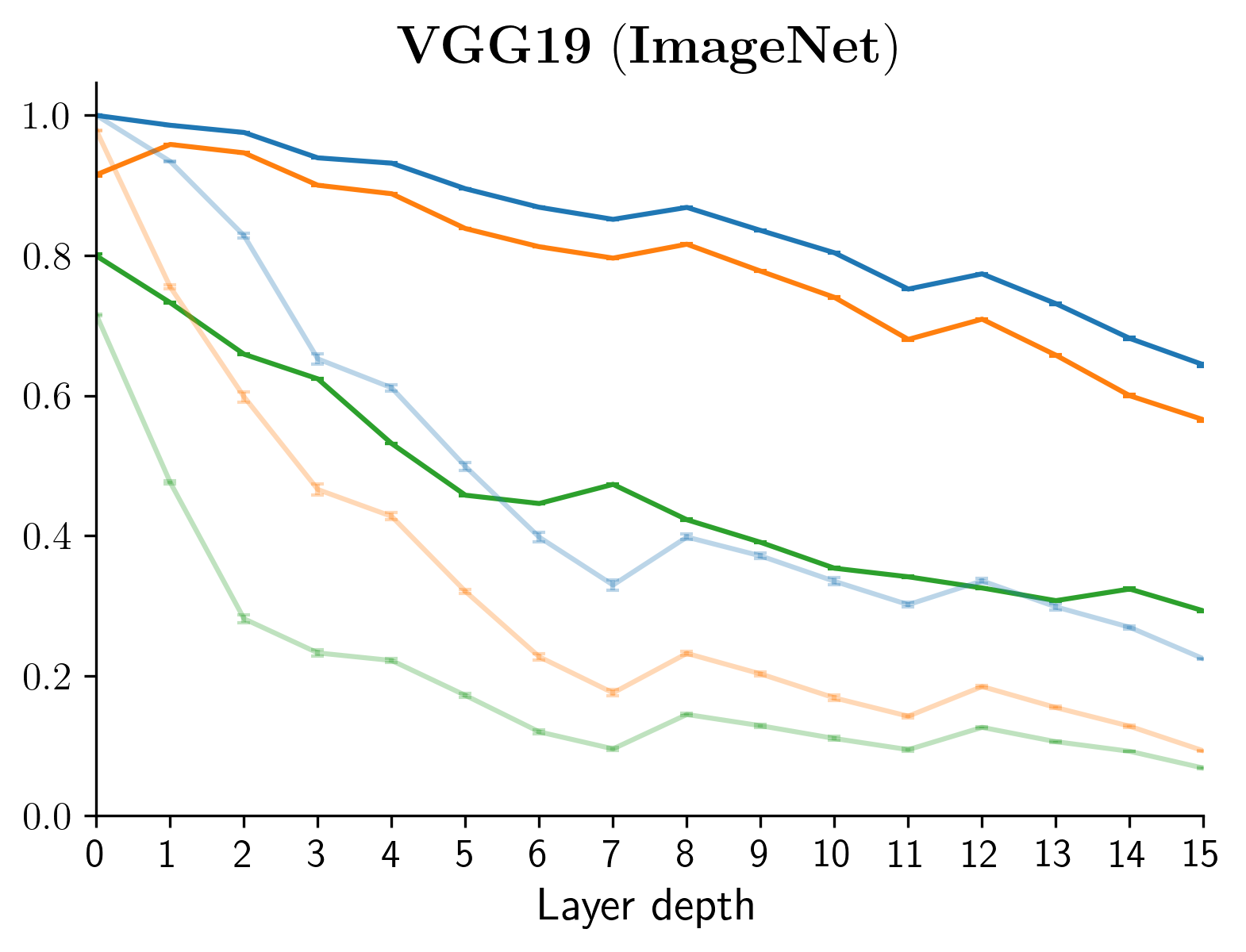}
    
    \caption{\textbf{Representational Convergence Across a Network Hierarchy.} We plot the evolution of alignment scores (computed between different seeds of the same network architecture) across the network hierarchy for four vision network architectures trained ImageNet. A consistent downward trend across layers indicates decreasing representational convergence as networks deepen. Alignment consistently follows the order: \textcolor{mplblue}{Linear} $>$ \textcolor{mplorange}{Procrustes} $>$ \textcolor{mplgreen}{Permutation}, reflecting the progressively stricter nature of the metrics. Lighter shades of the same color denote alignment for random networks. Notably, \textcolor{mplorange}{Procrustes} transformations align representations nearly as well as \textcolor{mplblue}{Linear} transformations, suggesting that most variability is due to rotations rather than more complex transformations. Even \textcolor{mplgreen}{Permutation} scores---despite their strictness---achieve substantial alignment, indicating a strong one-to-one correspondence between neurons across seeds, which points to stable, convergent neuron-level representations. Error bars represent the standard deviation computed across $5$-fold cross-validation.}
    \label{fig:network-hierarchy}
\end{figure*}

\paragraph{How convergence varies with network depth.}
When comparing representational convergence across the network hierarchy for different seeds of the same architecture, we observe that convergence is strongest in the earliest layers and gradually diminishes in deeper layers (Fig.~\ref{fig:network-hierarchy}). This pattern is consistent across all three metrics and across networks trained on ImageNet. The high alignment in early layers likely arises because they capture fundamental, low-frequency features (\emph{e.g.,} edges, corners, contrast) that are universal across representations~\citep{rahaman2019spectral, bau2017network, zeiler2014visualizing}. In contrast, deeper layers, while still showing significant alignment $(> 0.5)$, exhibit greater variability due to their sensitivity to specific training conditions and noise. We also contrast this result with randomly initialized (untrained) networks as a baseline. We find that alignment scores across all metrics are consistently lower in untrained networks compared to their trained counterparts. This difference is especially pronounced in deeper layers: for instance, using the Procrustes metric, the mean alignment increases by $145.26\%$ in early layers (depth $<0.5$) and by $493.84\%$ in deeper layers (depth $>0.5$) following training. This trend also holds when comparing networks with different architectures, underscoring the robustness of hierarchical convergence across diverse models. Interestingly, a similar hierarchical trend is observed in human brain responses to visual stimuli (Appendix~\ref{sec: nsd-comparisons}).

% training evolution
%\input{./figures-code/inter-model-proc}
\begin{figure*}[htbp!]
    \centering
    % procrustes
    \includegraphics[width=.16\textwidth]{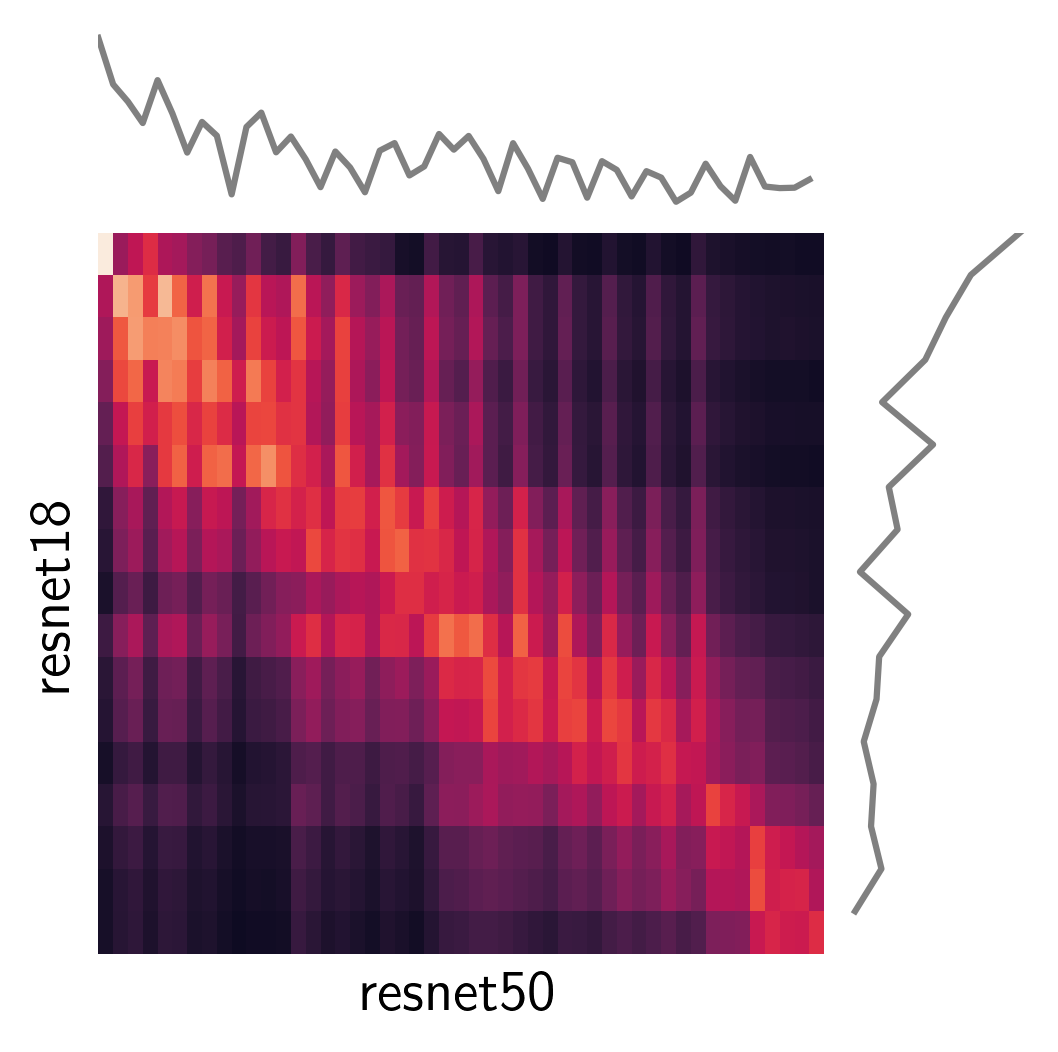}\hfill
    \includegraphics[width=.16\textwidth]{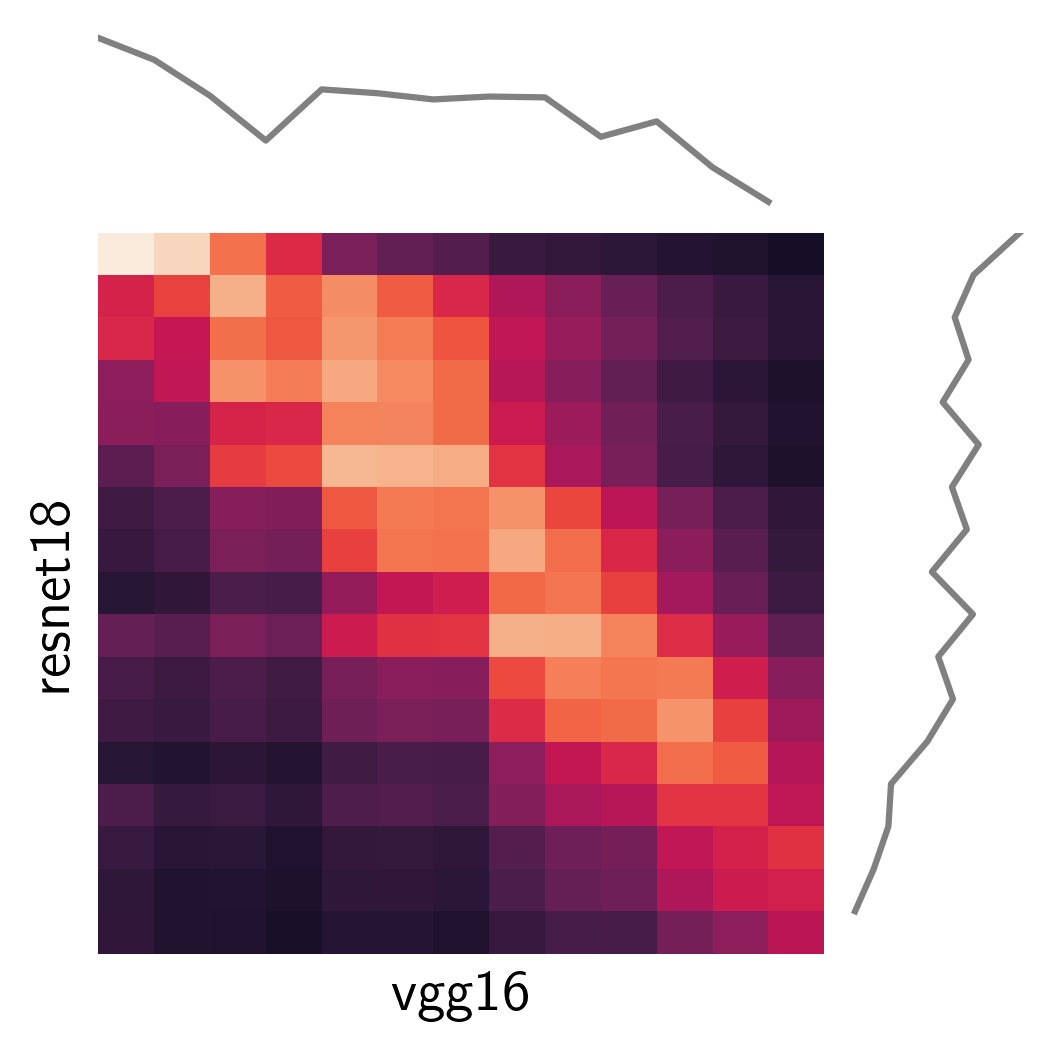}\hfill
    \includegraphics[width=.16\textwidth]{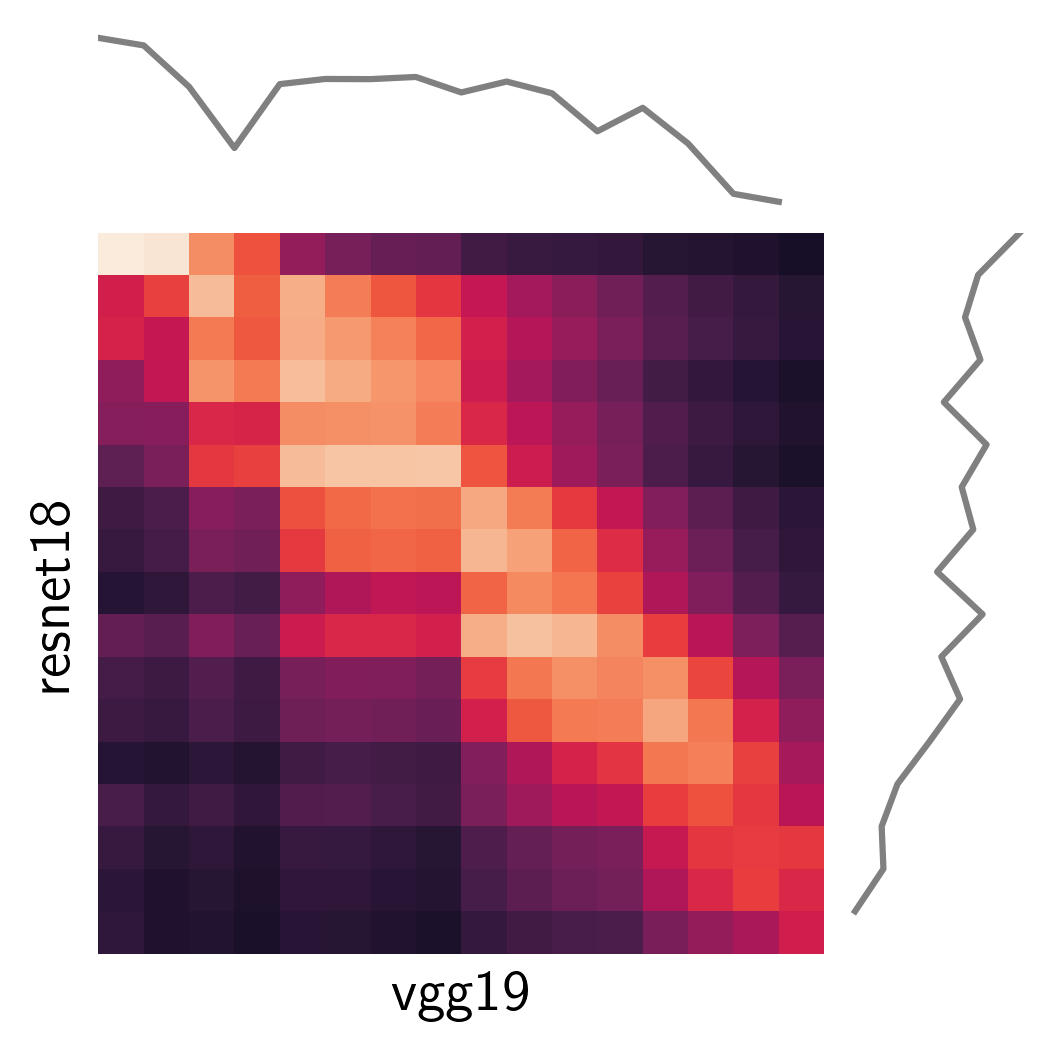}\hfill 
    \includegraphics[width=.16\textwidth]{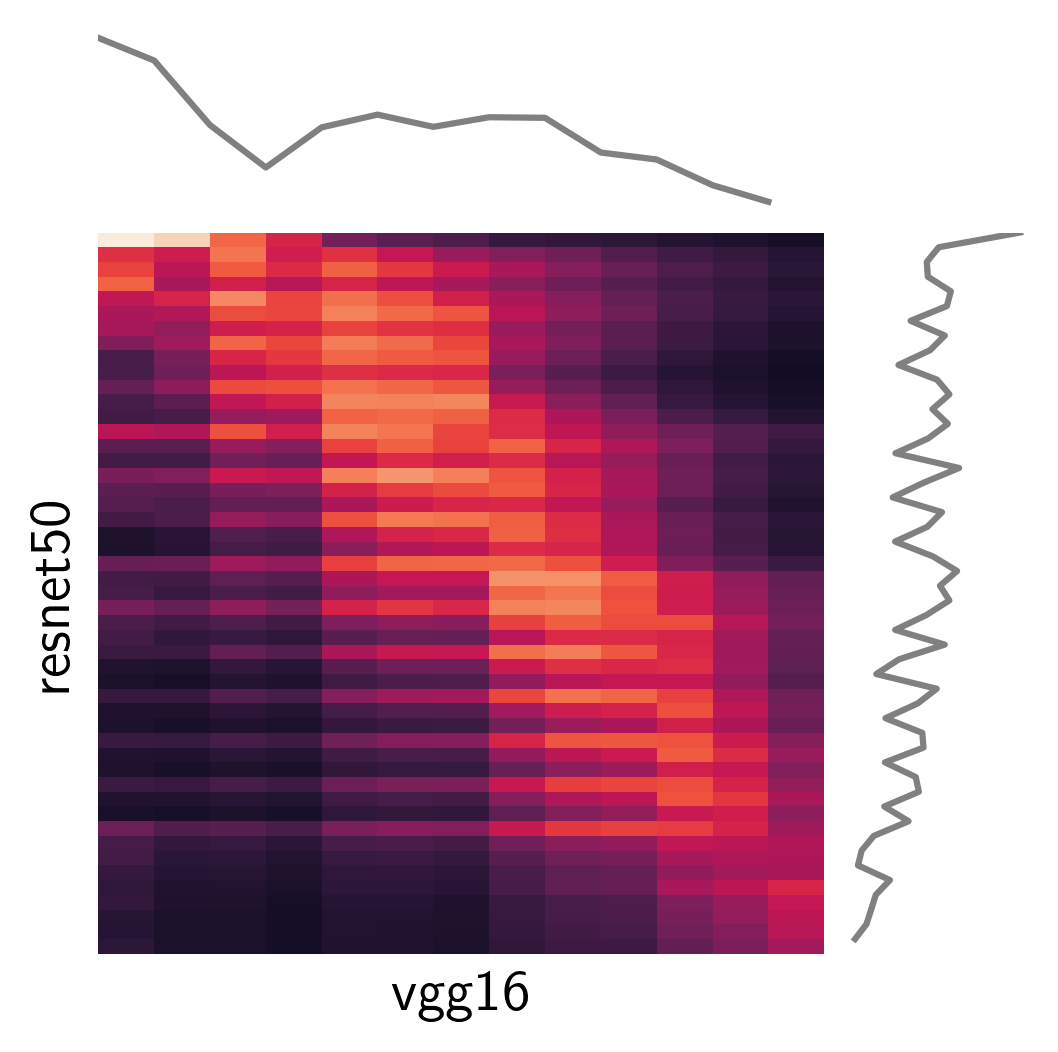}\hfill
    \includegraphics[width=.16\textwidth]{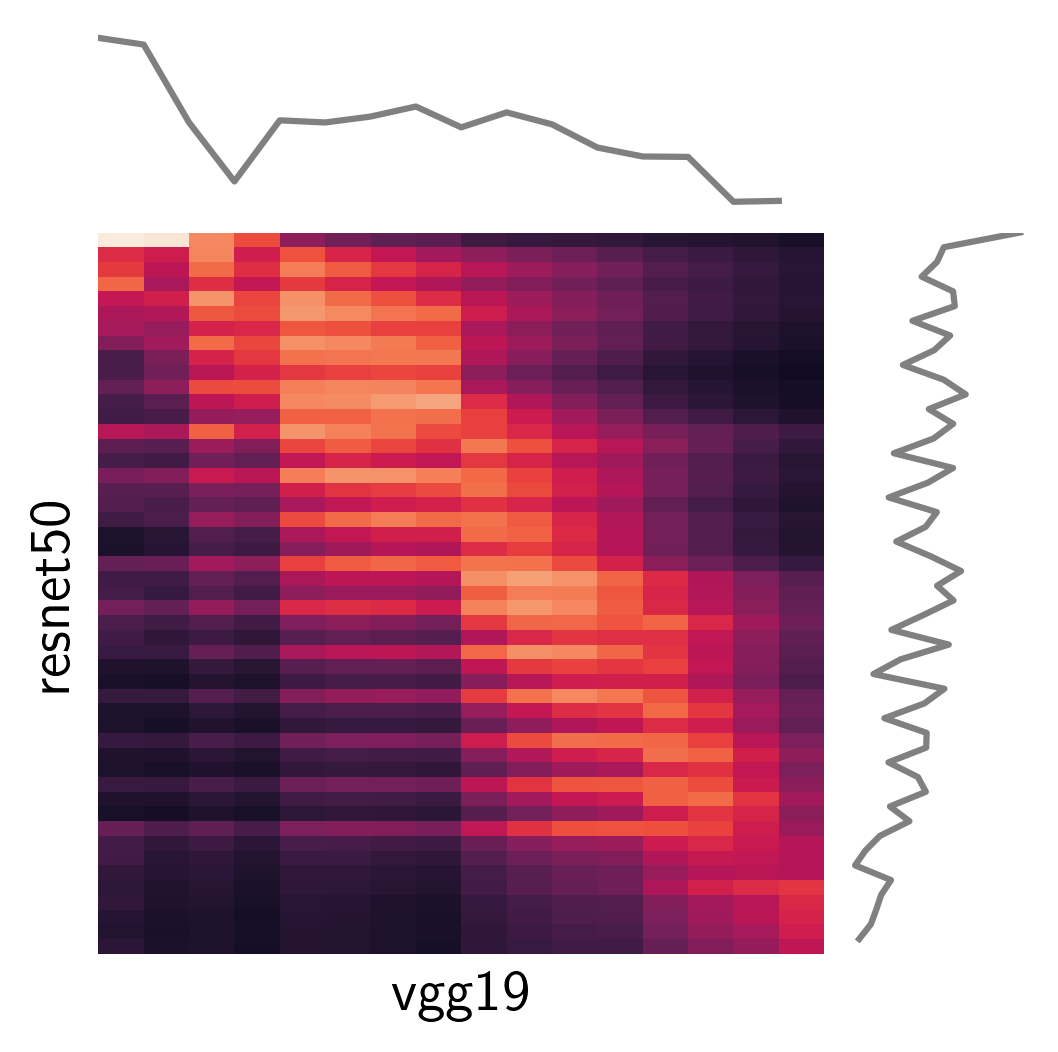}\hfill 
    \includegraphics[width=.16\textwidth]{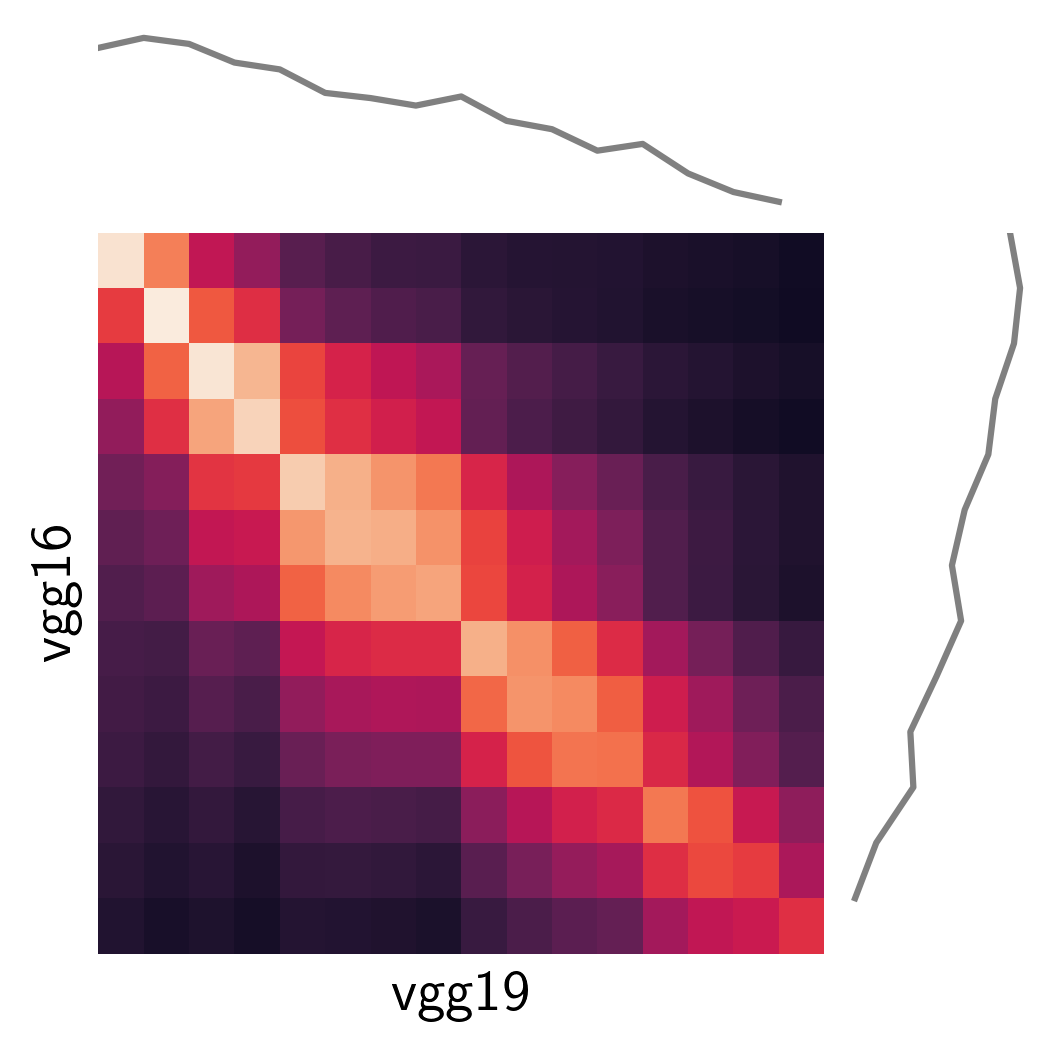}\\

    % soft-matching
    \includegraphics[width=.16\textwidth]{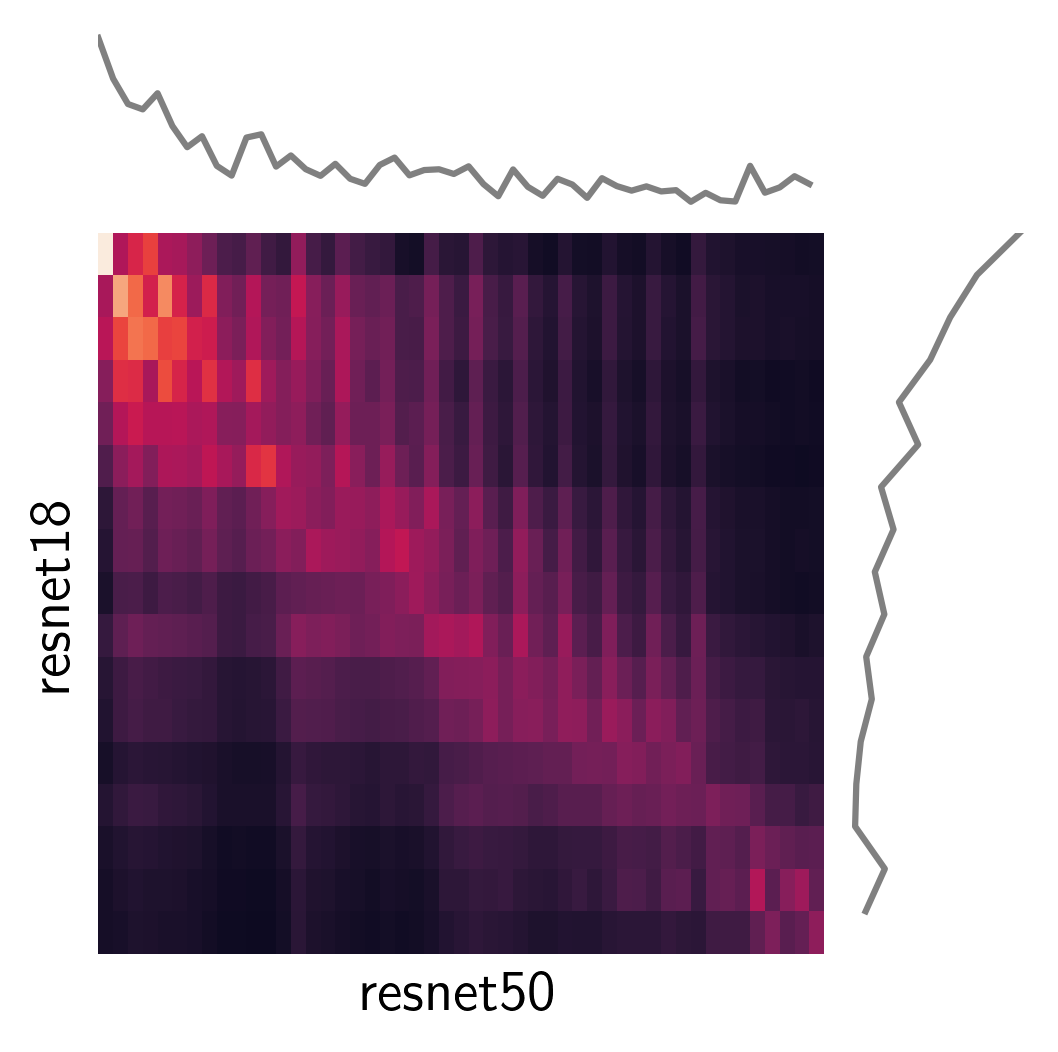}\hfill
    \includegraphics[width=.16\textwidth]{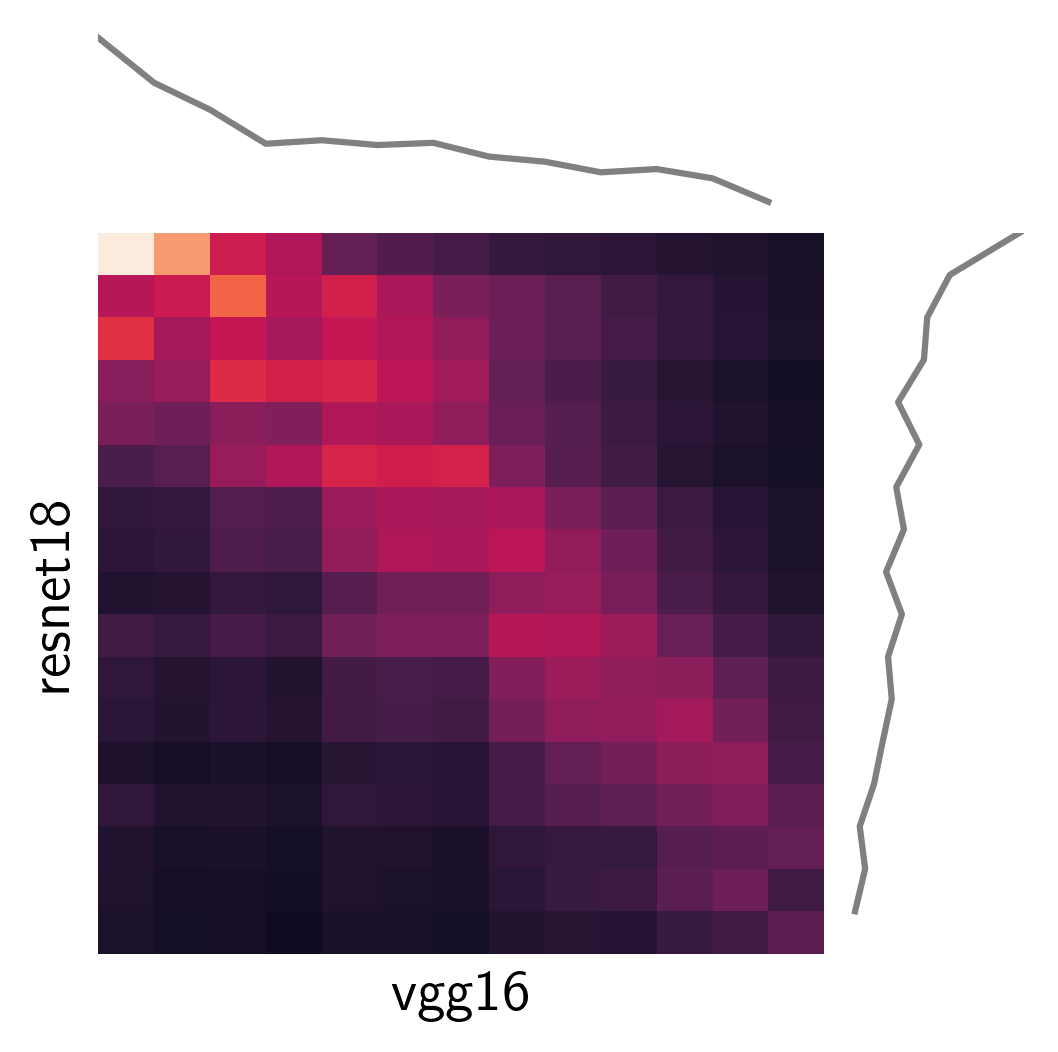}\hfill
    \includegraphics[width=.16\textwidth]{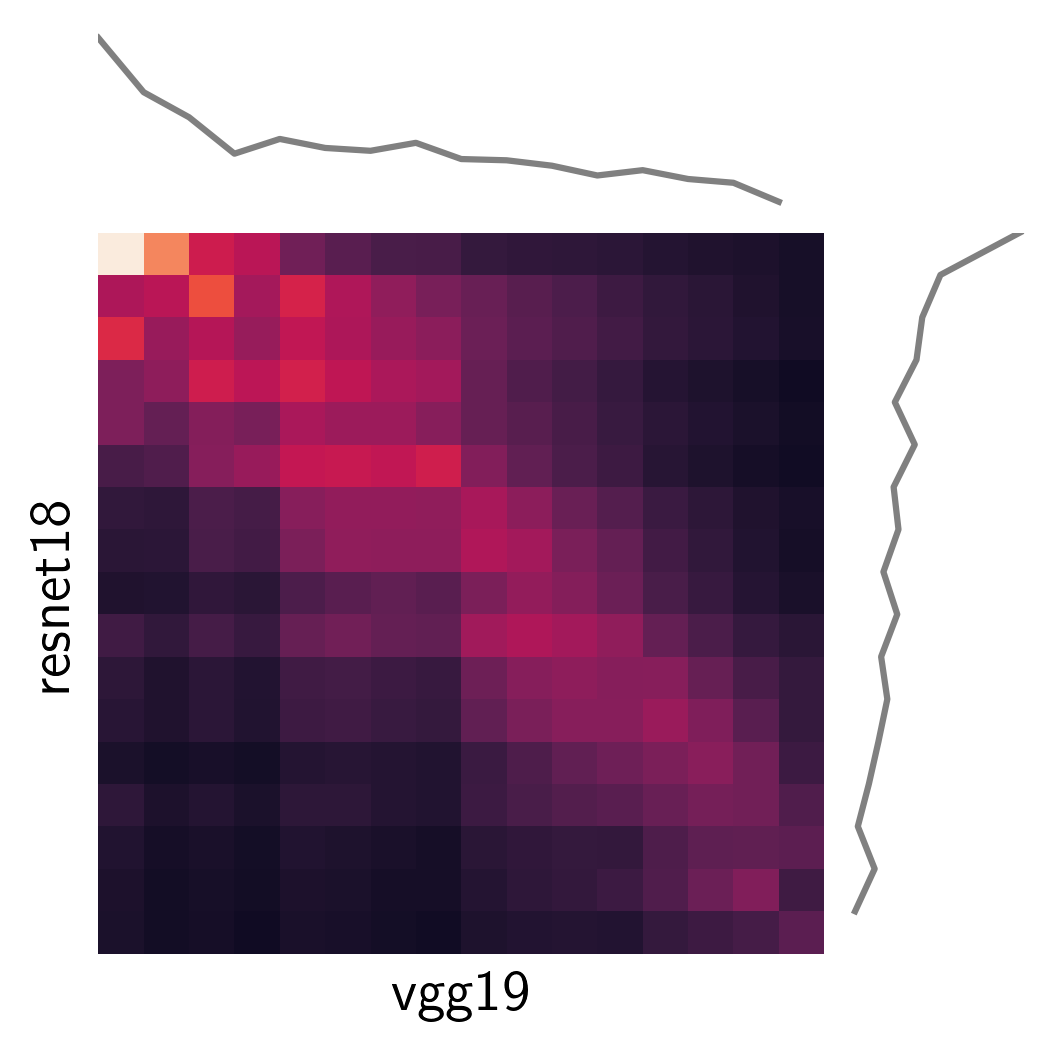}\hfill 
    \includegraphics[width=.16\textwidth]{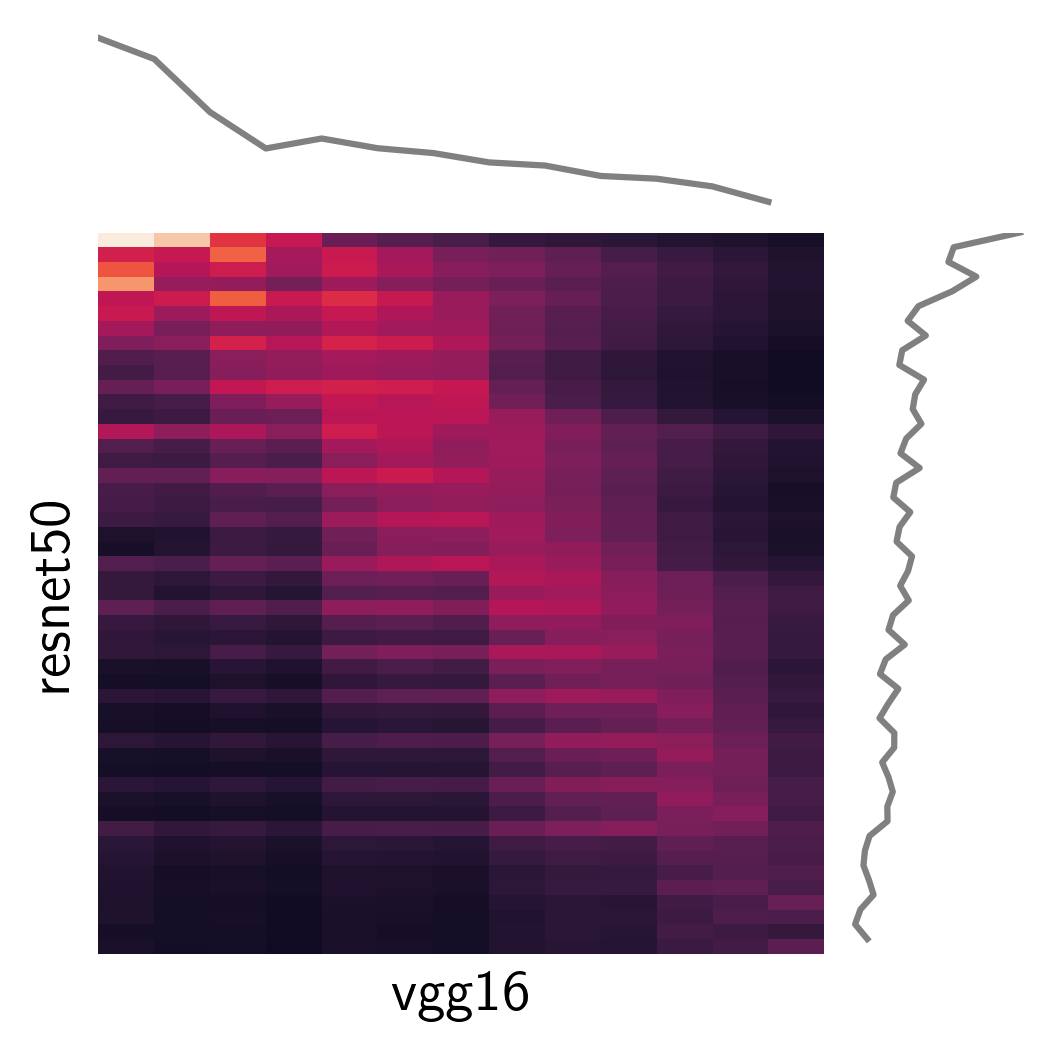}\hfill
    \includegraphics[width=.16\textwidth]{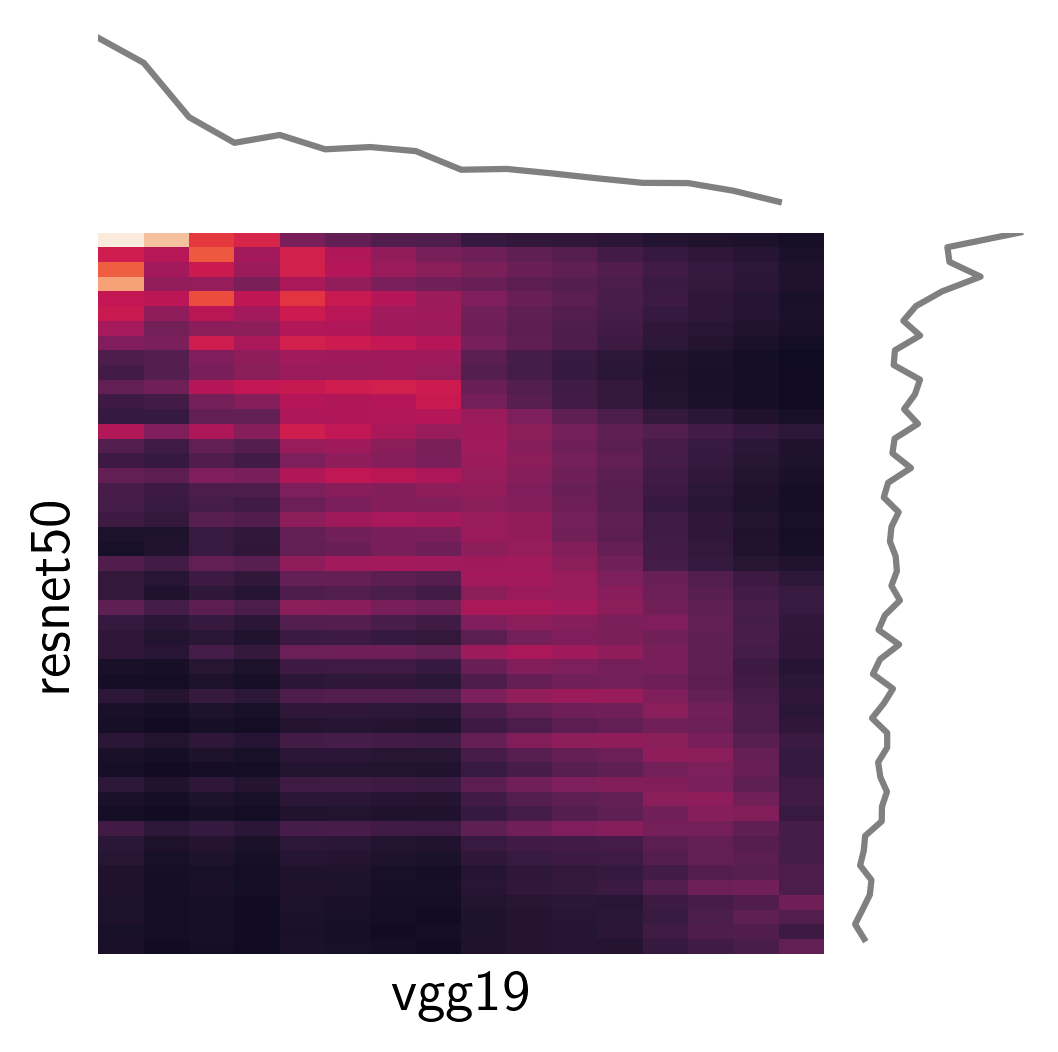}\hfill 
    \includegraphics[width=.16\textwidth]{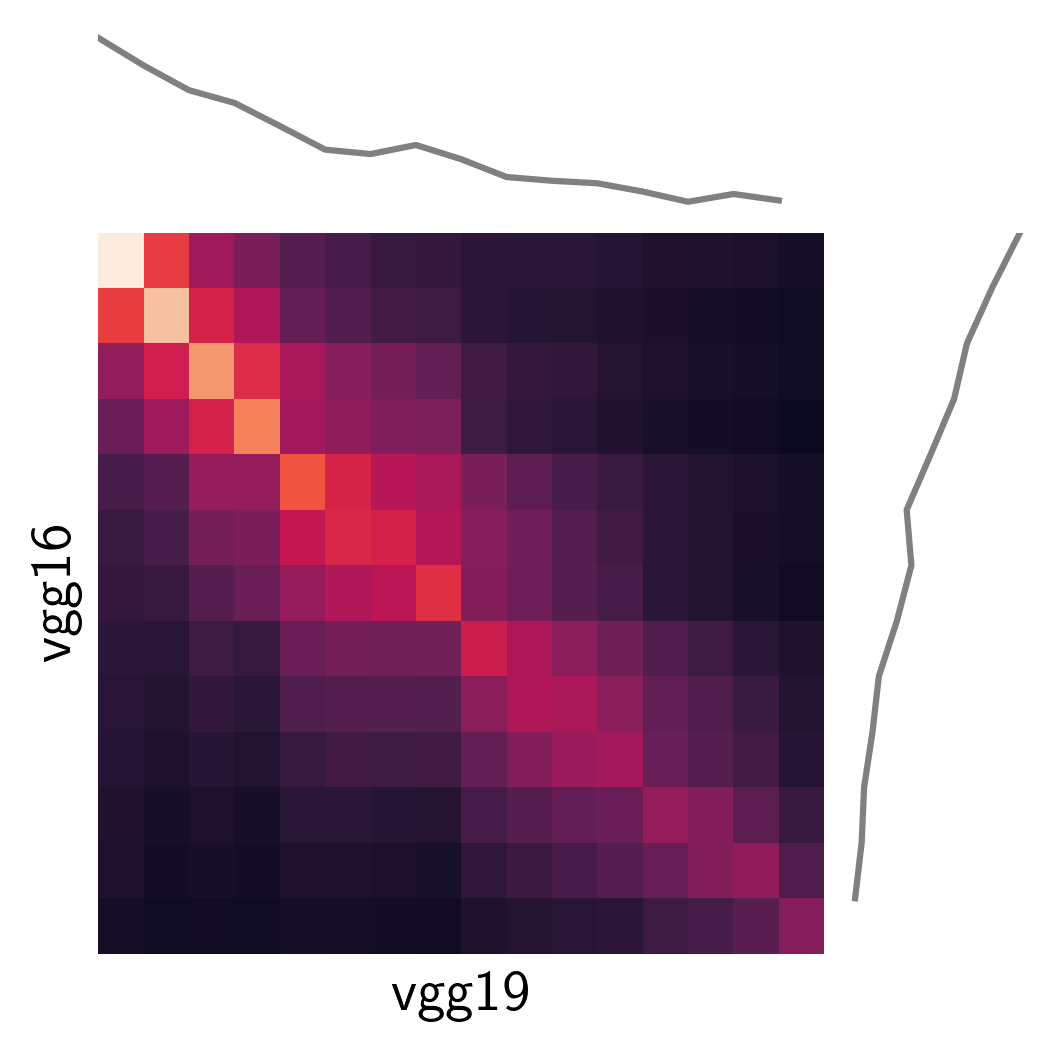}
    
    \caption{\textbf{Inter-Model Comparisons.} We consider all pairs of vision models, and for each pair, compute the alignment scores between every pair of layers using the orthogonal Procrustes (\textbf{Top}) and Soft-Matching (\textbf{Bottom}) metric trained on ImageNet. Gray line plots denote the \textbf{maximum} alignment value for each network over rows (right line) and columns (top line). A common trend that is observed here is the consistent relationships between layers of CNNs trained with different architectures.}
    \label{tab:procrustes_inter_model}
\end{figure*}
\paragraph{Minimal transformations needed to align representations.}
Across all layers, we find that alignment scores increase as the mapping functions become more permissive (\textcolor{mplgreen}{Permutation} $\rightarrow$ \textcolor{mplorange}{Procrustes} $\rightarrow$ \textcolor{mplblue}{Linear}), as expected. However, linear mappings provide only modest improvements over Procrustes correlations, indicating that rotational transformations are sufficient to capture the majority of alignment information. This suggests that the added flexibility of linear mappings---such as scaling and shearing---does not substantially enhance alignment beyond what is achieved with Procrustes transformations. Importantly, because Procrustes is symmetric, this result highlights that alignment reflects a deeper similarity in the geometric structure of representations, rather than merely the ability of one representation to predict another.

\paragraph{Simple permutations achieve significant alignment.}  
\label{par: privileged-axes}
Despite the strict constraints imposed by permutation-based alignment, Permutation scores achieve surprisingly high alignment levels, indicating a strong one-to-one correspondence between individual neurons across network instances---suggesting that convergent learning extends down to the single-neuron level, even without allowing for more flexible transformations.

\begin{wraptable}{r}{8cm}
    \resizebox{\textwidth}{!}{
        \begin{tabular}{cccc}
        \toprule
        Model & Native (Min / Max) & Rotated (Min / Max) & Difference ($\%$) (Min / Max)\\
        \midrule
            ResNet18 & $0.202 \ / \ 0.721$ & $0.154 \ / \ 0.581$ & $10.47\% \ / \ 201.84\%$ \\
            ResNet50 & $0.225 \ / \ 0.791$ & $0.143 \ / \ 0.739$ & $2.61\% \ / \ 180.75\%$ \\
            VGG16 & $0.288 \ / \ 0.746$ & $0.181 \ / \ 0.599$ & $21.72\% \ / \ 81.12\%$ \\
            VGG19 & $0.292 \ / \ 0.799$ & $0.171 \ / \ 0.562$ & $41.95\% \ / \ 89.68\%$ \\
        \bottomrule
        \end{tabular}
    }
    \caption{\textbf{Sensitivity of Permutation Scores to Representational Axes.} For each ImageNet-trained network we apply a random rotation to the network’s unit basis, recompute permutation-based alignment scores, and summarize the results across convolutional layers. Columns report the minimum and maximum alignment scores observed over layers in the native basis and in the rotated basis. The final column gives the percentage change in alignment after rotation. Rotations almost always reduce alignment, indicating the distinguished nature of the representational axes for all layers in networks.}
    \label{tab:perm-rotated}
\end{wraptable}

To further probe this result and assess the depth of convergent learning, we tested the sensitivity of permutation alignment to changes in the representational basis. 

Specifically, we applied a random rotation matrix $\bm{Q} \in \mathbb{R}^{n \times n}$ to the converged basis of a neural representation, where $n$ is the number of neurons in a given layer. The rotation matrix was sampled from a Haar distribution via a $QR$ decomposition, ensuring that all orthogonal matrices were equally likely. We then recomputed the Permutation score after applying this rotation.

We conducted this analysis by taking response matrices from two identical DCNNs (initialized with different random seeds) at a given convolutional layer, $\{\bm{X}_1, \bm{X}_2\} \in \mathbb{R}^{m \times n}$, where $m$ represents the number of stimuli. We applied the random rotation $\bm{Q}$ to one network’s responses and computed the resulting permutation-based correlation score, $s_{\text{perm}}(\bm{X}_1\bm{Q}, \bm{X}_2)$. This process was repeated across all convolutional layers, with the alignment differences summarized in Table~\ref{tab:perm-rotated}.

These rotations consistently reduced alignment, with a drop between $\sim 10-202\%$ for ResNet$18$ across all layers, for instance. This significant decrease highlights that the learned representations are \emph{not} rotationally invariant and the specific bases in which features are encoded is meaningfully preserved across networks. In other words, convergent learning aligns not just the overall representational structure but also the specific axes along which features are encoded. This observation echoes recent findings by~\citep{khosla2024privileged}, who report the existence of privileged axes in biological systems as well as the penultimate layer representations of trained artificial networks. %Interestingly, Vision Transformers (ViTs) exhibit no comparable axis preservation (Appendix \ref{sec: vit-alignment}), pointing to a fundamental architectural split: CNNs converge on a shared neuron-level coordinate system, whereas ViTs organise features in a basis that varies across runs.  

\paragraph{Hierarchical correspondence holds across metrics.}
\label{par: inter-network}
Previous studies have shown that for architecturally identical networks trained from different initializations, the layer most similar to a given layer in another network is the corresponding architectural layer~\citep{kornblith2019similarity}. However, this finding has primarily been supported using metrics invariant to affine transformations (\emph{e.g.,} CCA, SVCCA). Here, we extend this result by showing that stricter metrics---such as Procrustes and soft-matching scores---also reveal the same hierarchical correspondence (Fig.~\ref{tab:procrustes_inter_model}), even when comparing networks with different architectures. This suggests that the hierarchical alignment of representations is a fundamental property of neural networks, robust to both architectural differences and the choice of alignment metric. Moreover, our results show that both representational shape (captured by Procrustes) and neuron-level tuning (captured by soft-matching) follow similar alignment patterns, reinforcing the consistency of this hierarchical organization across different levels of representational analysis.
%Previous studies have shown that, for architecturally identical networks trained from different initializations, the most similar layer in one network to a given layer in another is the corresponding architectural layer~\citep{kornblith2019similarity}. However, this finding has primarily been supported using metrics invariant to affine transformations (\emph{e.g.,} CCA, SVCCA). Here, we extend this result by showing that stricter metrics---such as Procrustes and soft-matching scores---also reveal the same hierarchical correspondence (Figs.~\ref{tab:procrustes_inter_model}, ~\ref{tab:sm_inter_model}), even when comparing networks with different architectures. This suggests that the hierarchical alignment of representations is a fundamental property of neural networks, robust to both architectural differences and the choice of alignment metric. Moreover, our results show that both representational shape (captured by Procrustes) and neuron-level tuning (captured by soft-matching) follow similar alignment patterns, reinforcing the consistency of this hierarchical organization across different levels of representational analysis.

\subsection{Evolution of convergence over training}
\label{sec: convergence-dynamics}
We next explore how representational convergence evolves during the training process. Specifically, we compute Procrustes alignment scores between pairs of networks over training epochs.
% training evo
\begin{figure*}[htbp!]
    \centering
    \includegraphics[width=\textwidth]{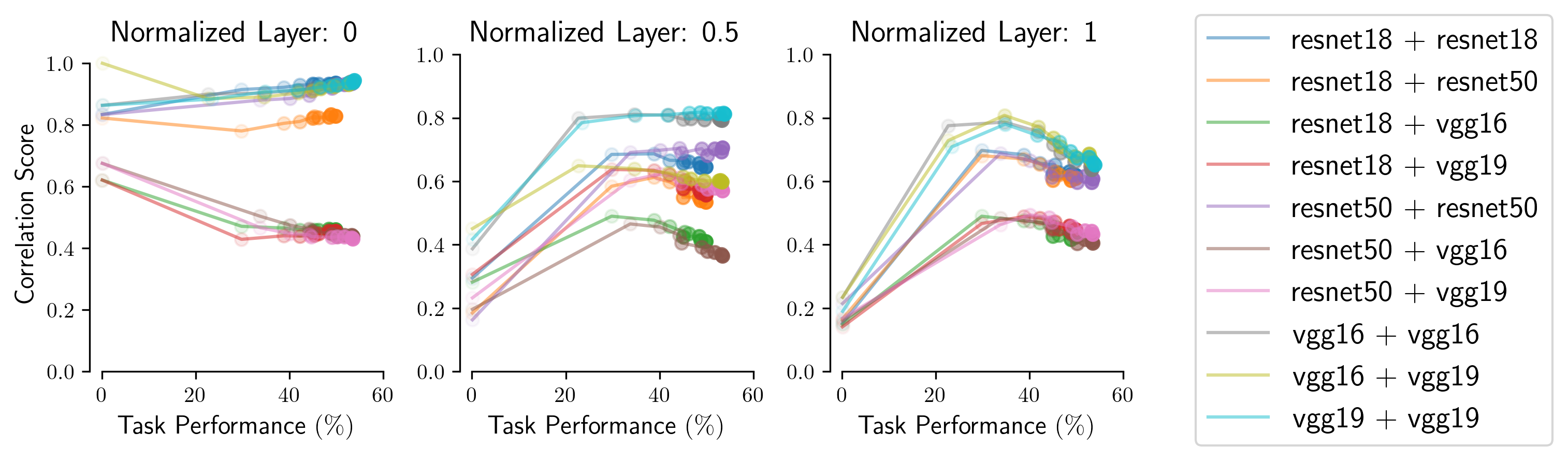}
    \caption{\textbf{Representational Alignment Through Training Evolution.}  We visualize the evolution of Procrustes alignment between network pairs during task optimization on ImageNet. Lighter shades indicate earlier epochs, progressively darkening with later epochs. The plots span from epoch $0$ (untrained) to epoch $10$, with task performance improving over time. Epoch progression can be inferred from the increasing task performance along the $x$-axis.}
    \label{fig:training-evo}
\end{figure*}
As shown in Fig.~\ref{fig:training-evo}, a striking pattern emerges: the majority of representational convergence occurs within the first epoch---long before a networks task performance peaks. This rapid early convergence suggests that factors independent of task optimization drive much of the representational alignment (see also Appendix~\ref{sec: training-rep-trained}). Shared input statistics, architectural biases, and early training dynamics seem to play a dominant role in shaping representational convergence, overshadowing the influence of the final task-specific solution.
%As shown in Fig.~\ref{fig:training-evo}, a striking pattern emerges: the majority of representational convergence occurs within the first epoch---long before networks approach optimal task performance. This rapid early convergence suggests that factors independent of task optimization drive much of the representational alignment. Shared input statistics, architectural biases, and early training dynamics seem to play a dominant role in shaping learned representations, overshadowing the influence of the final task-specific solution.
This observation challenges prevailing hypotheses, such as the \emph{contravariance principle}~\citep{cao2021explanatory} and \emph{task generality hypothesis}~\citep{huh2024platonic}, which argue that networks align because only a narrow solution subspace yields high performance. Instead, our findings point to inductive biases and early learning processes as the dominant forces shaping representational convergence.
%This observation challenges prevailing hypotheses that attribute convergence to task constraints, such as the \emph{contravariance principle}~\citep{cao2021explanatory} and \emph{task generality hypothesis}~\citep{huh2024platonic}, which posits that networks converge towards a limited subspace of solutions capable of achieving high task performance. However, the early emergence of representational alignment---well before high task accuracies---implies that convergence is instead driven by inductive biases, input statistics, and early training dynamics.

%Our early training convergence parallels the \emph{``silent alignment''} effect~\citep{atanasov2021neural}---networks can align with the target function early in training, before significant loss reduction. However, a key difference is that they study output-level alignment while we examine internal representations across different models.
%Our observed early alignment is consistent with the \emph{``silent alignment''} effect~\citep{atanasov2021neural}, where networks align with the target function early in training, before significant loss reduction. However a key difference is that they study output-level alignment while we examine internal representations across different models.

Our findings echo those of~\citep{frankle2020early}, who showed that most representational re-organization happens within the first few hundred iterations---well before any substantive task learning. Perturbing weights during this brief window (by re-initializing or shuffling them) severely impairs eventual accuracy, indicating the formation of early, data-dependent structure. Together, these observations reinforce the view that shared input statistics—not task labels---are the principal drivers of early representational convergence.

Early convolutional layers show little training-induced change on representational similarity (even reducing in some cases). This phenomenon arises because early layers compute near-linear mappings of the input, even in untrained networks, allowing them to align well with simple linear transformations. Subsequent training only makes minor adjustments that slightly disrupt initial alignment, though they remain highly similar overall. 
%This result highlights that the convergence of early-layer representations is not driven by the specifics of the training process. 
%For the earliest convolutional layers, we find that training has minimal impact on representational similarity and can even reduce it in some cases. This phenomenon arises because early layers compute a largely linear function of the input, even in untrained networks, allowing them to align well with simple linear transformations. As training progresses, these layers may undergo minor adjustments that slightly disrupt this initial alignment, though they remain highly similar overall. This result highlights that the convergence of early-layer representations is not driven by the specifics of the training process. 

%\input{./figures-code/pearson-ood}
\subsection{Convergence across distribution shifts}
\label{sec:ood-results}
Having established alignment on in-distribution images, we next probed its robustness to distribution shift. We applied $17$ out-of-distribution (OOD) variants from ~\citep{geirhos2018imagenet} to ImageNet-trained CNNs. Dataset specifics are given in Appendix~\ref{sec: ood-dataset-details}.

% In the previous sections, we examined representational alignment under identical input training distributions. However, a critical question remains: does representational convergence persist under distribution shifts? To explore this, we analyzed the internal representations of ImageNet-trained DCNNs when exposed to OOD stimuli. We used $17$ OOD datasets from~\citep{geirhos2018imagenet}, all sharing the same $16$ coarse labels as ImageNet~\citep{deng2009imagenet}, allowing for a controlled comparison of representational alignment under varying distributional shifts. The specifics of each OOD dataset has been described in Sec.~\ref{sec: ood-dataset-details}.
%In the previous sections, we examined representational alignment under the same input distributions used for training. However, a critical question remains: does representational convergence persist under distribution shifts? To explore this, we analyzed the internal representations of ImageNet-trained DCNNs when exposed to OOD stimuli. We used $17$ OOD datasets from~\citep{geirhos2018imagenet}, all sharing the same $16$ coarse labels as ImageNet~\citep{deng2009imagenet}, allowing for a controlled comparison of representational alignment under varying distributional shifts. The specifics of each OOD dataset has been described in Sec.~\ref{sec: ood-dataset-details}.
%\begin{figure}
%    \centering
%    \includegraphics[scale=0.33]{figures/ood-divergence/ood_amplification.png}
%    \caption{\textbf{Amplifying Model Differences.}}
%    \label{fig:enter-label}
%\end{figure}

\begin{figure}[htbp!]
    \centering
    \includegraphics[width=\textwidth]{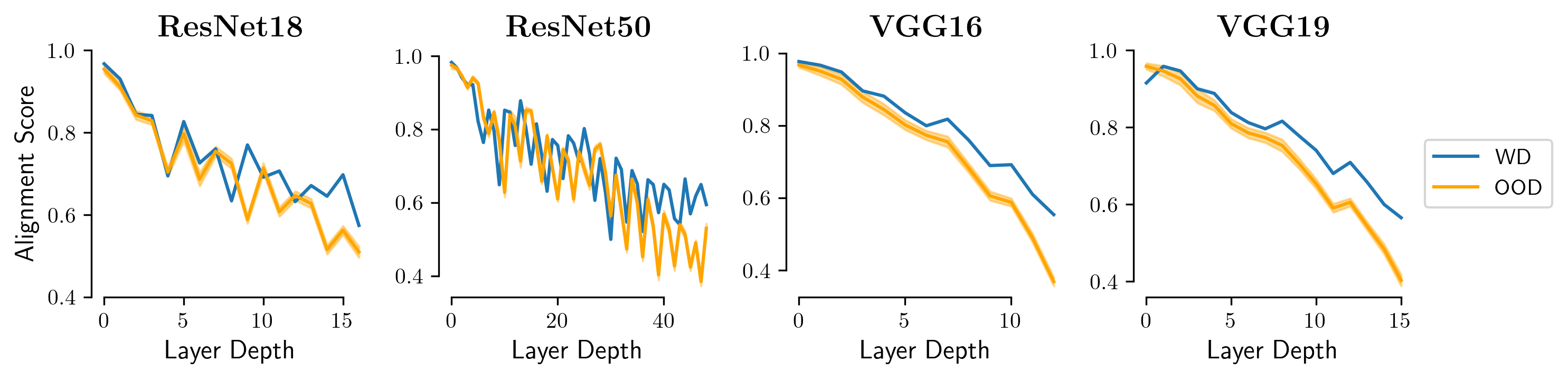}
    \caption{\textbf{Procrustes Score-based Alignment Between Networks Sharing the Same Architecture but Trained With Different Random Seeds, Plotted as a Function of Layer Depth.} Alignment is measured using within-distribution (WD) stimuli (ImageNet test set) and out-of-distribution (OOD) stimuli, with OOD values averaged across $17$ datasets. Error bars represent the standard error computed across the $(n = 17)$ OOD datasets.}
    \label{fig:ood-amplification}
\end{figure}

%\input{8pager/figures-code/pearson-ood}

\begin{figure*}[htbp!]
    \resizebox{\textwidth}{!}{
    \begin{tabular}{cc}
        \raisebox{0.5cm}{
        \includegraphics[height=0.15\textheight]{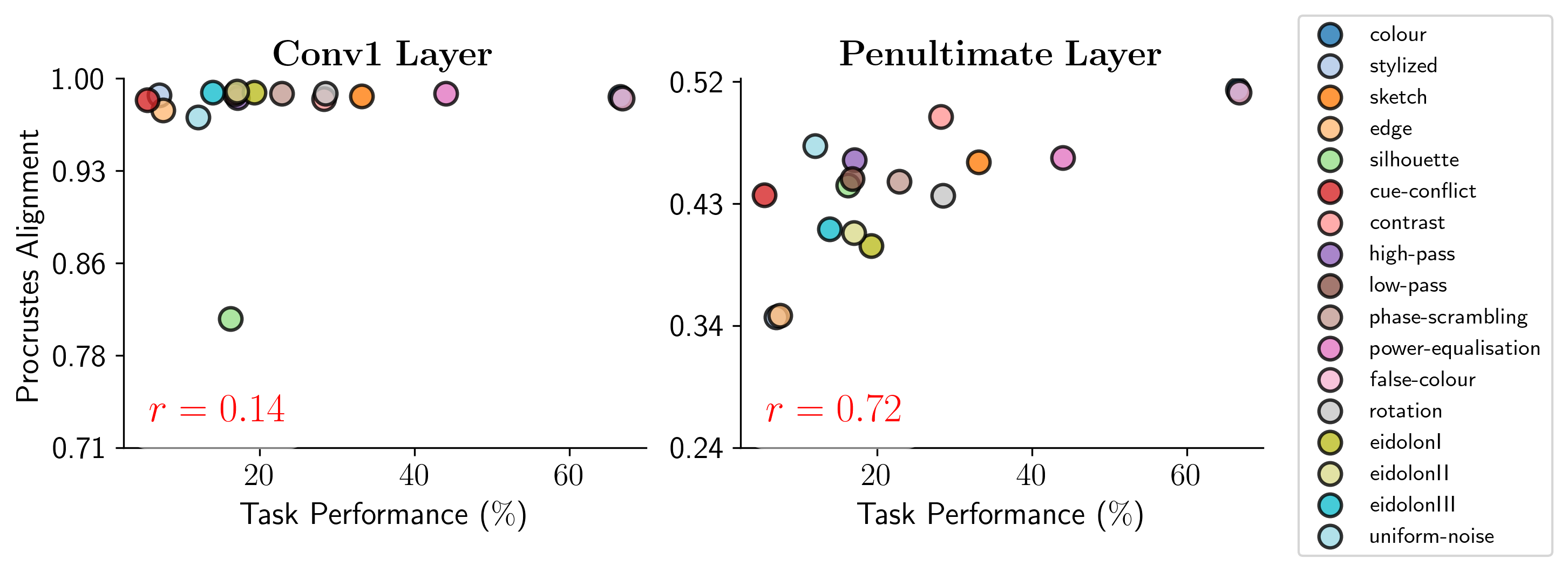}%
        }
        & \includegraphics[width=0.4\textwidth]{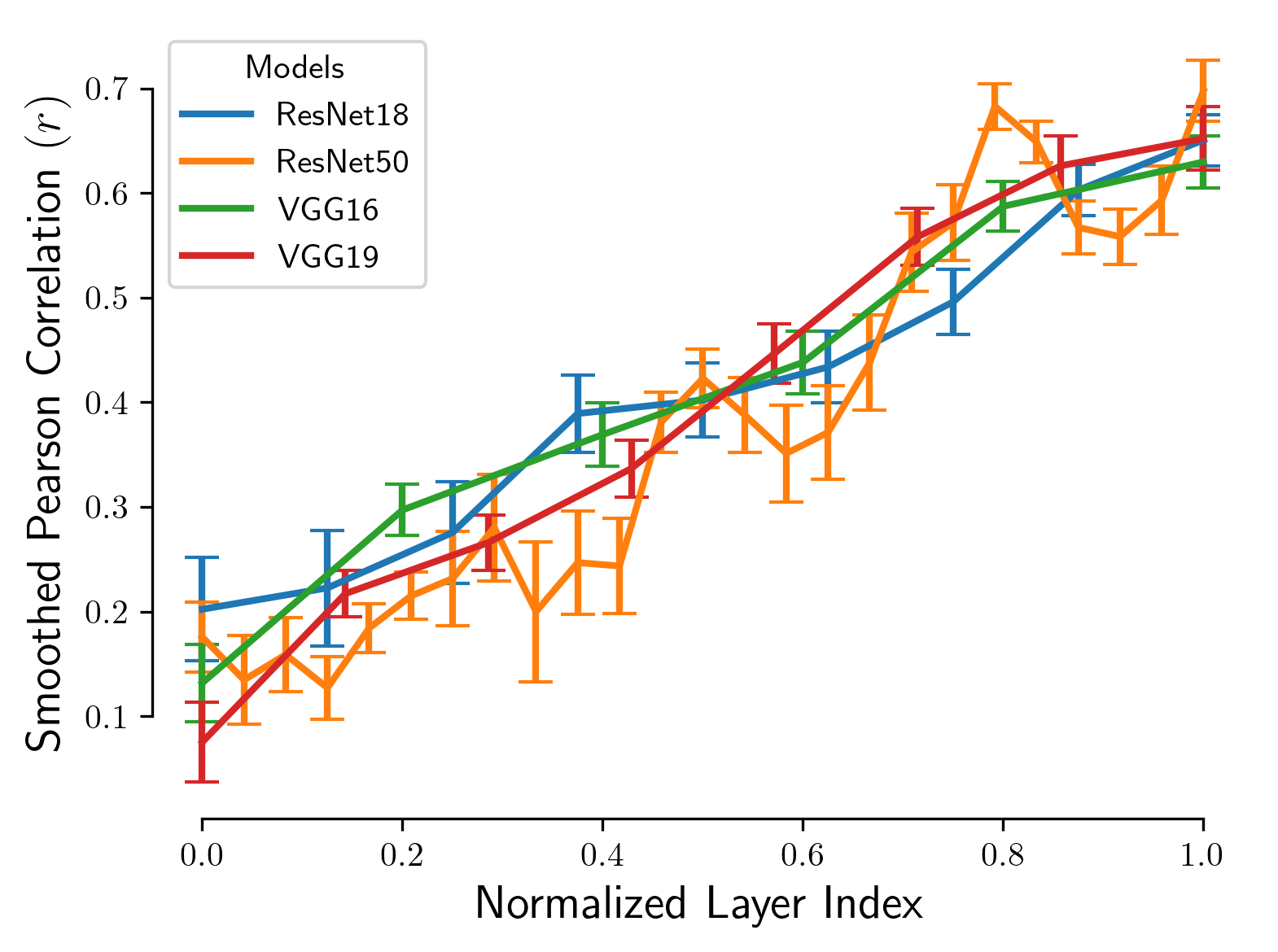}\\
    \textbf{(a)} & \textbf{(b)} 
    \end{tabular}
    }
    \caption{\textbf{Convergence on OOD Inputs.} \textbf{(a)} Procrustes alignment vs. task performance of ResNet$50$ models on each of the $17$ datasets for the first convolutional layer \textbf{(Left)} and the penultimate \textbf{(Right)} layer. \textbf{(b)} Correlation between these variables as a function of network depth (normalized by each models depth). 
    % Here, the error bars represent the standard deviation of $r$ after excluding one dataset at a time (i.e., using $n=16$ remaining datasets).
    }
    \label{fig: ood-convergence}
\end{figure*}

% ood amplification
Using the Procrustes metric, we compute representational alignment on these datasets and observed a consistent pattern: OOD inputs amplify differences in the later layers of the networks, while early layers maintain comparable alignment levels between in-distribution and OOD stimuli (Fig.~\ref{fig:ood-amplification}). We hypothesize that this pattern arises because early layers capture universal features (\emph{e.g.,} edges, textures) that remain nearly identical across distributions, whereas later layers encode task-specific features that are sensitive to distributional shifts, thus amplifying the divergence between models.
%We computed representational alignment across these datasets using the Procrustes metric and observed a consistent pattern: OOD inputs amplify differences in the later layers of the networks, while early layers maintain comparable alignment levels between in-distribution and OOD stimuli (Fig.~\ref{fig:ood-amplification}). We hypothesize that this pattern arises as a result of early layers capturing basic, universal features (\emph{e.g.,} edges, corners, textures) that remain nearly identical across distributions, whereas later layers encode more task-specific features that are more sensitive to distributional shifts, thus amplifying the divergence between models.
We also found a strong correlation between representational alignment and OOD accuracy: datasets where models maintained higher accuracy show strong alignment and vice versa. This correlation is minimal in early layers, but progressively increases with network depth across all architectures (Fig.~\ref{fig: ood-convergence}); analogous trends appear in other networks Fig.~\ref{fig:ood-convergence-appendix}.
%Moreover, we found a strong correlation between representational alignment in later layers and the networks' classification accuracy on the OOD datasets. Datasets where models maintained higher accuracy showed stronger alignment, whereas datasets with lower accuracy exhibited weaker alignment. This correlation was notably weaker in early layers but increased progressively with network depth across all architectures (Fig.~\ref{fig: ood-convergence}). We extend these analyses to other vision networks in Fig.~\ref{fig:ood-convergence-appendix}.

These results have several important implications. First, early-layer alignment is remarkably stable---across random initializations and in-distribution and OOD inputs—indicating that these layers encode broadly transferable features that serve as a common scaffold for higher processing. Because this scaffold endures under distribution shift, one could plausibly improve OOD generalisation by fine-tuning only the later layers.  Second, these findings inform model-brain comparisons. Though diverse architectures and learning objectives yield similar brain predictivity~\citep{conwell2024large}, the amplified divergence of later-layer representations under OOD conditions suggests that OOD stimuli could be especially useful to distinguish and select between models whose representations closely mirror the brain.
%\begin{figure}
%    \centering
%    \includegraphics[scale=0.5]{figures/ood-divergence/pearson_ood_all_layers.png}
%    \caption{\textbf{Pearson Correlation Across Network Hierarchy.}}
%    \label{fig: corr-ood-network-hierarchy}
%\end{figure}

%These results have several important implications. First, the stability of early-layer alignment across distributions suggests that these layers encode generalizable features that are consistent across both network initializations and input distributions. This highlights their role as a shared foundation for higher-level processing, which becomes more specialized and sensitive to distribution shifts in later layers. Second, these findings inform model-brain comparisons. Prior studies have shown that diverse architectures and learning objectives can yield similar brain predictivity~\citep{conwell2024large}. However, the observed amplification of representational divergence in later layers under OOD conditions suggests that using OOD stimuli could be an effective strategy for distinguishing between models and identifying more brain-like models. 

\subsection{Representational Alignment for Self-Supervised Networks}
\label{sec: self-sup-alignment}

\begin{figure}[htbp!]
    \centering
    \includegraphics[width=\textwidth]{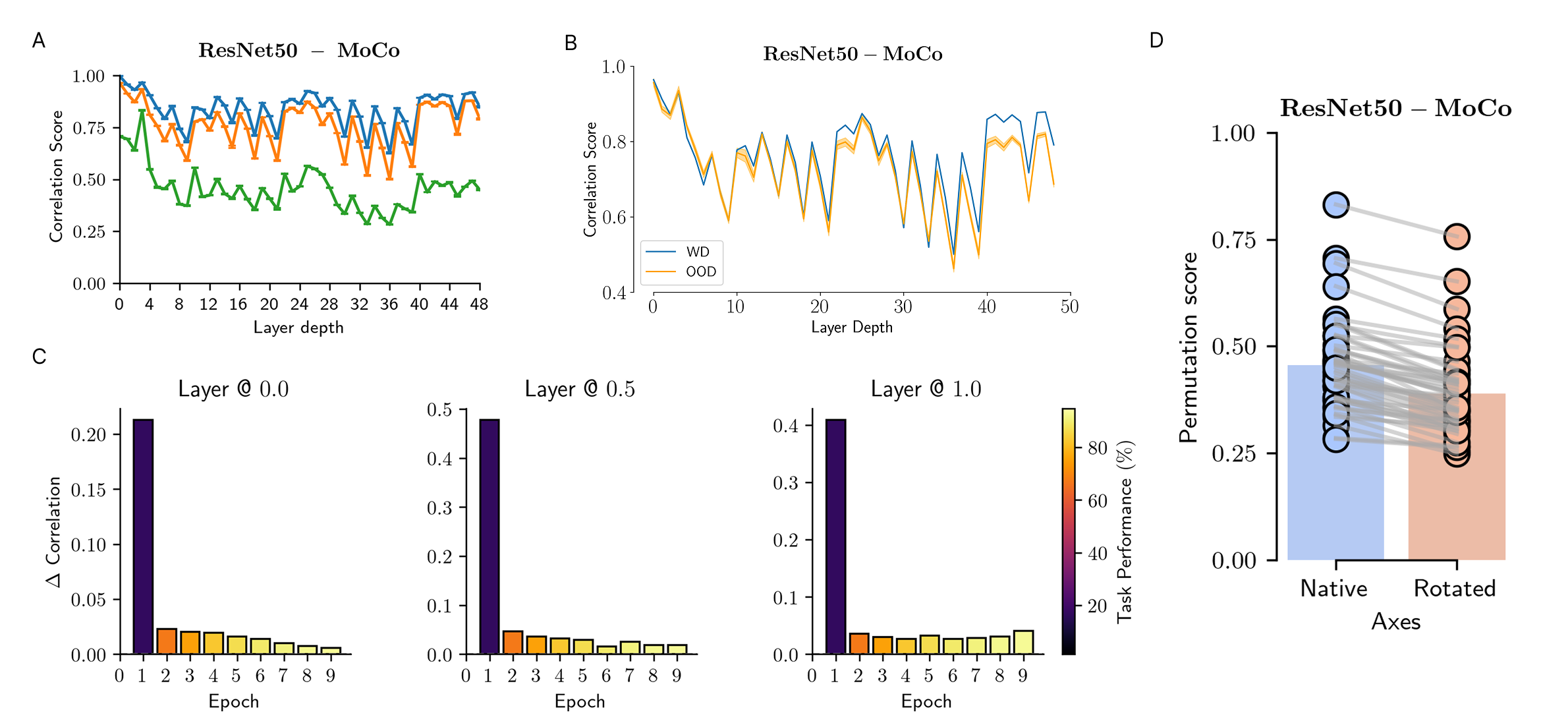}
    \caption{\textbf{Representational Alignment with Self-Supervised Networks.} \textbf{(A)} Representational alignment across layers of a pair of MoCo-trained ResNet$50$ models on ImageNet. Error bars indicate standard deviation from $5$-fold cross-validation. \textbf{(B)} We plot the Procrustes alignment between MoCo models evaluated on in-distribution (ImageNet) and out-of-distribution (Stylized ImageNet~\citep{geirhos2018imagenet}) stimuli. Error bars show standard error across all $(n=17)$ OOD datasets. \textbf{(C)} Change in Procrustes alignment score with every training epoch. Colors in the bar-plot indicate the top-$1$ accuracies during training. \textbf{(D)} Permutation alignment in the native and randomly rotated basis. Each dot corresponds to a convolutional ResNet$50$ layer. Bars indicate mean alignment. Rotation reduces mean alignment by $14.87\%$.}
    \label{fig:self-supervised}
\end{figure}

We next test whether our findings generalize to self-supervised learning. All analyses were replicated on models trained with a Momentum Contrast (MoCo) objective~\citep{he2020momentum} (training details in Appendix~\ref{sec: network-training}). 
(1) \textbf{Across hierarchy:} convergence is strongest in early layers and tapers with depth (Fig.~\ref{fig:self-supervised}-A); \textcolor{mplblue}{Linear} and \textcolor{mplorange}{Procrustes} scores remain comparable, suggesting inter-model variability reflects rotations/reflections. 
(2) \textbf{Across distribution shifts:} early layers maintain alignment for both in-distribution and OOD inputs, while deeper layers diverge (Fig.~\ref{fig:self-supervised}-B), indicating that self-supervised training, like supervised learning, produces stable, general early filters and increasingly distribution-sensitive higher representations.
(3) \textbf{Across training:} most alignment emerges within the first epoch, well before optimal task performance (Fig.~\ref{fig:self-supervised}-C). 
(4) \textbf{Permutation sensitivity:} applying a random rotation substantially reduces permutation alignment (Fig.~\ref{fig:self-supervised}-D), confirming convergence to a privileged basis regardless of learning paradigm.

% \paragraph{Alignment Over Training.}
% \textcolor{red}{Fig.~\ref{fig:self-supervised}-C shows how a Procrustes-based correlation score evolves between a pair of MoCo-trained networks across training. As in the supervised regime (Sec.~\ref{sec: convergence-dynamics}), the bulk of representational alignment appears in the first epoch---well before optimal task performance.}
% %In Fig.~\ref{fig:self-supervised}-C, we show how the correlation score computed using the Procrustes alignment metric for a pair of networks evolves over training. Consistent with our findings in the supervised regime (Sec.~\ref{sec: convergence-dynamics}), the bulk of representational alignment occurs in the first epoch---well before task performance has peaked.

% \paragraph{Sensitivity To Permutation Alignment.} 
% \textcolor{red}{Applying a random orthogonal rotation to the converged basis substantially drops permutation alignment (Fig.~\ref{fig:self-supervised})---consistent with supervised networks. This indicates that regardless of the learning paradigm, networks converge on a privileged coordinate basis.}
% %We apply a random rotation to the converged basis of MoCo-trained models. Consistent with our earlier findings, this perturbation leads to a significant drop in permutation alignment (Fig.~\ref{fig:self-supervised}-D). These results suggest that regardless of the learning paradigm, networks converge to a privileged basis.

\subsection{Representational Convergence in Vision Transformers}
\label{sec: vit-alignment}
%\textcolor{red}{We next analyze representational convergence in Vision Transformers (ViTs). All model and training details are in Appendix~\ref{sec: network-training}. For each layer, we use the [\texttt{CLS}] (class) token as a proxy for a layers representational geometry\footnote{Our underlying assumption here hinges on the fact that the [\texttt{CLS}] token aggregates the \emph{``information content''} of all patch embeddings via the self-attention operator. In addition, computing representational similarity for [\texttt{CLS}] vectors avoids having to deal with pooling patch similarities into a single representative score.}.}
We next analyze representational convergence in ViTs. All models and training details are in Appendix~\ref{sec: network-training}. We analyze the [\texttt{CLS}] (class) token representations at each layer. This is because the [\texttt{CLS}] token aggregates the \emph{``information content''} of all patch embeddings via the self-attention operator, making it a useful proxy to study representational geometry. In addition, computing representational similarity for [\texttt{CLS}] vectors avoids having to deal with pooling patch similarities into a single representative score.
\begin{figure}[htbp!]
    \centering
    \includegraphics[width=0.7\textwidth, keepaspectratio]{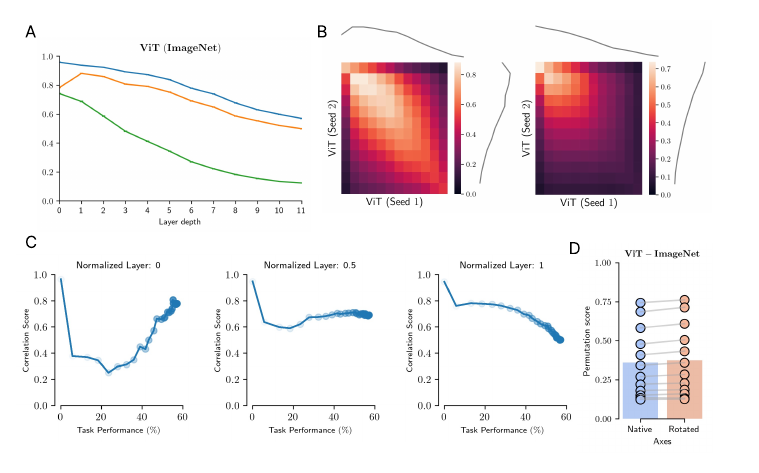}
    \caption{\textbf{Representational Convergence in Vision Transformers.} \textbf{(A)} We plot the evolution of alignment scores of three metrics (\textcolor{mplblue}{Linear}, \textcolor{mplorange}{Procrustes}, \textcolor{mplgreen}{Permutation}) computed between different seeds of the same ViT, which was trained on ImageNet. Error bars denote the standard deviation across $5$-fold cross validation. \textbf{(B)} We plot the inter-model orthogonal Procrustes (\textbf{left}) and permutation (\textbf{right}) scores for all layer pairs. Gray line plots denote the maximum alignment value over rows (right line) and columns (top line). \textbf{(C)} We visualize the evolution of the orthogonal Procrustes score at different checkpoints, ranging from epochs $0$ (untrained) to $25$. Darker colors correspond to epoch progression. \textbf{(D)} We rotate the converged basis of ViTs by a random rotation matrix and recompute the permutation scores for models trained ImageNet. Each dot represents a layer in the ViT. The permutation alignment remains approximately constant in both cases.}
    \label{fig:vit-alignment}
\end{figure}

\textbf{Results.} Analyses parallel to DCNNs reveal three key findings: 
(1) \textbf{Depth-wise convergence:} alignment tapers with network depth across metrics (Fig.~\ref{fig:vit-alignment}-A); \textcolor{mplorange}{Procrustes} and \textcolor{mplblue}{Linear} scores remain comparable, indicating rotations/reflections explain most inter-seed variability, and layer-wise patterns mirror DCNNs (Fig.~\ref{fig:vit-alignment}-B, left). 
(2) \textbf{No privileged axes:} independent ViT runs do not share a common basis—permutation scores between native and rotated axes are statistically indistinguishable across layers (Fig.~\ref{fig:vit-alignment}-D; Appendix~\ref{sec: basis-alignment-cnn-vit}). 
(3) \textbf{Early plateau:} alignment stabilizes after the first epoch, except for a \texttt{[CLS]} embedding artifact due to uniform positional encoding, indicating late-stage training does not drive convergence (Fig.~\ref{fig:vit-alignment}-C).

\subsection{Representational Convergence in Language Models}
\label{sec: lang-alignment}
We analyze representation convergence in language models using sentence embeddings from the Semantic Textual Similarity Benchmark (STSB)~\citep{cer2017semeval}. Our analyses include two primary comparisons: \textbf{(i) Same architecture:} multiple \textbf{Pythia-160m} instances differing only by random seed, allowing us to analyze variability in representational spaces solely from stochastic training factors \textbf{(ii) Cross-architecture:} \textbf{Pythia-70m} vs. \textbf{Pythia-160m}, to probe the effect of network depth on learned representations.
%In this section, we rigorously analyze representational convergence in different language models. Specifically, our analyses includes:
%\begin{itemize}
%    \item \textbf{Same-Architecture Comparisons:} We compare multiple instances of the \textbf{Pythia-160m} model, which share an identical architecture but differ in the initialization of random seeds. This allows us to assess the variability in representational spaces resulting solely from stochastic training factors.
%    \item \textbf{Cross-Architecture Comparisons:} In addition to seed variation, we compare models with different architectural configurations, namely the \textbf{Pythia-70m} and the \textbf{Pythia-160m} models. The primary architectural difference lies in network depth (i.e., number of layers), offering insights into how differences in model capacity and depth impact the learned representations.
%\end{itemize}

%\noindent
%All alignment metrics are computed using the representations of all unique sentences from the Semantic Textual Similarity Benchmark (STSB) dataset~\citep{cer2017semeval}.

\begin{figure*}[t]%[!htbp]
    \centering
    \includegraphics[width=0.85\textwidth]{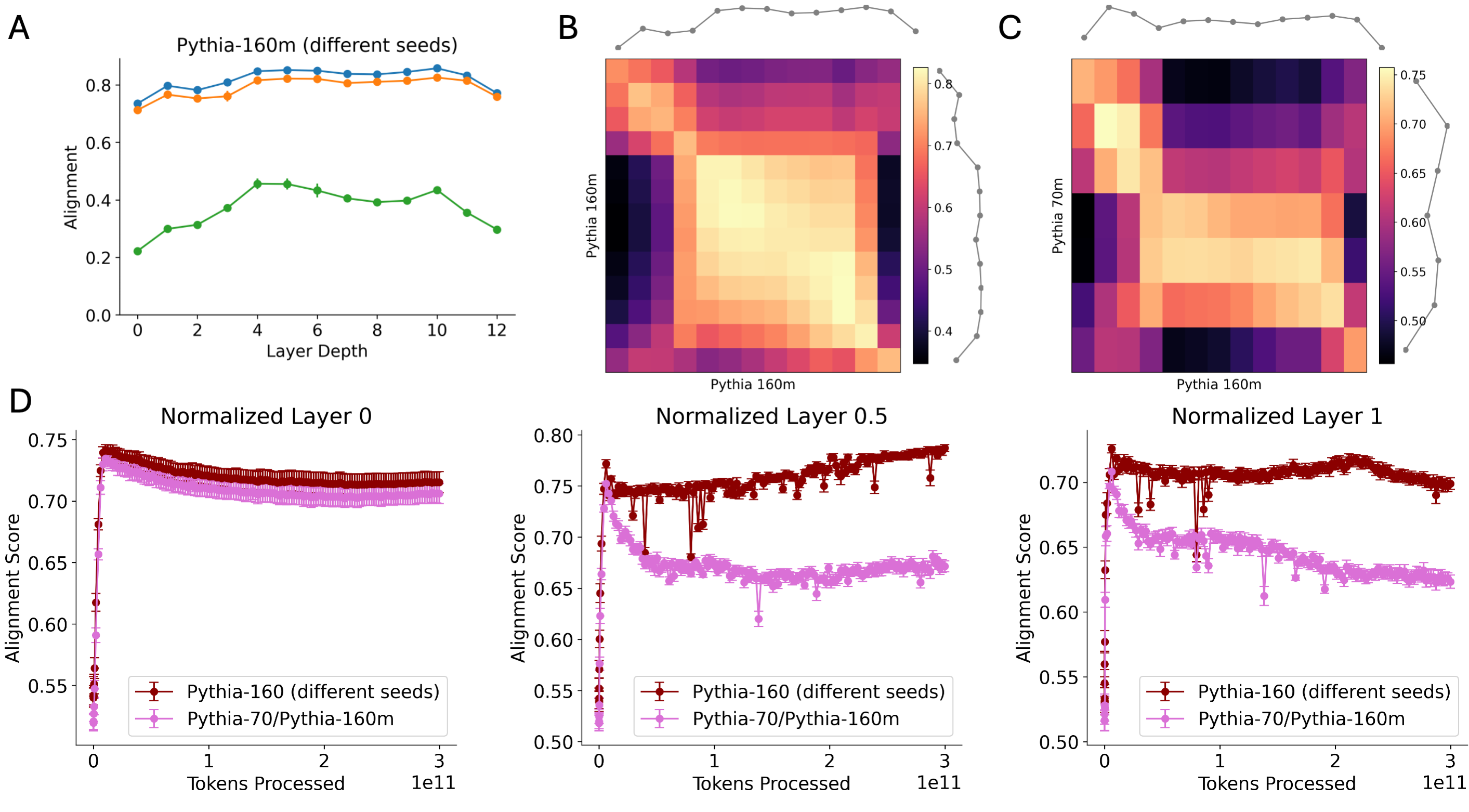}
    \caption{\textbf{Representational Convergence in Language Models.} \textbf{(A)} Alignment scores across all layers for Pythia-160m models trained from different seeds evaluated on STSB. \textcolor{mplorange}{Procrustes} align representations nearly as well as \textcolor{mplblue}{linear} transformations, mimicking the trend observed for vision models. \textbf{(B)} We compute the alignment scores using the Procrustes metric for every pair of layer in a seed-pair of Pythia-160m models. \textbf{(C)} Same as \textbf{(B)}, but for models with different architectures (varying depth). \textbf{(D)} Procrustes alignment between model pairs over training checkpoints.}
    \label{fig:lang}
\end{figure*}

\paragraph{Methodology.} For our representational similarity analysis, we perform two sets of computations: \textbf{(i) Final Checkpoint:} we compute the similarity between every pair of layers for all models at their final checkpoints using the same metrics as in the case of vision models (Linear, Procrustes, Permutation). \textbf{(ii) Intermediate Checkpoint:} across $154$ intermediate checkpoints, we analyze layer representations at normalized depths $0$ (beginning), $0.5$ (middle), $1$ (end) in the network. We track changes in the Procrustes similarity over training.

%\textbf{1. Final Checkpoint Analysis:}  
%For each model at its final checkpoint, we compute the similarity between every pair of layers using the three metrics used for vision models, namely the Permutation, Procrustes and Linear Alignment measures. 
%
%\textbf{2. Intermediate Checkpoint Analysis:}  
%Across the training run, we analyze 154 intermediate checkpoints sampled during the training process of the Pythia suite. At each checkpoint, for each model, we extract layer representations at normalized indices corresponding to the beginning $(0)$, middle $(0.5)$, and end $(1)$ of the network. The similarity between these layers is then computed using the \textbf{Procrustes} measure. \\
%
\noindent

\textbf{Results.} Comparing representations across language models reveals three key patterns: 
(1) \textbf{Metric-dependent alignment:} alignment follows \textcolor{mplblue}{Linear} $>$ \textcolor{mplorange}{Procrustes} $>$ \textcolor{mplgreen}{Permutation} (Fig.~\ref{fig:lang}-A); \textcolor{mplorange}{Procrustes} and \textcolor{mplblue}{Linear} scores are nearly identical, indicating that rotations/reflections explain most variability across models, and high alignment persists across layers. 
(2) \textbf{Hierarchical correspondence:} layers at similar depths align most closely both within (Fig.~\ref{fig:lang}-B) and across architectures (Fig.~\ref{fig:lang}-C). 
(3) \textbf{Rapid convergence:} alignment emerges early (layers at indices $0$, $0.5$, $1$), peaks around $8$–$10$B tokens, then plateaus or slightly declines (Fig.~\ref{fig:lang}-D) even as next-token prediction improves up to $300$B tokens~\citep{biderman2023pythia}.

\section{Discussion}
We present a large-scale account of convergent learning, showing how representational alignment between independently trained networks depends on depth, training time and distribution shifts. Using alignment metrics with varying degrees of transformation invariances for capturing representational similarity, we provide a more nuanced view of convergent learning.
%This study fills critical gaps in our understanding of convergent learning, offering a comprehensive analysis of how representational alignment between independently trained networks varies across network depth, training, and distribution shifts. We systematically explored how different alignment metrics---with varying levels of transformation invariance---capture representational similarities, providing a more nuanced view of convergent learning than previous work.

Nonetheless, certain limitations remain. While we show the early emergence of alignment during training, we do not precisely quantify \emph{when} within the first epoch this convergence occurs. Our empirical evidence for early alignment is striking and raises a crucial question: what drives such rapid convergence across diverse models? Although a full theoretical account is beyond our current scope, future work using tools like Neural Tangent Kernels (NTKs)~\citep{jacot2018neural}, and its extension to CNNs~\citep{arora2019exact}---under simplifying assumptions like linearity---may help elucidate the temporal dynamics of representational convergence.
%Nonetheless, certain limitations remain. While we show the early emergence of alignment during training, we do not precisely quantify \emph{when} within the first epoch this convergence occurs. Though our empirical evidence for early alignment is striking, it raises a crucial question: what drives such rapid convergence across diverse models? Although a full theoretical account is beyond our current scope, future work using tools like Neural Tangent Kernels (NTKs)~\citep{jacot2018neural}, and its extension to CNNs~\citep{arora2019exact}---under simplifying assumptions like linearity---may help elucidate the temporal dynamics of representational convergence.
Another limitation stems from the alignment metrics themselves. Although our metrics reveal an early stabilization of alignment over training, this could reflect the limitations of the metrics rather than the absence of representational changes. Prior work~\citep{bo2024evaluating} has shown that certain alignment metrics may fail to capture subtle, task-relevant shifts in representations. It remains possible that alternative metrics could reveal gradual alignment changes over training that are invisible to the methods employed here.

Finally, our method for computing alignment uses the center pixel of each feature map, enforcing a strict spatial correspondence between representations. This may underestimate alignment, especially in cases where features are slightly shifted spatially. While one could incorporate spatial shifts into alignment computations, they are computationally intensive and beyond the scope of this work. However, prior research has shown that optimal spatial shifts in many convolutional layers are typically negligible~\citep{williams2021generalized}, justifying our approximation. Nonetheless, scalable methods that account for such spatial variability remain an important direction for future research.

% author checklist
%\input{./checklist}
%\input{./supplementary}

\bibliographystyle{abbrv}
\bibliography{Styles/bibliography}

\appendix
\setcounter{secnumdepth}{2}
\renewcommand{\thesection}{A\arabic{section}}
\renewcommand{\thefigure}{A\arabic{figure}}
\renewcommand{\thetable}{A\arabic{table}}
\setcounter{figure}{0}  
\setcounter{table}{0}
\setcounter{section}{0}
\newpage
\section*{Appendix}

%\section{Network Training}
\section{Training Details}
\label{sec: network-training}
Consider a model type denoted by $M$. We train a pair of models, $\{M_1, M_2\}$ initialized with two different random seeds. We initialize the models using a uniform Xavier distribution~\citep{glorot2010understanding}. This setup ensures that the two models are identical in architecture and achieve comparable task performance, allowing us to isolate the effects of stochastic variations in the SGD process (such as initialization differences and input order). By comparing the representations from these models, we can quantify the minimal set of transformations required to align them. Below, we outline the specific training parameters for each of the model types presented in Sec.~\ref{sec: dcnn-results}.

\paragraph{Supervised Convolutional Neural Networks.} All models are trained from scratch on CIFAR100 or ImageNet for $100$ and $80$ epochs respectively. We save model weights at every epoch and additionally store the best-performing weights based on test-set performance for each dataset.

\paragraph{Self-Supervised Networks.} {We train a pair of networks (ResNet50 backbone) using a Momentum Contrastive (MoCo) objective~\citep{he2020momentum} initialized with $2$ different random seeds on ImageNet for $50$ epochs using a batch size of $256$.

\paragraph{Vision Transformers.} We analyze the \texttt{ViT-B/16} variant of Vision Transformers (ViTs)~\citep{dosovitskiy2020image} having patch size $16\times 16$ trained on ImageNet and CIFAR100. We train multiple models using different random seeds for $25$ epochs with batch size $32$.

\textbf{Language Models.} We analyze models from the \textbf{Pythia suite}~\citep{biderman2023pythia}, a collection of autoregressive language models trained with varying architectures and random seeds. These models were predominantly trained on the \textbf{Pile dataset}~\citep{gao2020pile}---a diverse and carefully curated corpus aggregating high-quality texts from sources such as academic publications, books, Wikipedia, and web-scraped data. This dataset provides a rich and heterogeneous distribution of language examples that supports robust learning of linguistic representations.

\section{Convergence with Network Depth Using Spearman's Rank-Order Correlation}
\label{sec: spearman-results}
For analyses presented thus far in Sec.~\ref{sec: dcnn-results}, we report alignment as a Pearson correlation. However, the use of such a metric could be susceptible to high-variance outlier dimensions. To address this possibility, we conduct an additional series of experiments, where we compute alignment scores using Spearman's rank-correlation. Concretely, for an optimal transformation matrix $\bm{M}$ obtained after using a specific alignment metric (linear, Procrustes or permutation) to align a representational pair $\{\bm{X}_i, \bm{X}_j\}$, we now report Spearman's rank coefficient between the aligned representations given by:
\begin{equation*}
    \texttt{Alignment} = \texttt{corr}(\bm{X}_i, \bm{MX}_j) = 1 - \frac{6\sum(\bm{X}_i - \bm{MX}_j)}{n(n^2 -1)}
\end{equation*}
where $n$ is the number of stimuli.

\begin{figure*}[htbp!]
    \centering
    \includegraphics[width=.25\textwidth]{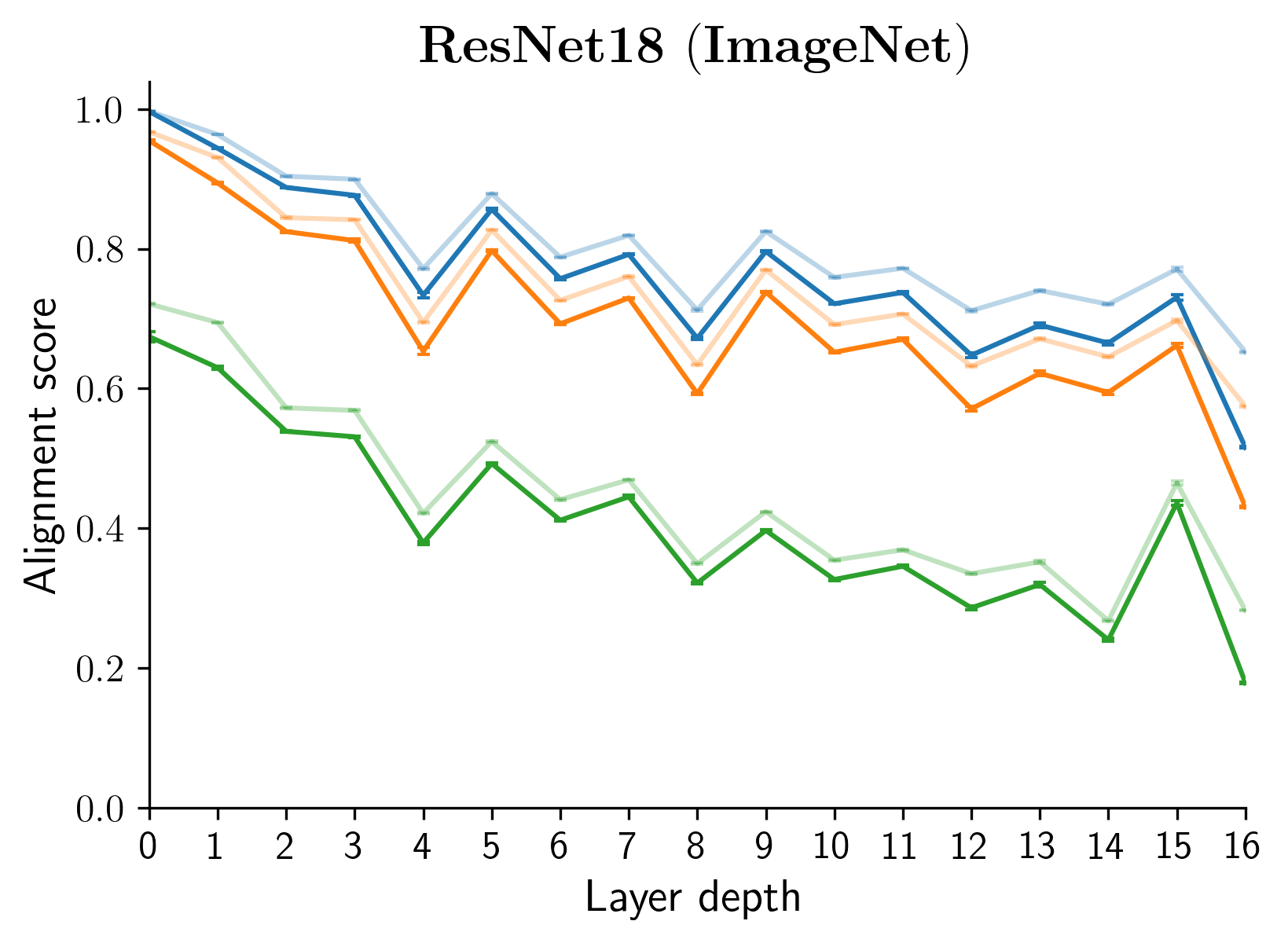}\hfill
    \includegraphics[width=.25\textwidth]{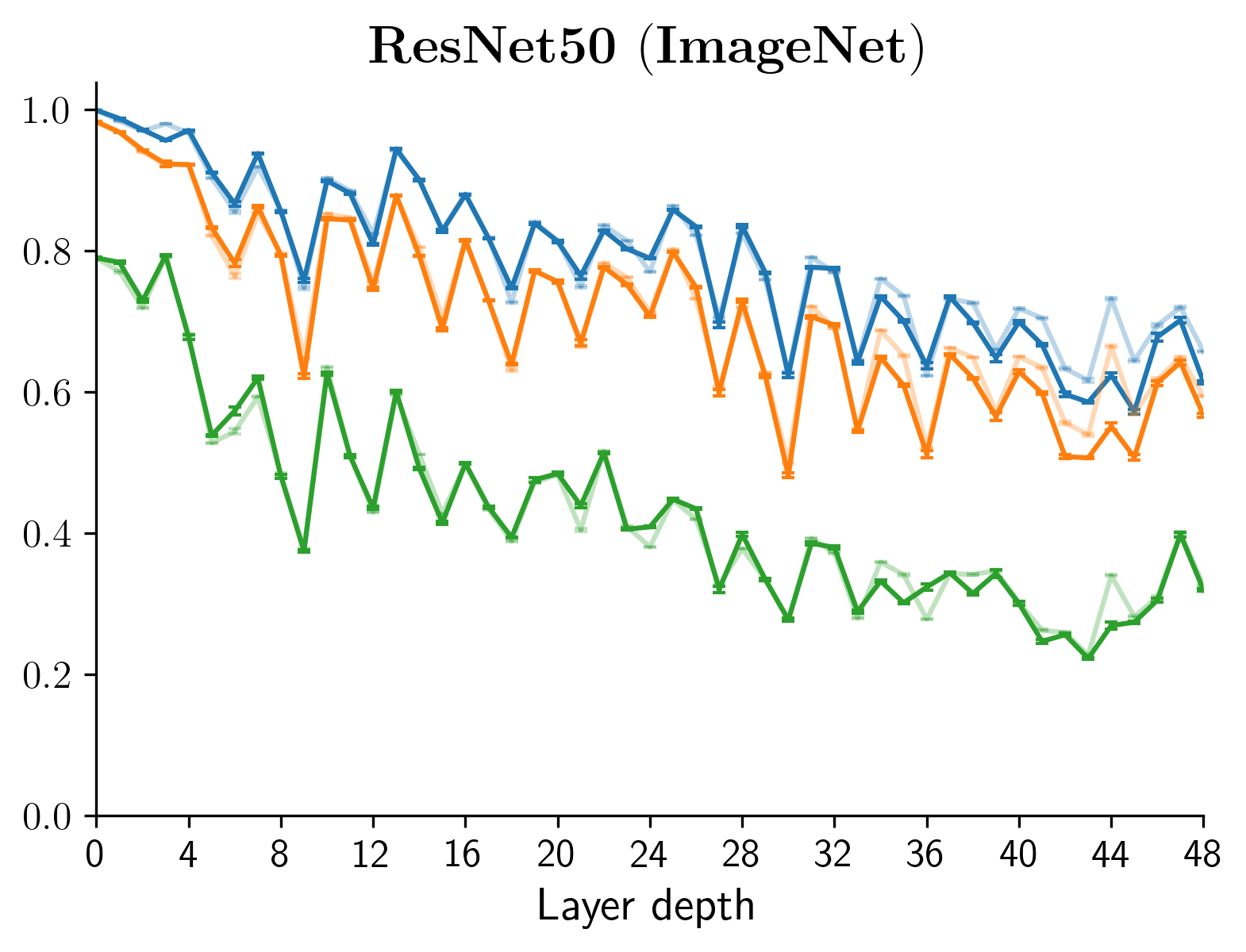}\hfill
    \includegraphics[width=.25\textwidth]{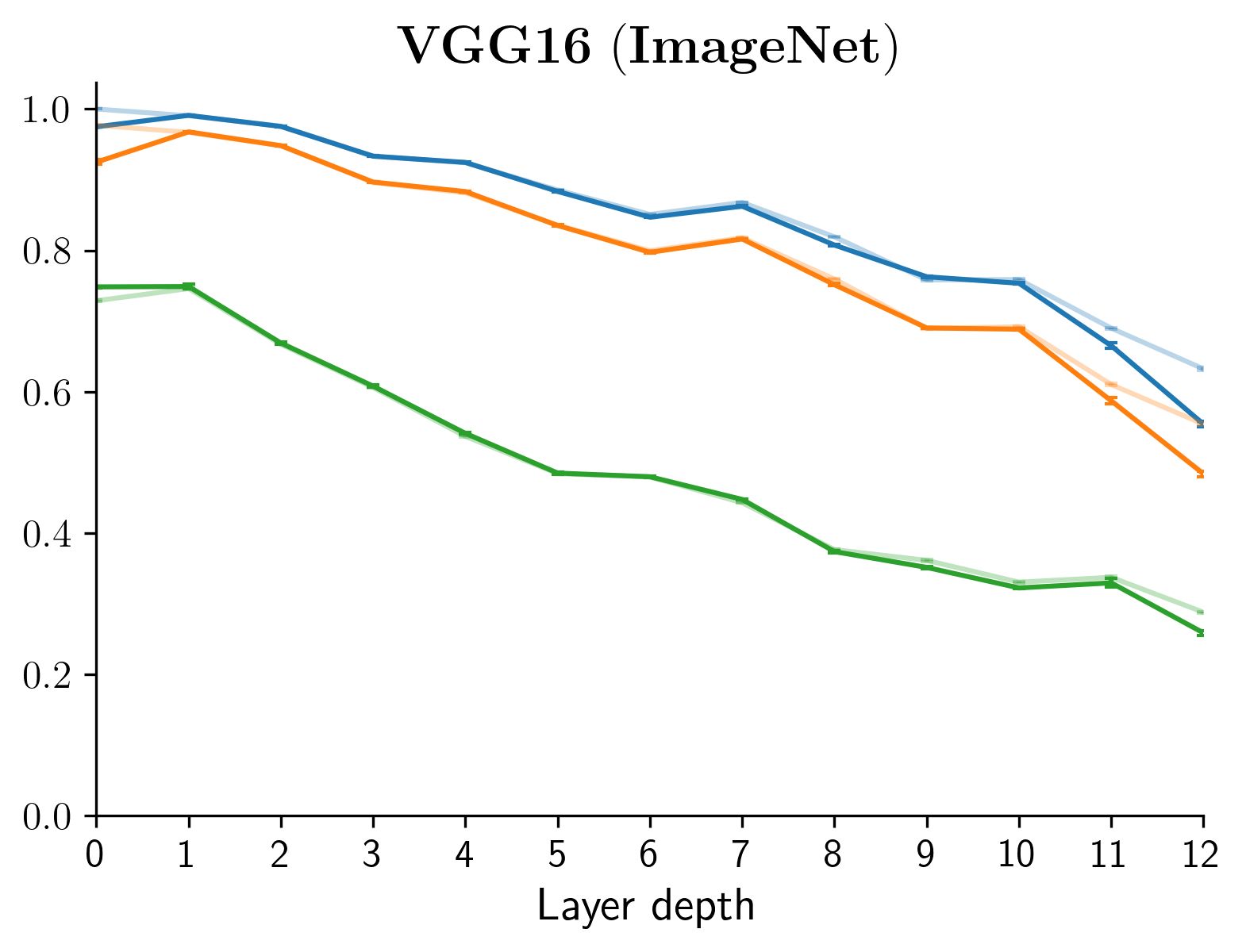}\hfill
    \includegraphics[width=.25\textwidth]{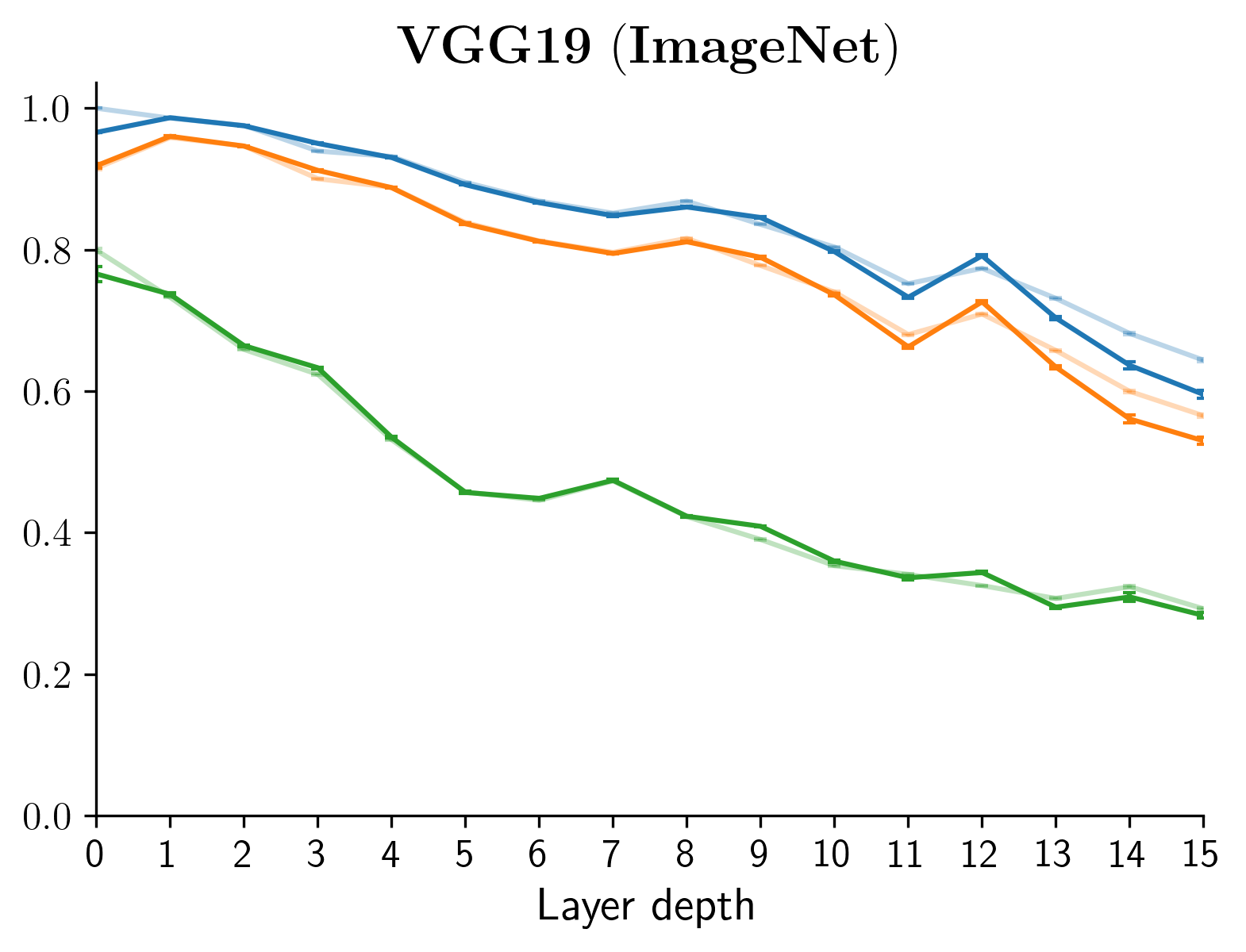}
    
    \caption{\textbf{Representational Convergence Across a Network Hierarchy Using Spearman's Rank Correlation.} We plot the evolution of alignment scores computed between different (ImageNet-trained) network seeds of the same architecture. Lighter shades of the same color denote the Spearman's ranked-correlation score, whereas darker shades indicates the Pearson correlation-based alignment score. We clearly observe that both correlation measures yield nearly identical results.}
    \label{fig:network-hierarchy-spearman}
\end{figure*}

As observed in Fig.~\ref{fig:network-hierarchy-spearman}, we note that the choice of correlation computation does not affect our conclusions.

\section{Basis Alignment in CNNs vs. Vision Transformers}
\label{sec: basis-alignment-cnn-vit}
In Sec.~\ref{sec: dcnn-results}, an interesting phenomenon emerges---a privileged basis set persists in DCNNs, whereas in case of ViTs, there is no clear evidence of a privileged solution axis. While a definitive mechanistic account remains an open area of investigation to answer this question, recent work~\citep{khosla2024privileged} offers compelling evidence that the emergence of basis alignment across CNNs and even between brains and CNNs—may be partly attributable to architectural choices, especially the presence of ReLU nonlinearities. To understand this constraint, we consider a representation of post-ReLU activations from a CNN, say $\bm{x}$. The ReLU operation ensures that all activations $\bm{x}\geq 0$, i.e.: non-negative. Now, if we apply a random rotation, say $\bm{Q}$ to these activations, we obtain a rotated basis set $\bm{y} = \bm{Qx}$, where $\bm{Q}$ is a rotation matrix. For both $\bm{x}$ and $\bm{y}$ to be valid post-ReLU activations, they must remain non-negative after the transformation. In other words, we must strictly have $\bm{y}\geq 0$. For this, the matrix $\bm{Q}$ must be a non-negative matrix. But this means that $\bm{Q}$ must be a permutation matrix, because every orthogonal matrix $\bm{Q}\in\mathbf{O}(N)$ with non-negative entries is necessarily a permutation matrix. Hence, it follows that $\bm{Q}$ can only permute (or shuffle) the activation units, rather than performing arbitrary rotations. Thus, the non-linearity induced by ReLU disrupts the rotational symmetry of the activation space, potentially explaining why different networks converge to similar bases. In contrast, Vision Transformers (ViTs) use GeLU nonlinearities in MLP layers. Moreover, the penultimate layer in ViTs often lacks \emph{any} nonlinearity. These architectural choices retain greater rotational freedom in the feature space, which likely explains the lack of axis alignment across transformer runs, as also confirmed by our results in Section~\ref{sec: vit-alignment}.

\section{Additional Results Using CIFAR100}
\label{sec: all-cifar100-results}
All analyses described in Sec.~\ref{sec: dcnn-results}, we demonstrate evidence for representational convergence along the following directions---hierarchy effects, sensitivity to solution bases, hierarchical correspondence and training-time dynamics on ImageNet. We conduct an identical set of experiments on CIFAR100, and observe that our findings generalize across these datasets.

\paragraph{Network Hierarchy.} We compare the representational convergence across a network hierarchy for different seeds of the same architecture using the CIFAR100 dataset in Fig.~\ref{fig:network-hierarchy-cifar100}.
    \begin{figure*}[htbp!]
        \centering
        % cifar100
        \includegraphics[width=.25\textwidth]{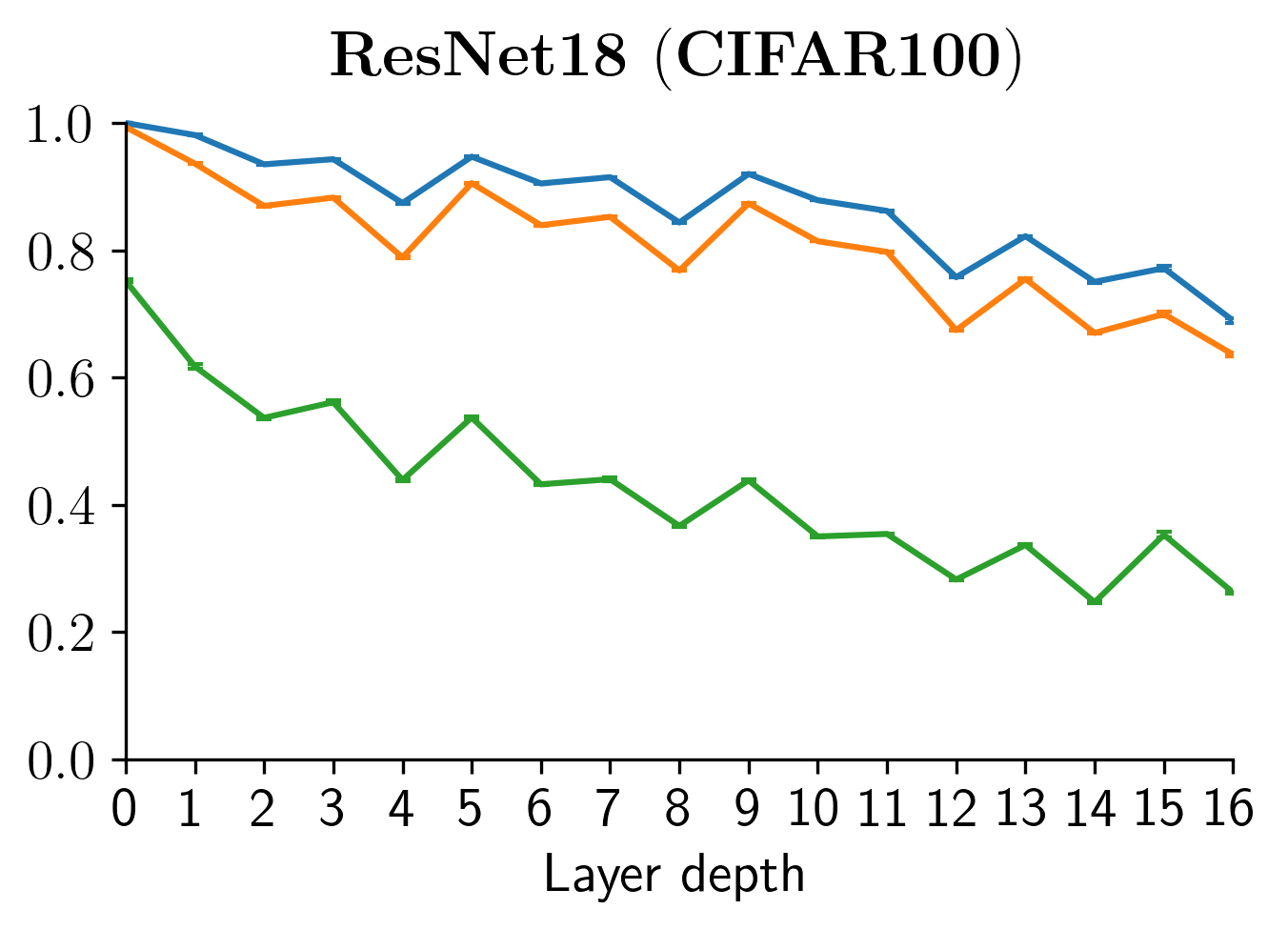}\hfill
        \includegraphics[width=.25\textwidth]{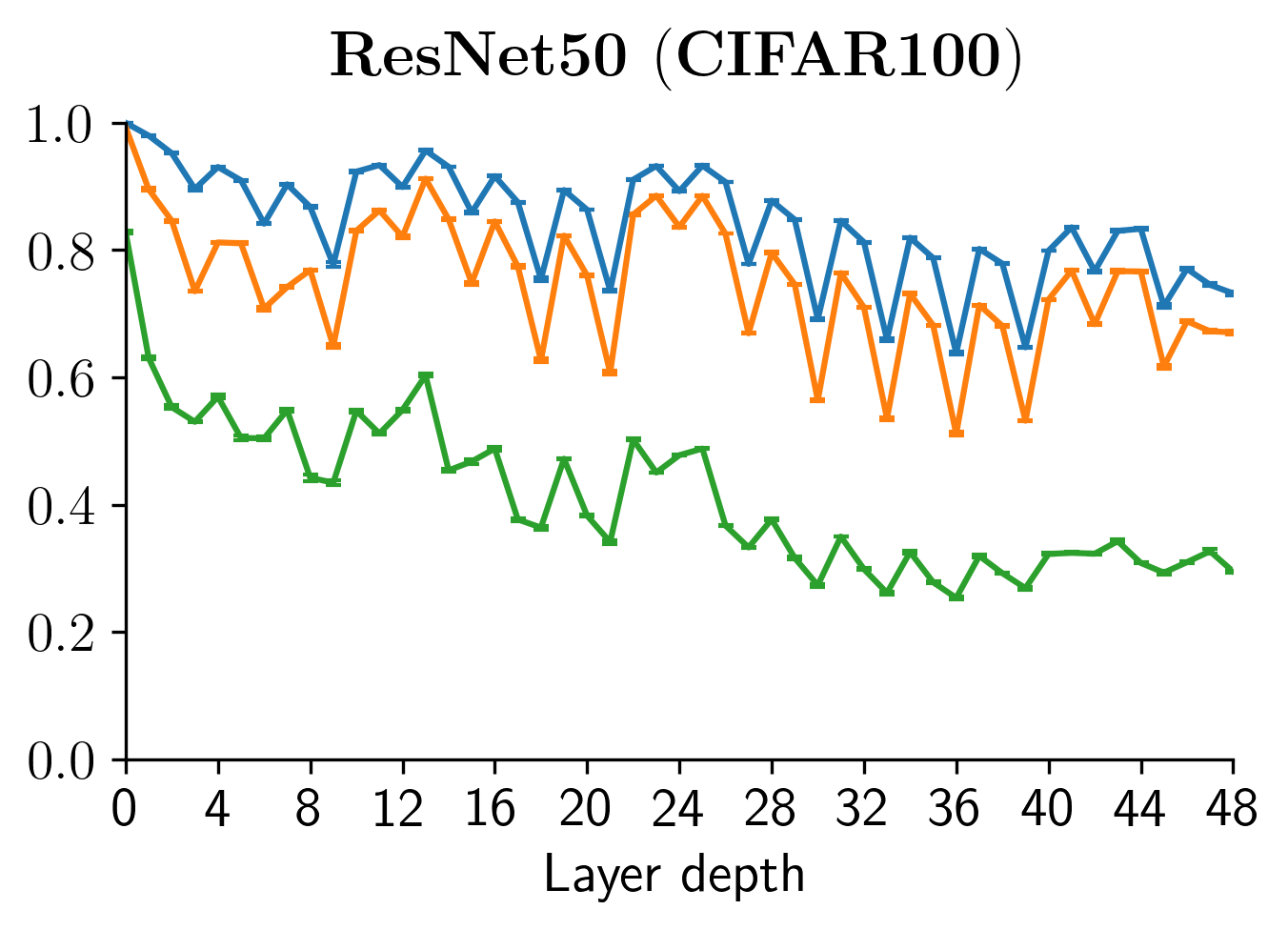}\hfill
        \includegraphics[width=.25\textwidth]{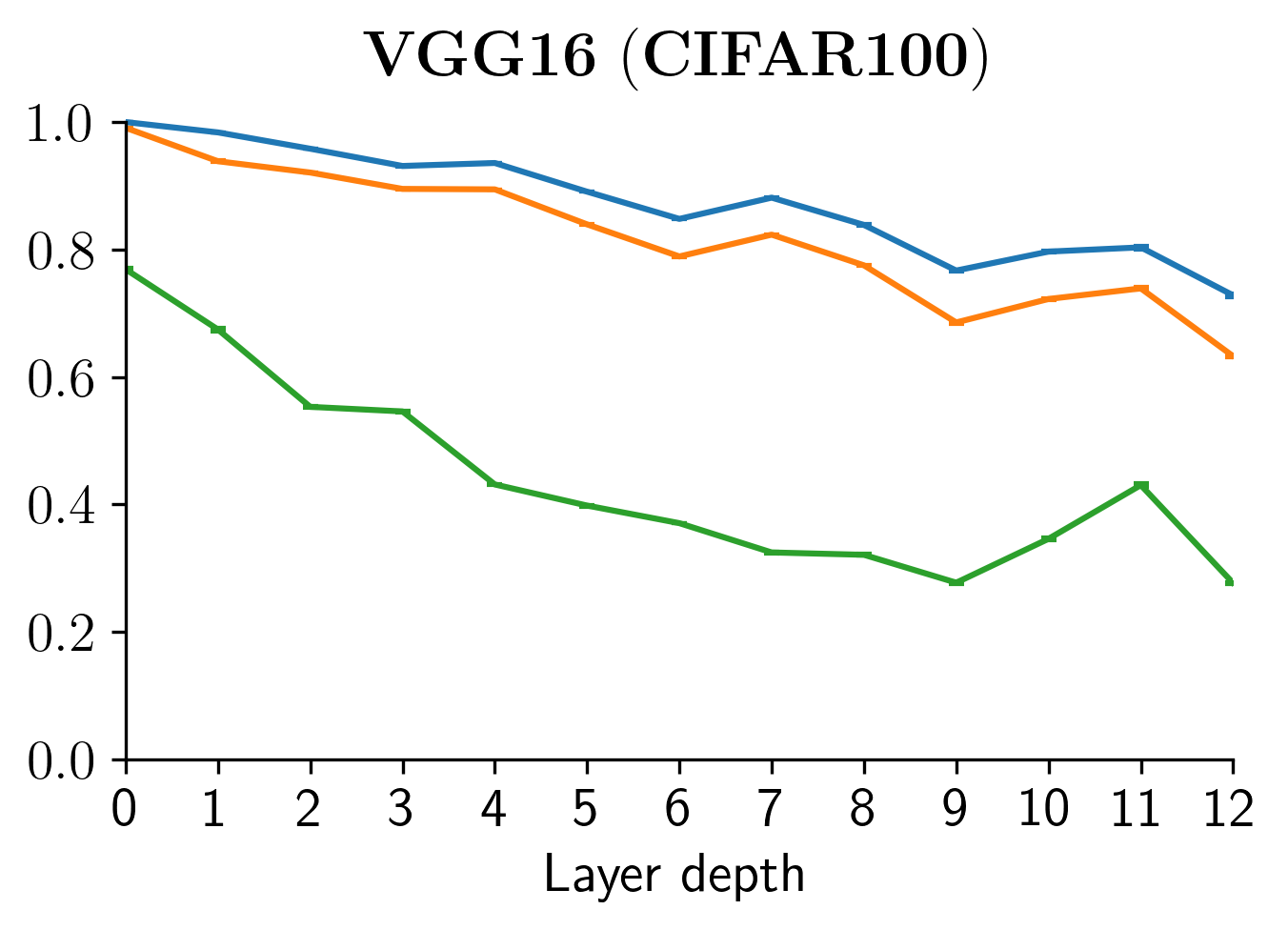}\hfill
        \includegraphics[width=.25\textwidth]{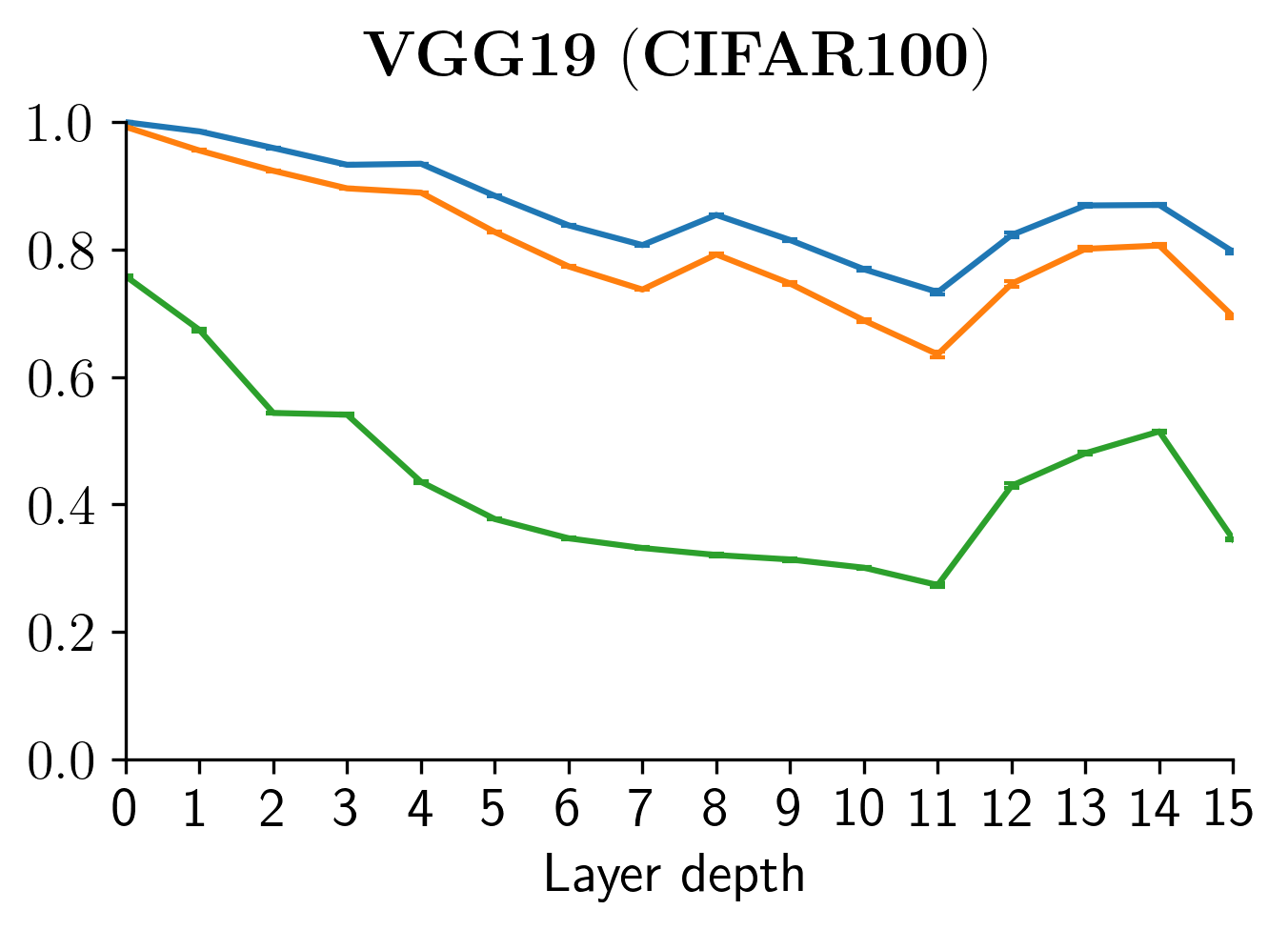}\\
       
        \caption{\textbf{Representational Convergence Across a Network Hierarchy Using CIFAR100.} Identical to ImageNet-trained networks, alignment follows the trend \textcolor{mplblue}{Linear} > \textcolor{mplorange}{Procrustes} > \textcolor{mplgreen}{Permutation}. Moreover, the \textcolor{mplblue}{Linear} and \textcolor{mplgreen}{permutation} alignment scores track each other closely, again, identical to ImageNet-trained networks.}
        \label{fig:network-hierarchy-cifar100}
    \end{figure*}
This holds an identical trend to those observed in ImageNet-trained networks---early layers show higher alignment, which tapers with network depth.

\paragraph{Sensitivity to Representational Axes.} Identical to the procedure applied to ImageNet-trained networks in Sec.~\ref{par: privileged-axes}, we apply a random rotation matrix $\bm{Q}$ to the converged basis of a neural representation of CIFAR100-trained networks.
    \begin{table}[htbp!]
        \centering
        \resizebox{\textwidth}{!}{
        \begin{tabular}{cccc}
        \toprule
        Model & Native (Min / Max) & Rotated (Min / Max) & Difference ($\%$) (Min / Max)\\
        \midrule
            ResNet18 & $0.247 \ / \ 0.752$ & $0.215 \ / \ 0.689$ & $6.40\% \ / \ 51.26\%$ \\
            ResNet50 & $0.254 \ / \ 0.828$ & $0.242 \ / \ 0.828$ & $-3.38\% \ / \ 35.29\%$ \\
            VGG16 & $0.277 \ / \ 0.769$ & $0.239 \ / \ 0.661$ & $5.66\% \ / \ 63.97\%$ \\
            VGG19 & $0.273 \ / \ 0.758$ & $0.231 \ / \ 0.684$ & $2.15\% \ / \ 35.36\%$ \\
        \bottomrule
        \end{tabular}
        }
        \vspace*{0.5em}
        \caption{\textbf{Sensitivity of Permutation Scores to Representational Axes on CIFAR100.}  For each CIFAR100-trained network we apply a random rotation to the network’s unit basis and recompute permutation alignment scores for all convolutional layers. Columns report the minimum and maximum alignment scores observed over layers in the native and rotated basis, and the final column gives the percentage change in alignment after rotation. Rotations reduce alignment, indicating that a privileged basis exists in trained networks independent of the training dataset.}
        \label{tab:perm-rotated-cifar100}   
    \end{table}

In Table~\ref{tab:perm-rotated-cifar100}, we note that alignment consistently decreases across all models after rotating the solution basis, identical to our observation on ImageNet networks.

\paragraph{Hierarchical Correspondence.} We plot the heatmap of Procrustes and Soft-Matching alignment scores for all layer and network pairs using CIFAR100.
\begin{figure*}[htbp!]
    \centering
    % procrustes
    \includegraphics[width=.16\textwidth]{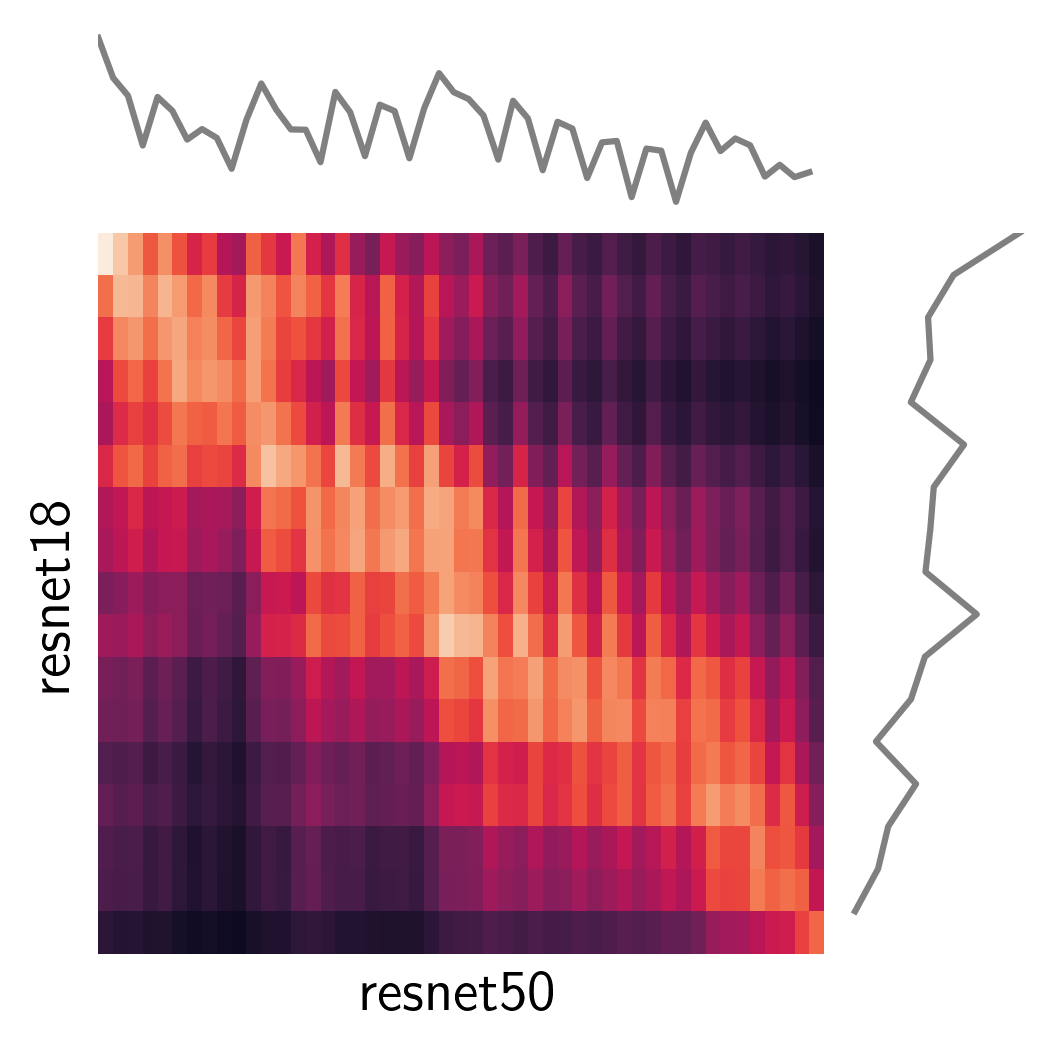}\hfill
    \includegraphics[width=.16\textwidth]{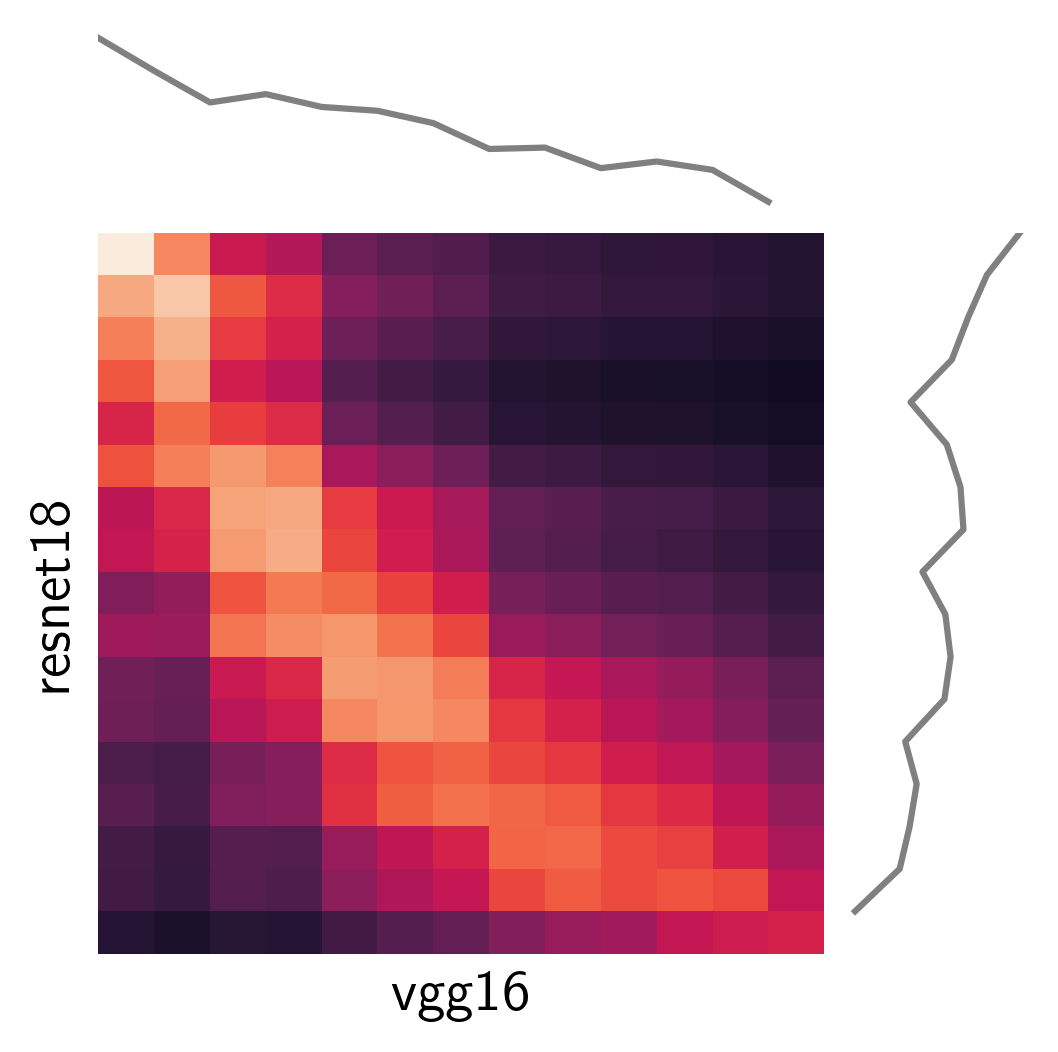}\hfill
    \includegraphics[width=.16\textwidth]{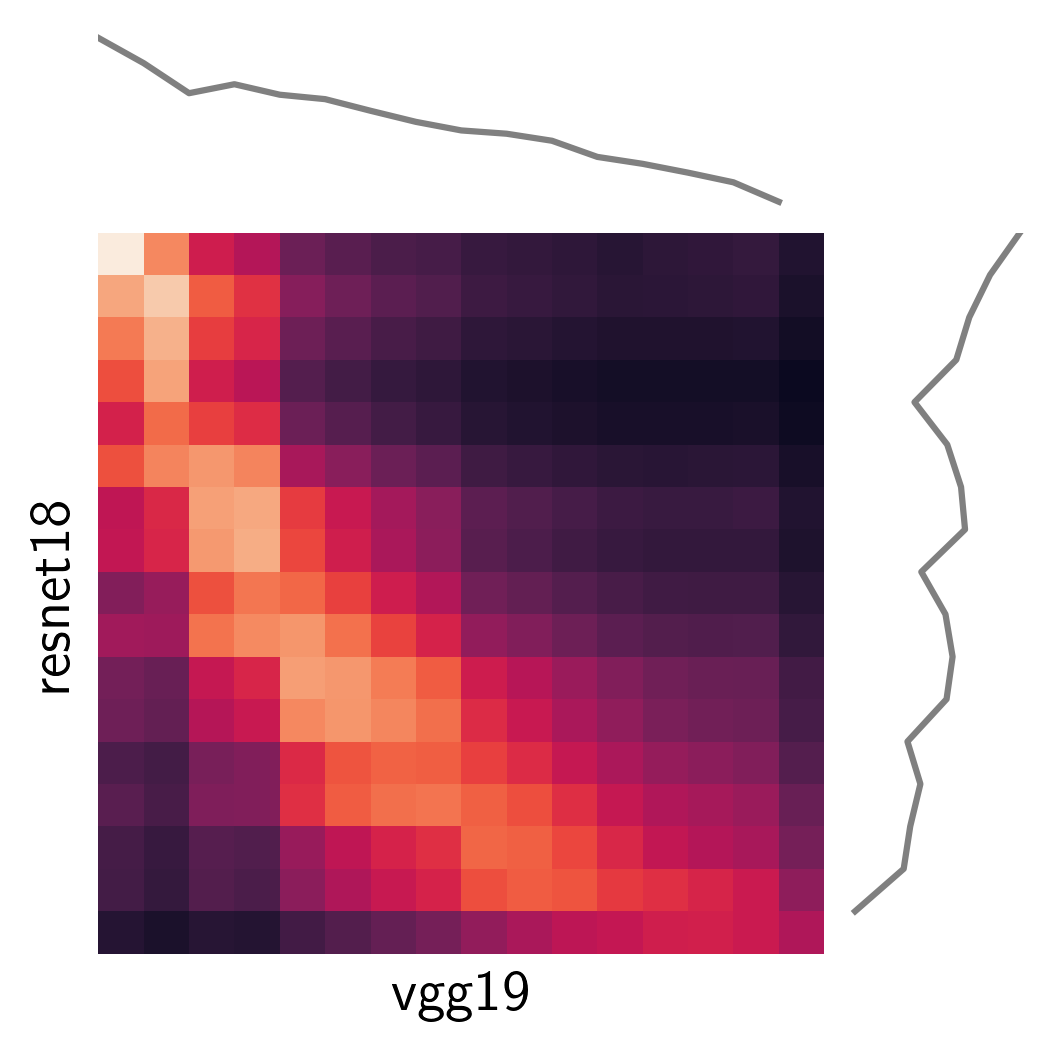}\hfill 
    \includegraphics[width=.16\textwidth]{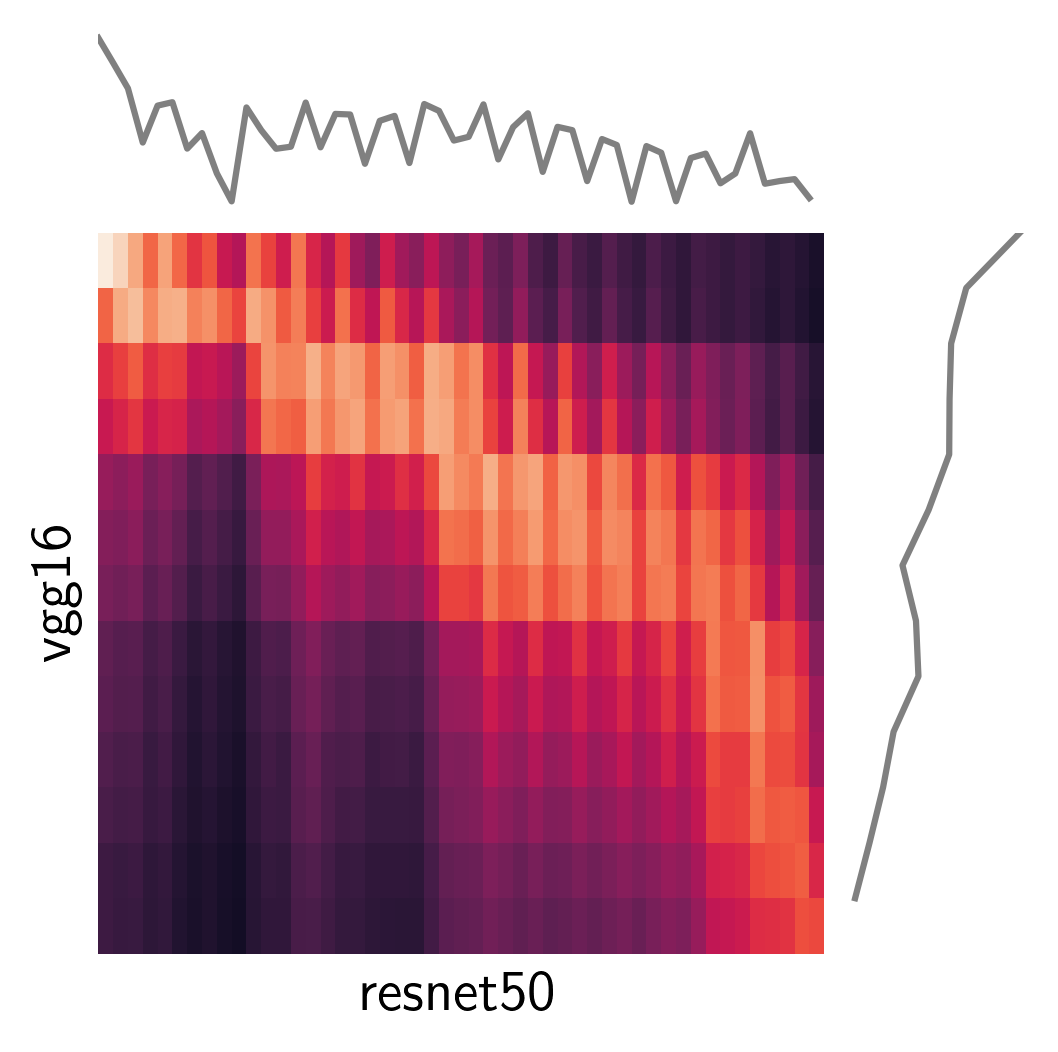}\hfill
    \includegraphics[width=.16\textwidth]{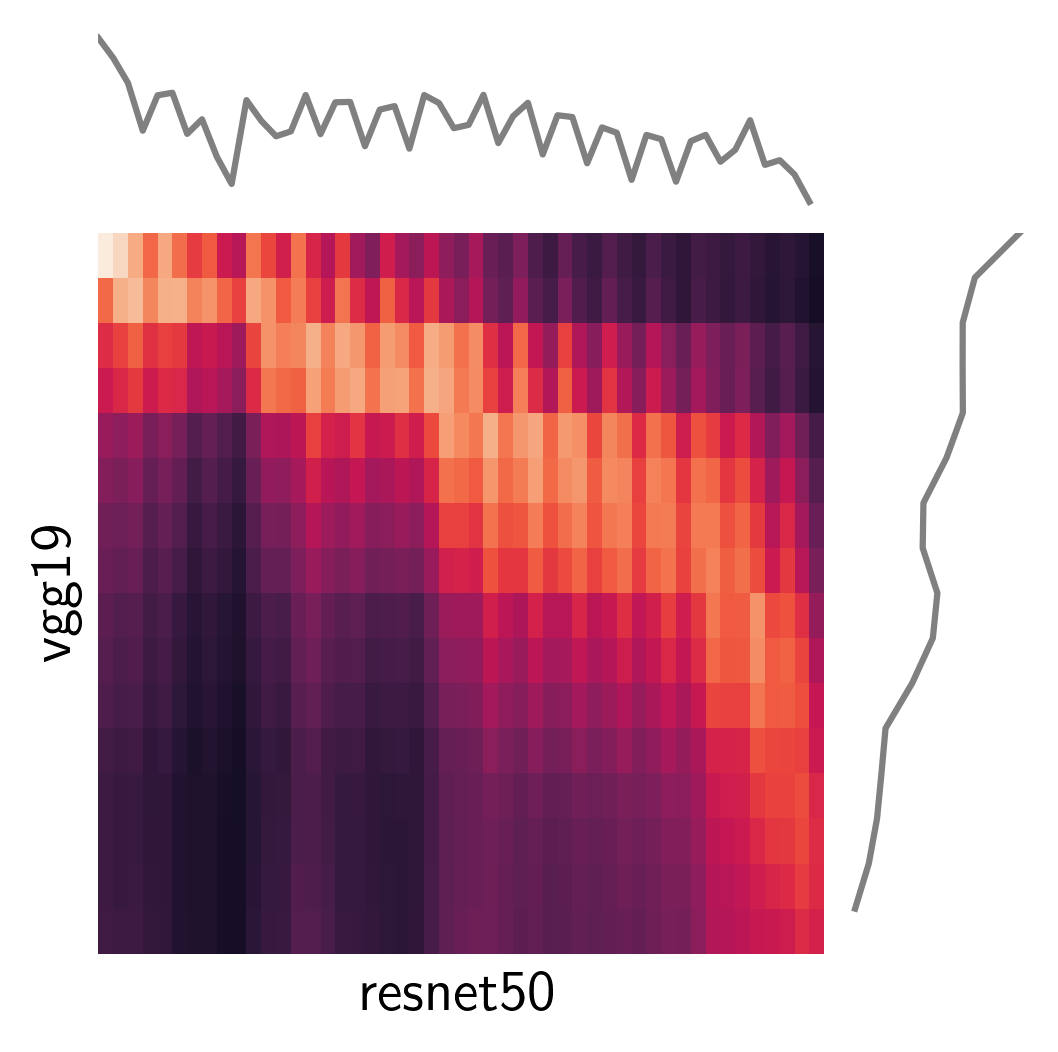}\hfill 
    \includegraphics[width=.16\textwidth]{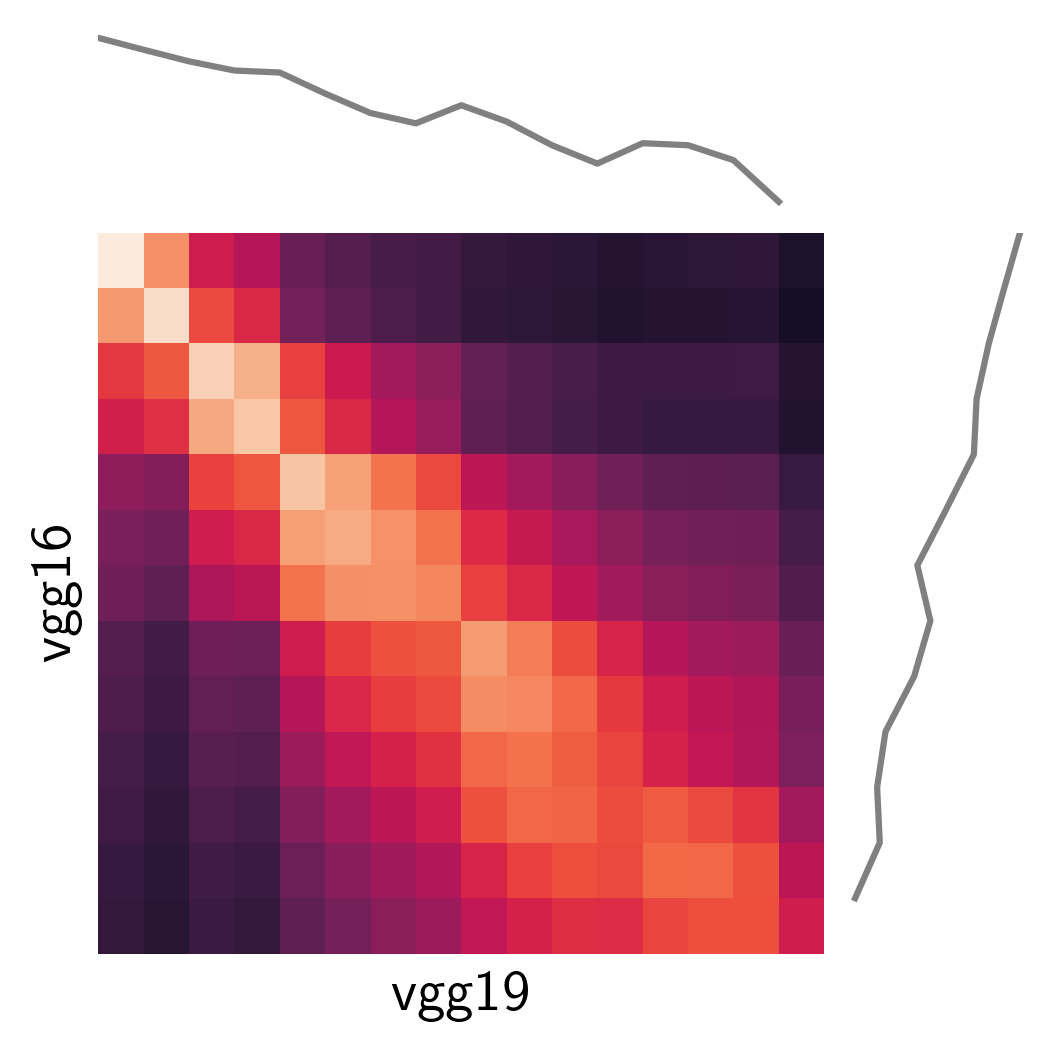}\\
    
    % soft-match
    \includegraphics[width=.16\textwidth]{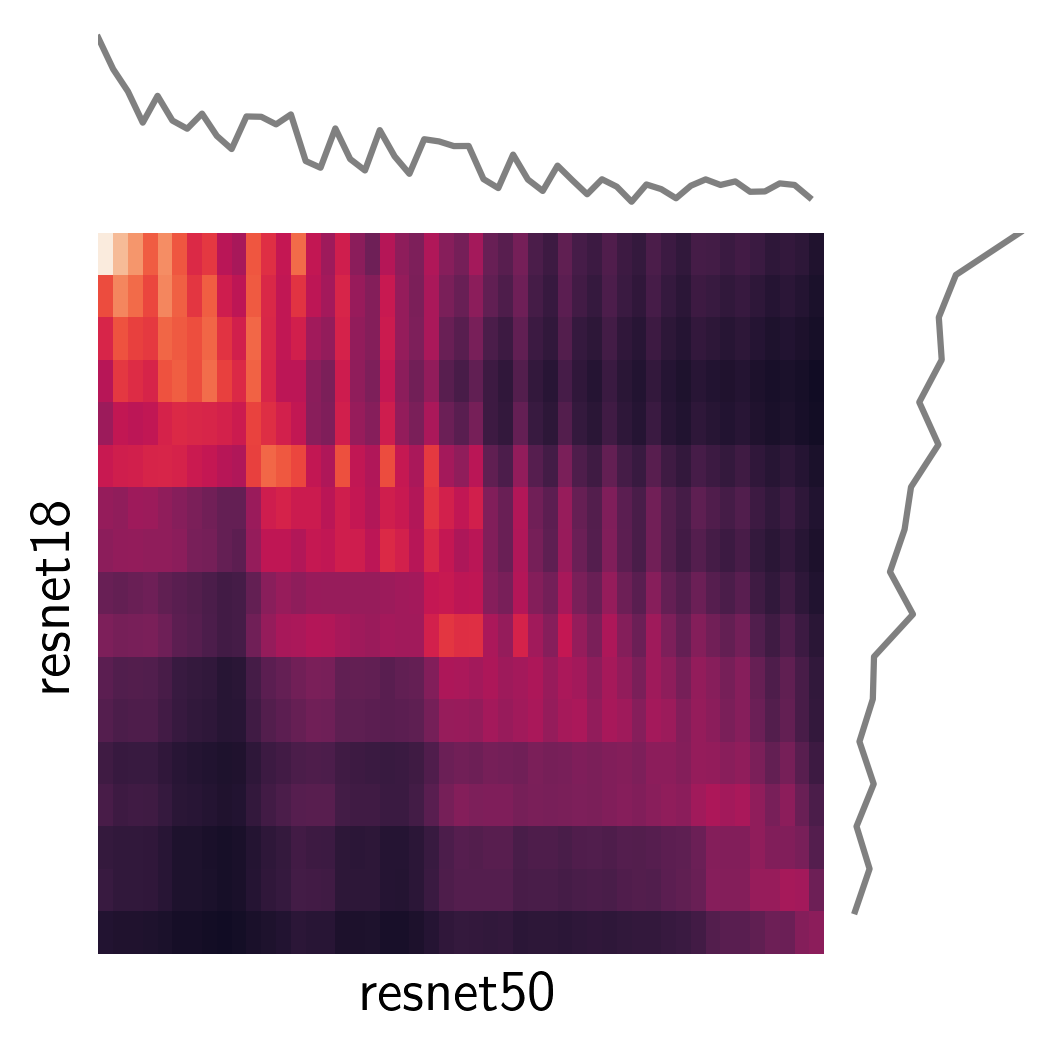}\hfill
    \includegraphics[width=.16\textwidth]{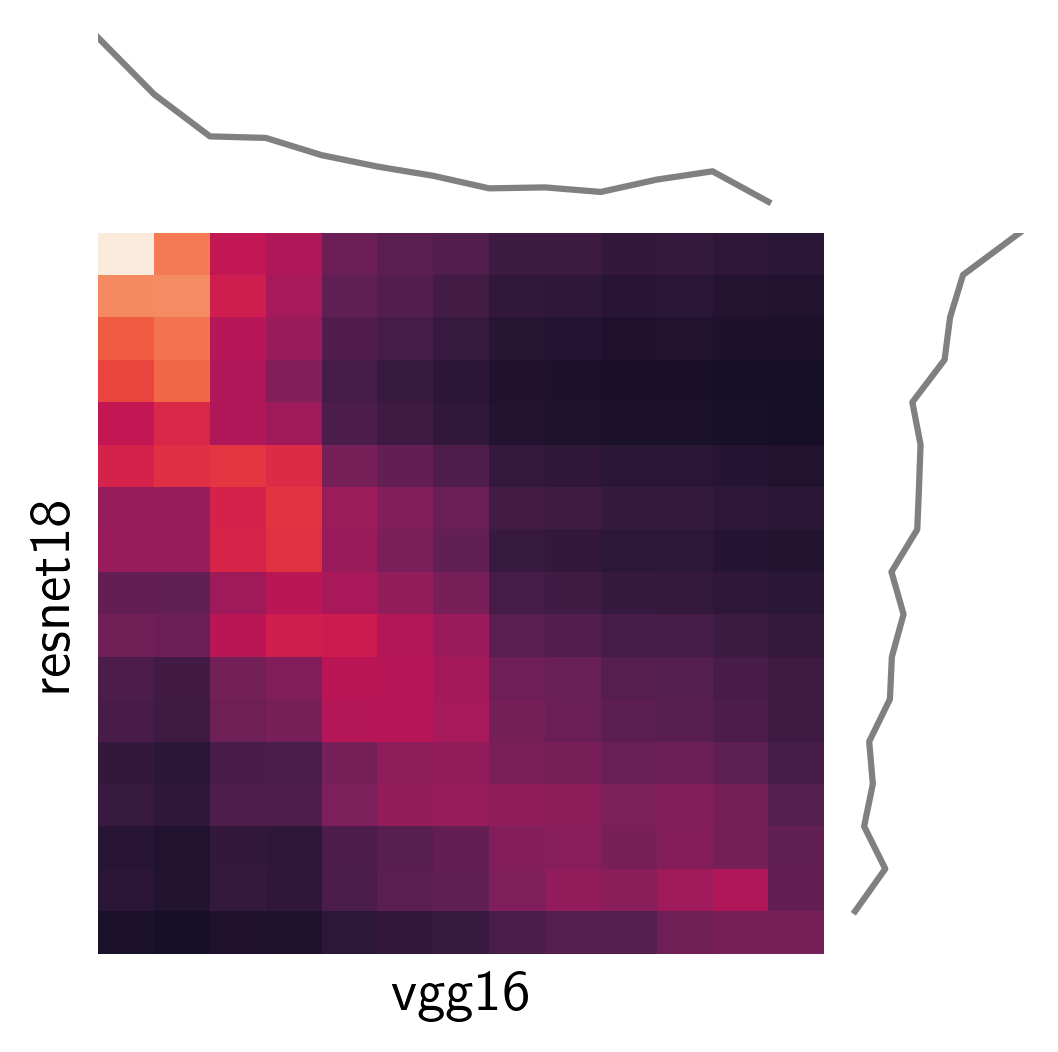}\hfill
    \includegraphics[width=.16\textwidth]{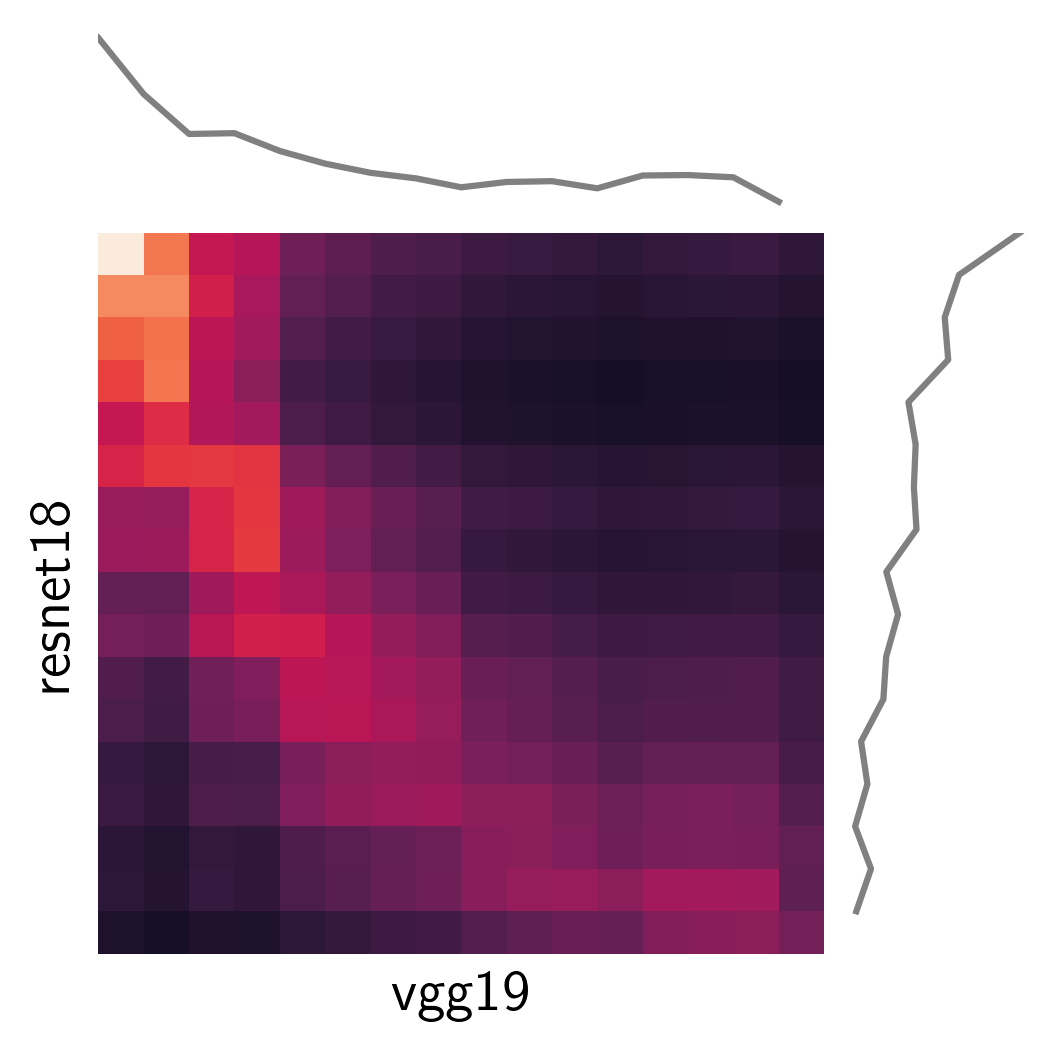}\hfill 
    \includegraphics[width=.16\textwidth]{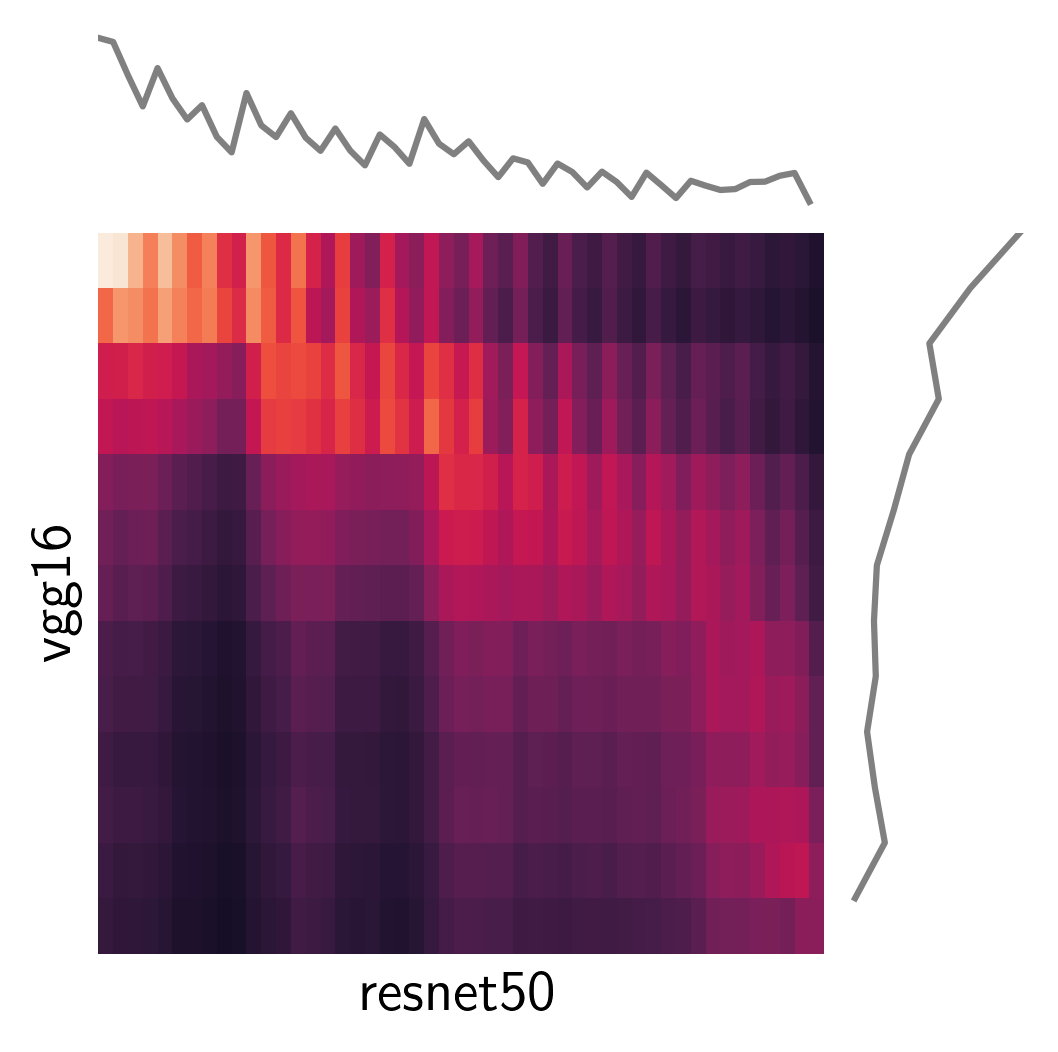}\hfill
    \includegraphics[width=.16\textwidth]{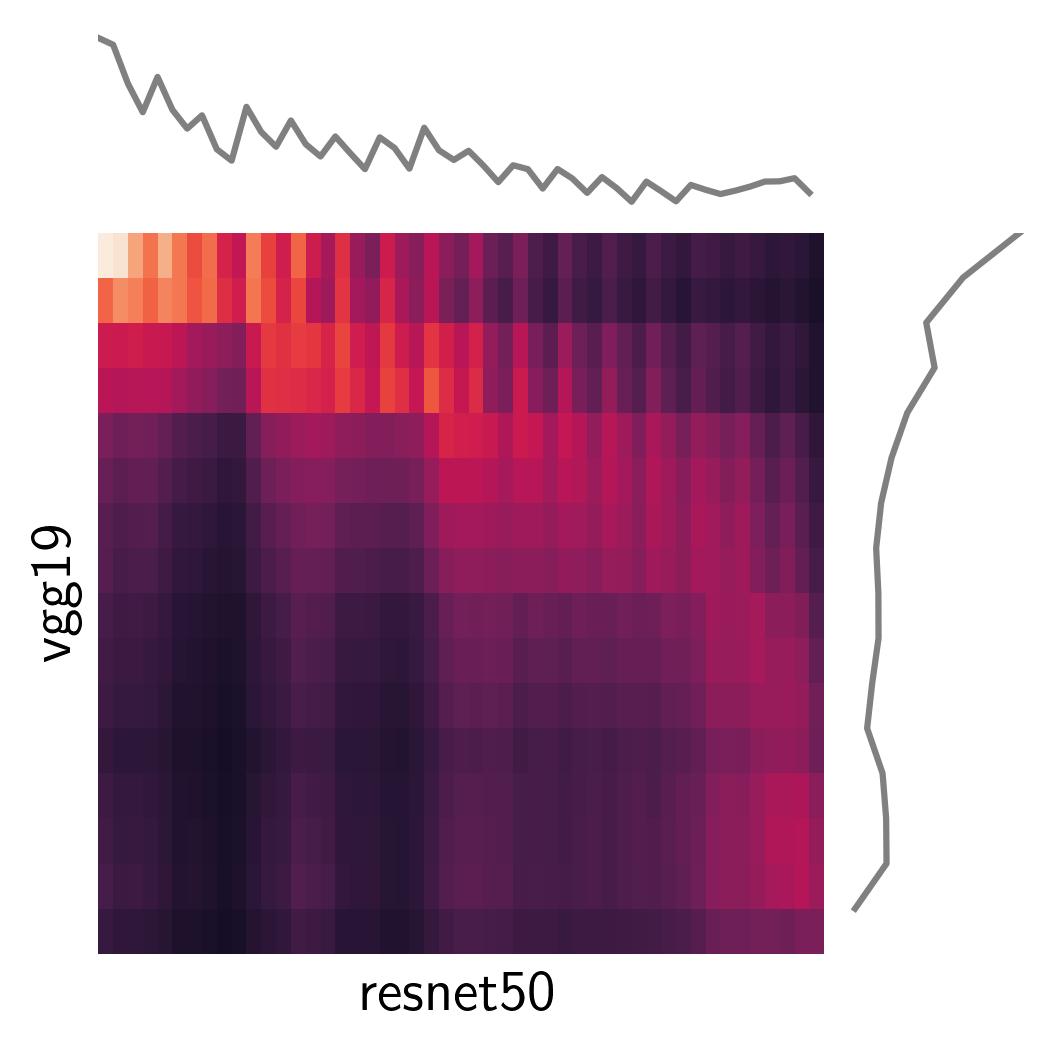}\hfill 
    \includegraphics[width=.16\textwidth]{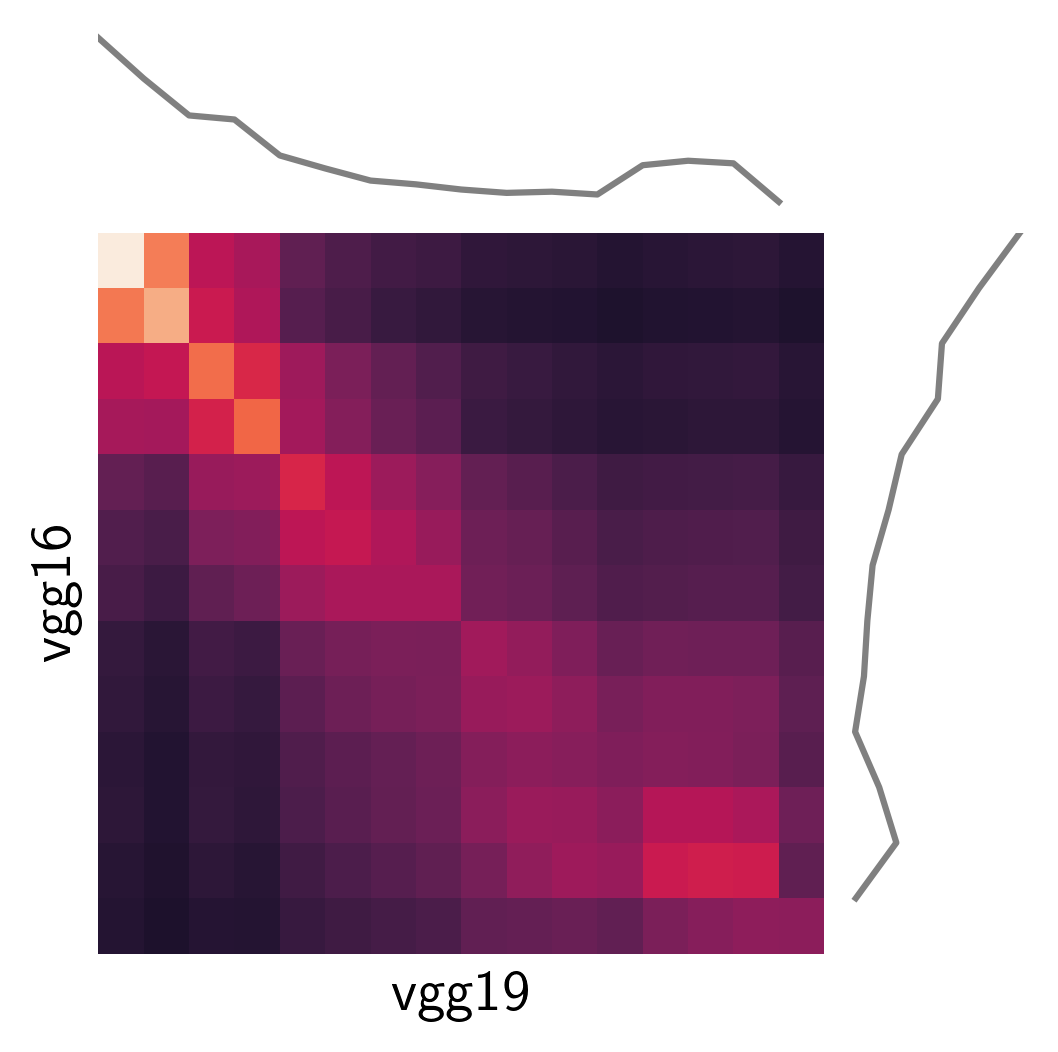}
    
    \caption{\textbf{Inter-Model Comparisons on CIFAR100.} We consider all pairs of vision models, and compute the alignment scores between every pair of layers using the orthogonal Procrustes (\textbf{Top}) and Soft-Matching (\textbf{Bottom}) metric trained on CIFAR100. Gray line plots denote the maximum alignment value for each network over rows and columns.}%We consider all pairs of vision models, and for each pair, compute the alignment scores between every pair of layers using the orthogonal Procrustes metric trained on ImageNet. Gray line plots denote the \textbf{maximum} alignment value for each network over rows (right line) and columns (top line). A common trend that is observed here is the consistent relationships between layers of CNNs trained with different architectures.}
    \label{tab:procrustes_inter_model}
\end{figure*}
For both alignment metrics, we observe a hierarchical correspondence---layers at approximately similar depths in a network pair are more highly aligned than dissimilar depths.

\paragraph{Convergence over Training.} We plot the Procrustes alignment scores between all network pairs trained on CIFAR100 from epochs $0$ (untrained) through $10$ in Fig.~\ref{fig:cifar100-training-evo}.

\begin{figure}[htbp!]
    \centering
    \includegraphics[width=\linewidth]{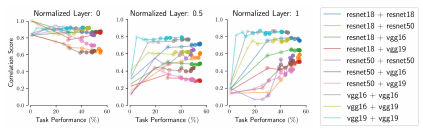}
    \caption{\textbf{Representational Alignment Through Training CIFAR100 Networks.} We plot the evolution of Procrustes alignment between network pairs during training on CIFAR100. Lighter shades indicate earlier epochs, progressively darkening with later epochs. The plots range from epoch $0$ (untrained) to epoch $10$, with task performance improving over time. Epoch progression can be inferred from the increasing task performance along the $x$-axis. This trend is identical to the convergence dynamics seen in ImageNet training---bulk of the alignment occurs within the first epoch itself, after which alignment saturates or even slightly reduces in some cases.}
    \label{fig:cifar100-training-evo}
\end{figure}

Identical to ImageNet networks in Sec.~\ref{sec: convergence-dynamics}, we see that early convolutional layers have almost no alignment change, presumably due to the fact that early layers learn filters with approximately linear isometries. On the other hand, in later layers, we observe that the bulk of representational alignment happens in the first epoch itself, independent of network task performance.

\section{Out-Of-Distribution Datasets}
\label{sec: ood-dataset-details}
All OOD datasets were directly taken from~\citep{geirhos2018imagenet}, which share the same $16$ coarse labels as ImageNet. Concretely, this set consists of the following classes: Airplane, Bear, Bicycle, Bird, Boat, Bottle, Car, Cat, Chair, Clock, Dog, Elephant, Keyboard, Knife, Oven, Truck.\\

\noindent
Each of the $17$ stylized datasets are described below:
\begin{itemize}
    \item \textbf{Color:} Half of the images are randomly converted to grayscale, and the rest kept in their original colormap. 
    \item \textbf{Stylized:} Textures from one class are transferred to the shapes of another, ensuring that object shapes remain preserved.
    \item \textbf{Sketch:} Cartoon-style sketches of objects representing each class.  
    \item \textbf{Edges:} Generated from the original ImageNet dataset using the Canny edge detector to produce edge-based representations.
    \item \textbf{Silhouette:} Black objects on a white background generated from the original dataset.
    \item \textbf{Cue Conflict:} Images with textures that conflict with shape categories, generated using iterative style transfer~\citep{gatys2015neural}, where \textbf{Texture} dataset images serve as the style and \textbf{Original} dataset images as the content.
    \item \textbf{Contrast:} Image variants modified to different contrast levels.
    \item \textbf{High-Pass / Low-Pass:} Images processed with Gaussian filters to emphasize either high-frequency or low-frequency components.
    \item \textbf{Phase-Scrambling:} Images with phase noise added to frequency components, introducing varying levels of distortion from $0^\circ$ to $180^\circ$.
    \item \textbf{Power-Equalization:} The images were processed to normalize the power spectra across the dataset by adjusting all amplitude spectra to match the mean value.
    \item \textbf{False-Color:} The colors of the images were inverted to their opponent colors while maintaining constant luminance, using the DKL color space.
    \item \textbf{Rotation:} Rotated images ($0^\circ$, $90^\circ$, $180^\circ$, or $270^\circ$) to test rotational invariance.
    \item \textbf{Eidolon I, II, III:} The images were distorted using the Eidolon toolbox, with variations in the coherence and reach parameters to manipulate both local and global image structures for each intensity level.
    \item \textbf{Uniform Noise:} White uniform noise was added to the images in a varying range to assess robustness, with pixel values exceeding the bounds clipped to the range $[0, 255]$.
\end{itemize}

\section{Additional Results on Convergence Across Distribution Shifts}
We also computed Procrustes alignment for the remainder of vision networks at the first convolutional and penultimate layer to assess whether a similar phenomenon holds as described in Sec.~\ref{sec:ood-results}. Indeed, in Fig.~\ref{fig:ood-convergence-appendix}, we observe a similar trend that was observed earlier, i.e.: alignment mirrors task performance at higher network depths. 

\begin{figure*}[htbp!]
    \resizebox{\textwidth}{!}{
    \centering
    \begin{tabular}{cc}
        \includegraphics[width=0.5\textwidth]{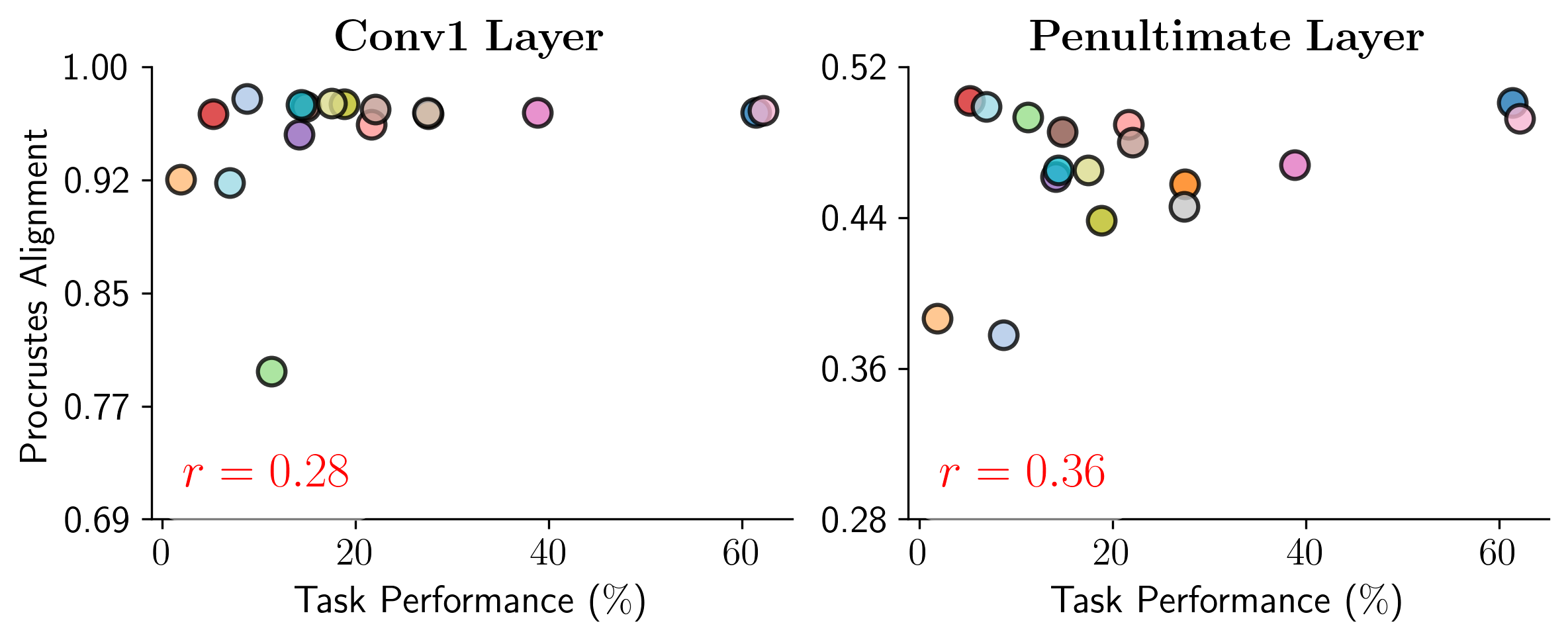} &      
        \includegraphics[width=0.5\textwidth]{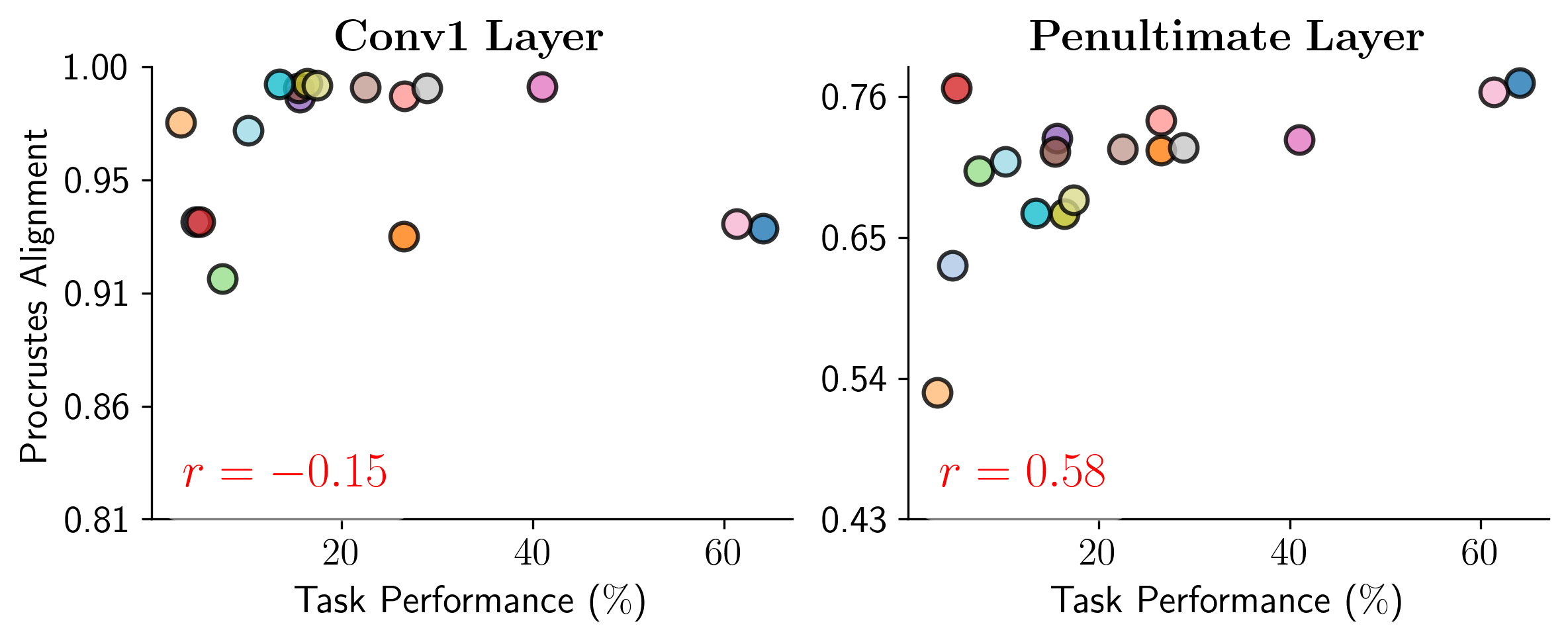} \\
        \textbf{(a)} ResNet18 & \textbf{(b)} VGG16 \\
        \multicolumn{2}{c}{\includegraphics[width=0.5\textwidth]{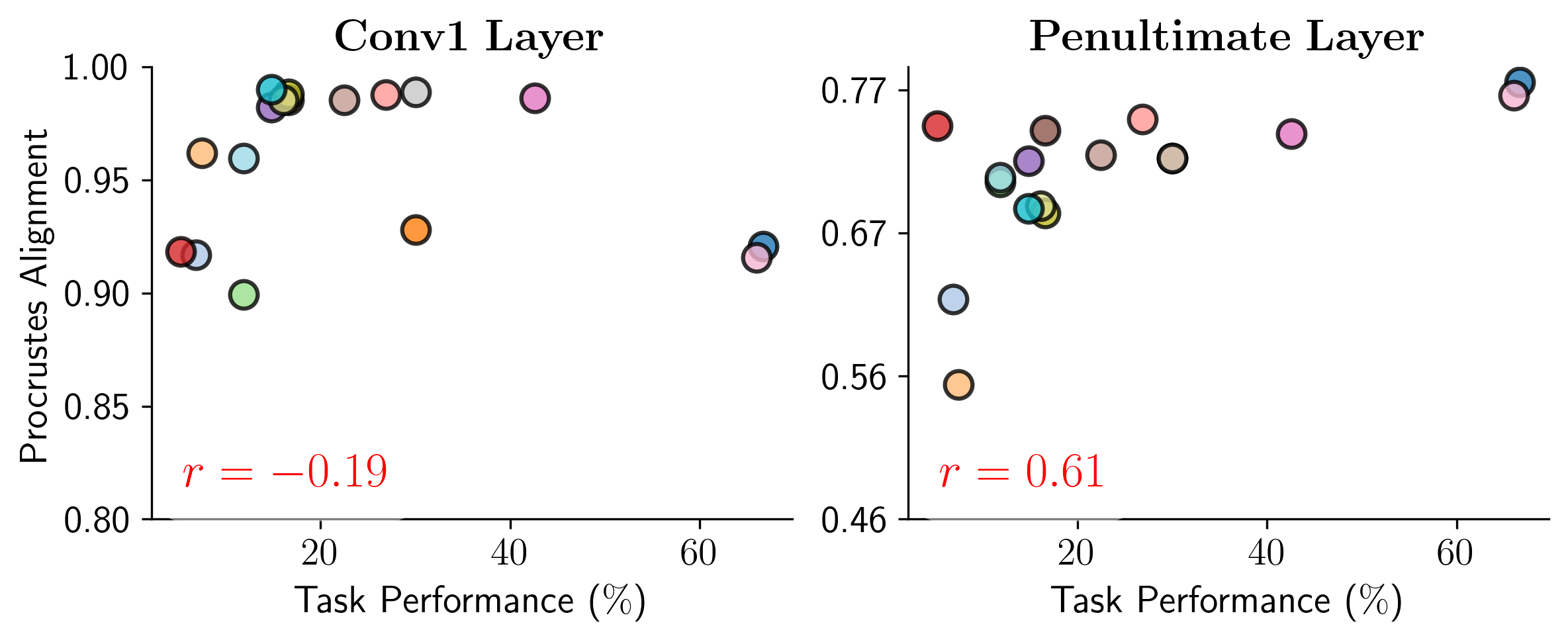}} \\
        \multicolumn{2}{c}{\textbf{(c)} VGG19}
    \end{tabular}
    }
    \caption{\textbf{Procrustes Alignment vs. Task Performance} We compute the Procrustes alignment of different network architectures on each of the $17$ datasets for the first convolutional layer \textbf{(Left)} and the penultimate \textbf{(Right)} layer from \textbf{(a)} - \textbf{(c)}.}
    \label{fig:ood-convergence-appendix}
\end{figure*}

\section{Representational Alignment Over Training}\label{sec: training-rep-trained}
In Section~\ref{sec: convergence-dynamics}, we compared networks trained for identical epochs and found that representational alignment plateaued within the first epoch. This rapid convergence, however, could still reflect networks following similar developmental trajectories driven by task optimization—essentially reaching high alignment early because they traverse a universal learning path toward the task solution.
\begin{figure}[htbp!]
    \centering
    \includegraphics[width=0.9\textwidth]{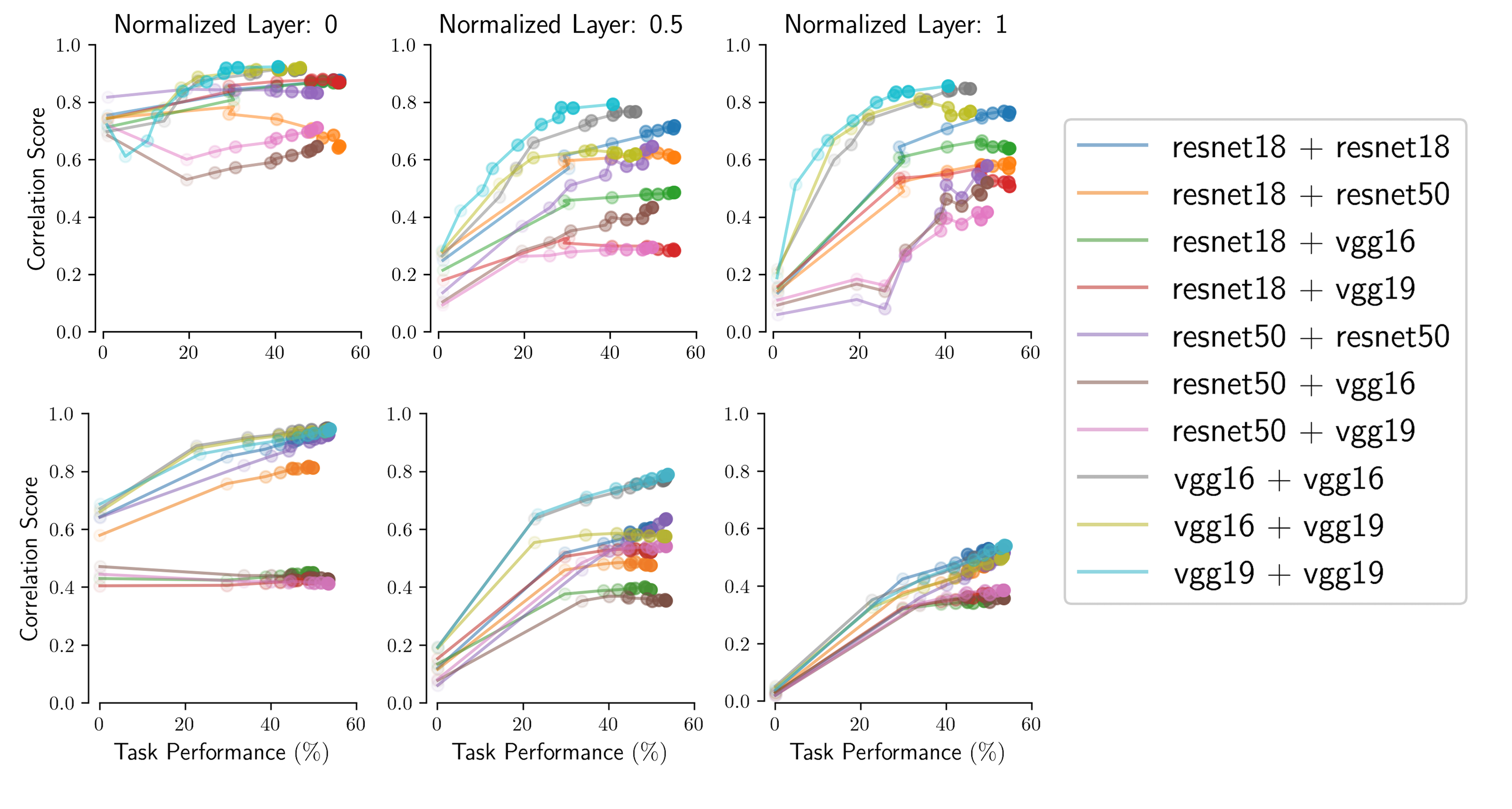}
    \caption{\textbf{  Evolution of Representational Alignment to a Fully Trained Network.} Procrustes alignment between each training checkpoint and the fully trained reference model, shown for CIFAR-$100$ (\textbf{Top}) and ImageNet (\textbf{Bottom}). Each marker is one epoch ($0 =$ untrained, $10 =$ ten epochs), with color lightening for early epochs and darkening as training progresses. Alignment climbs sharply within the first epoch and then levels off, while the earliest convolutional layers exhibit only minimal change---highlighting that most convergence occurs long before peak task performance is reached.}
    \label{fig: alignment-wrt-trained}
\end{figure}

To test whether task-optimization explains this phenomenon, we compared fully trained networks with networks at various intermediate training stages. Remarkably, high representational alignment still emerged predominantly within the first epoch, well before networks achieved optimal task performance (Fig.~\ref{fig: alignment-wrt-trained}). The earliest convolutional layers showed minimal change throughout training, consistent with learning approximately linear transformations for basic visual feature extraction.
Altogether, these results imply that representational convergence is driven by early optimization dynamics, not by attaining the final task solution.

\section{Comparisons to Brain Data}
\label{sec: nsd-comparisons}
In the following section, we apply our comparative analysis framework on brain data (Sec.~\ref{fig:network-hierarchy}). We analyze fMRI responses from four subjects (IDs $1$, $2$, $5$ and $7$) using data from the Natural Scenes Dataset (NSD)~\citep{allen2022massive}. In this dataset, each subject viewed $37,000$ \emph{naturalistic} images, with $1000$ images shared among all participants. For our analysis, we use these  $1000$ shared images to find how representational alignment between different subjects brains changes across the network hierarchy and to better understand the minimal sets of transformations needed to align two brains. We use the Soft-Matching score instead of the permutation alignment score since the number of recorded voxels is different across all subjects.\

We align responses from five key brain regions along the visual pathway: $\mathrm{V1}, \mathrm{V2}, \mathrm{V3}, \mathrm{V4},$ and the high-level ventral stream, arranged in approximate order of increasing visual processing depth. Regions $\mathrm{V1}$–$\mathrm{V4}$ are defined using the population receptive field (pRF) localizer scan session from the NSD, and the high-level ventral visual stream region is delineated according to the NSD streams atlas. All alignment values are normalized by the mean noise ceiling for each brain region, with noise ceilings computed following the standard procedure described in \citep{allen2022massive}, based on the variability in voxel responses across three repeat measurements per stimulus.

\begin{figure}[htbp!]
    \centering
    \includegraphics[width=0.5\textwidth]{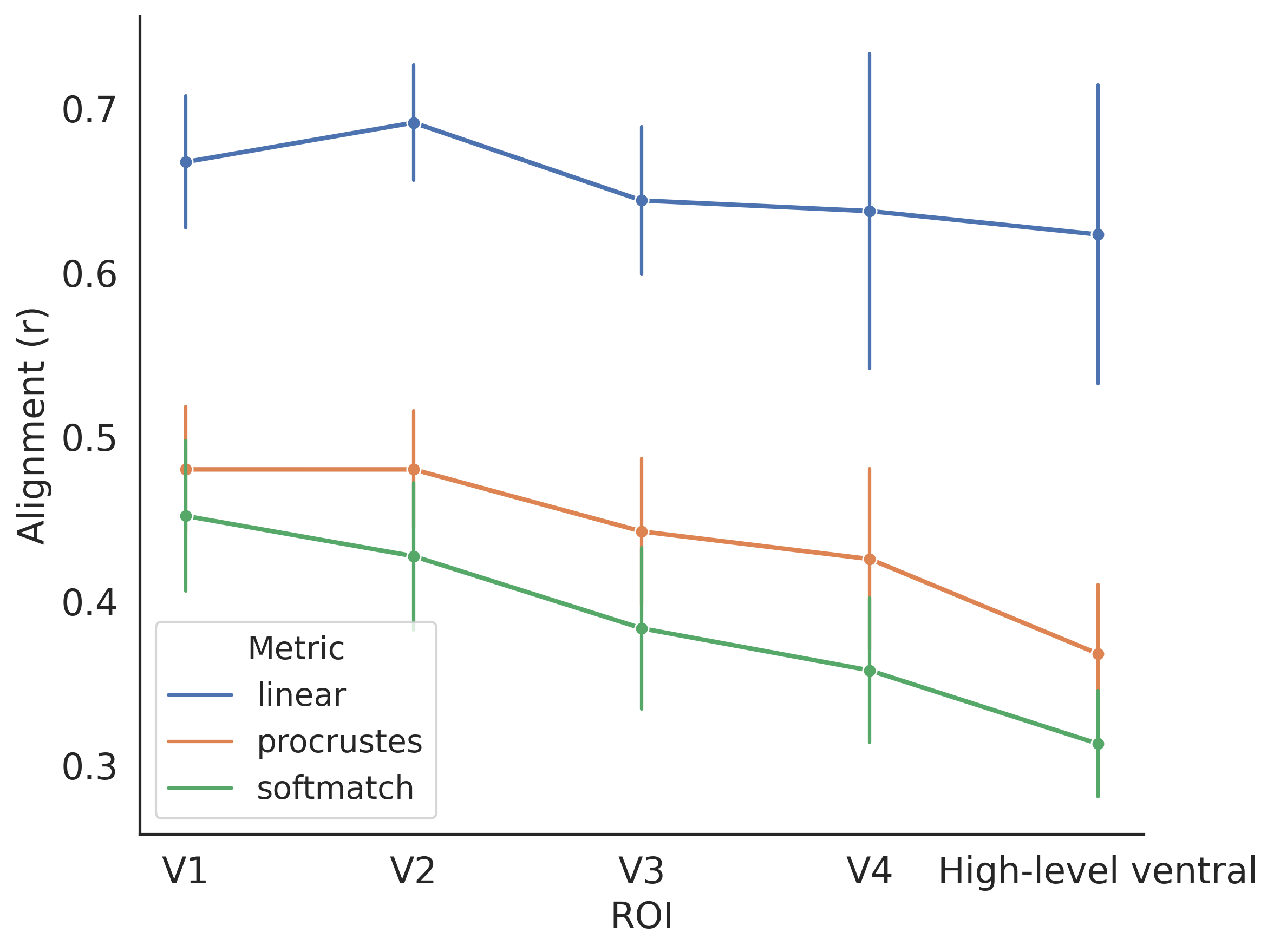}
    \caption{\textbf{Convergence Across the Visual Cortex.} Evolution of alignment scores computed between different NSD participants across the visual cortex hierarchy. Consistent with Fig.~\ref{fig:network-hierarchy}, alignment decreases along the \emph{depth} of the visual cortex. Notably, \textcolor{mplgreen}{Soft-Matching} achieves comparable alignment scores to \textcolor{mplorange}{Procrustes}, suggesting a strong, region-specific voxel correspondence across subjects. Error bars denote standard deviation across all $(n = 6)$ participant pairs.}
    \label{fig:nsd-core}
\end{figure}

We observe that consistent with network results, inter-subject alignment decreases with visual processing depth across all alignment metrics (Fig.~\ref{fig:nsd-core}). However, unlike the network results, the soft-match scores closely approximate Procrustes scores in these brain data, suggesting that voxel responses are already highly axis-aligned across subjects and thus leave little room for rotations to further improve alignment.  Notably, we also observe a substantial gap between Procrustes and linear alignment in the brain data, in contrast to ANNs where Procrustes closely approximates linear alignment. This discrepancy implies that inter-subject variability in human brains requires more flexible transformations (\emph{e.g.,} scaling or shearing) to achieve high alignment. Such variability could stem from individual differences in anatomical and functional organization, or from imperfect cross-subject ROI definitions.  

\section{Choosing Random Pixels in the Convolutional Map}
Throughout the manuscript, we use the central pixel from each convolutional feature map as a representative sample for alignment analyses. However, this begs a simple question---does spatial choice bias our results? To test this, we repeat the analyses using a random activation pixel for each model–seed pair in ImageNet-trained networks (Fig.~\ref{fig:network-hierarchy-random-pixel}).
\begin{figure*}[htbp!]
    \centering
    \includegraphics[width=.25\textwidth]{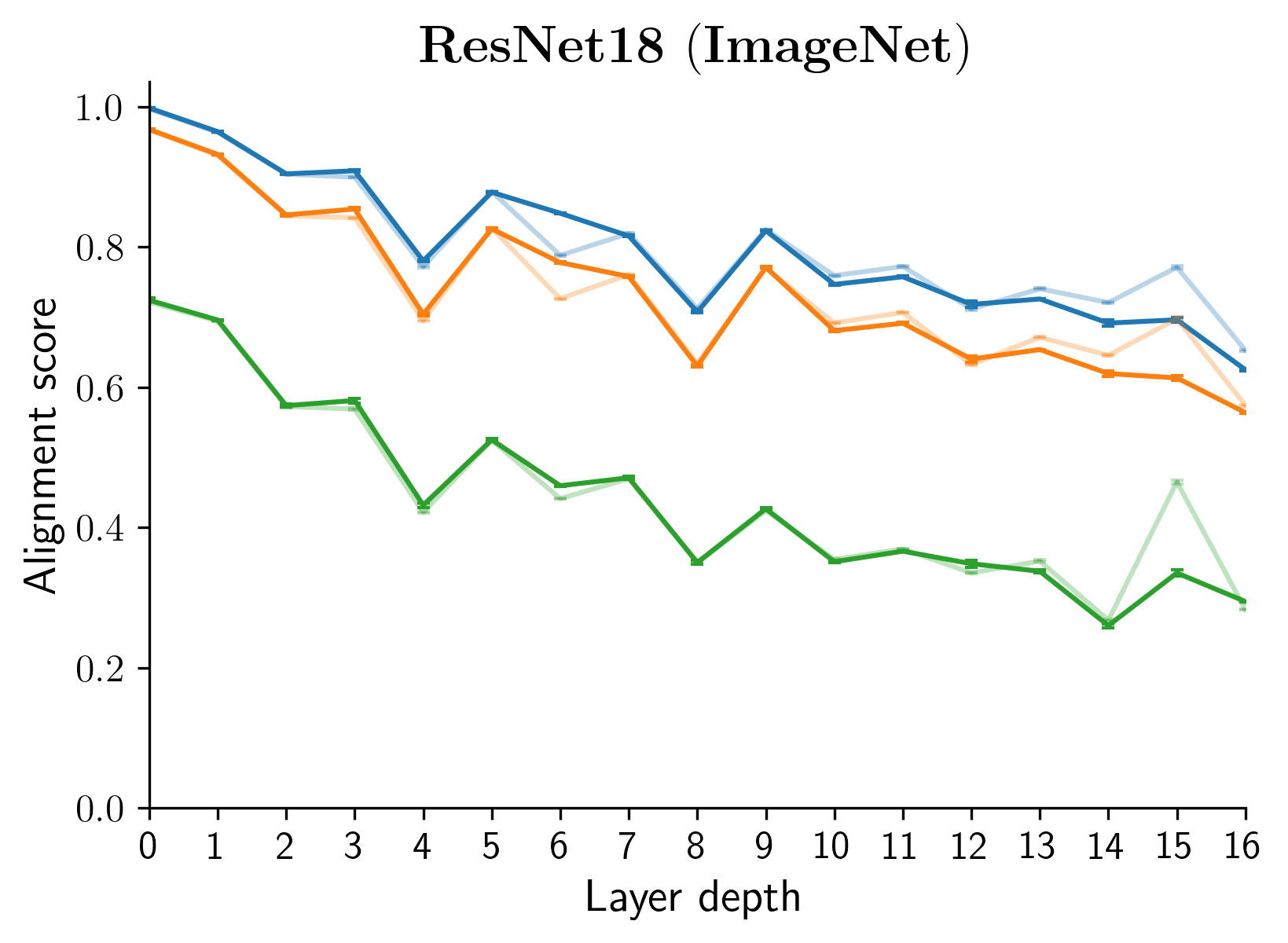}\hfill
    \includegraphics[width=.25\textwidth]{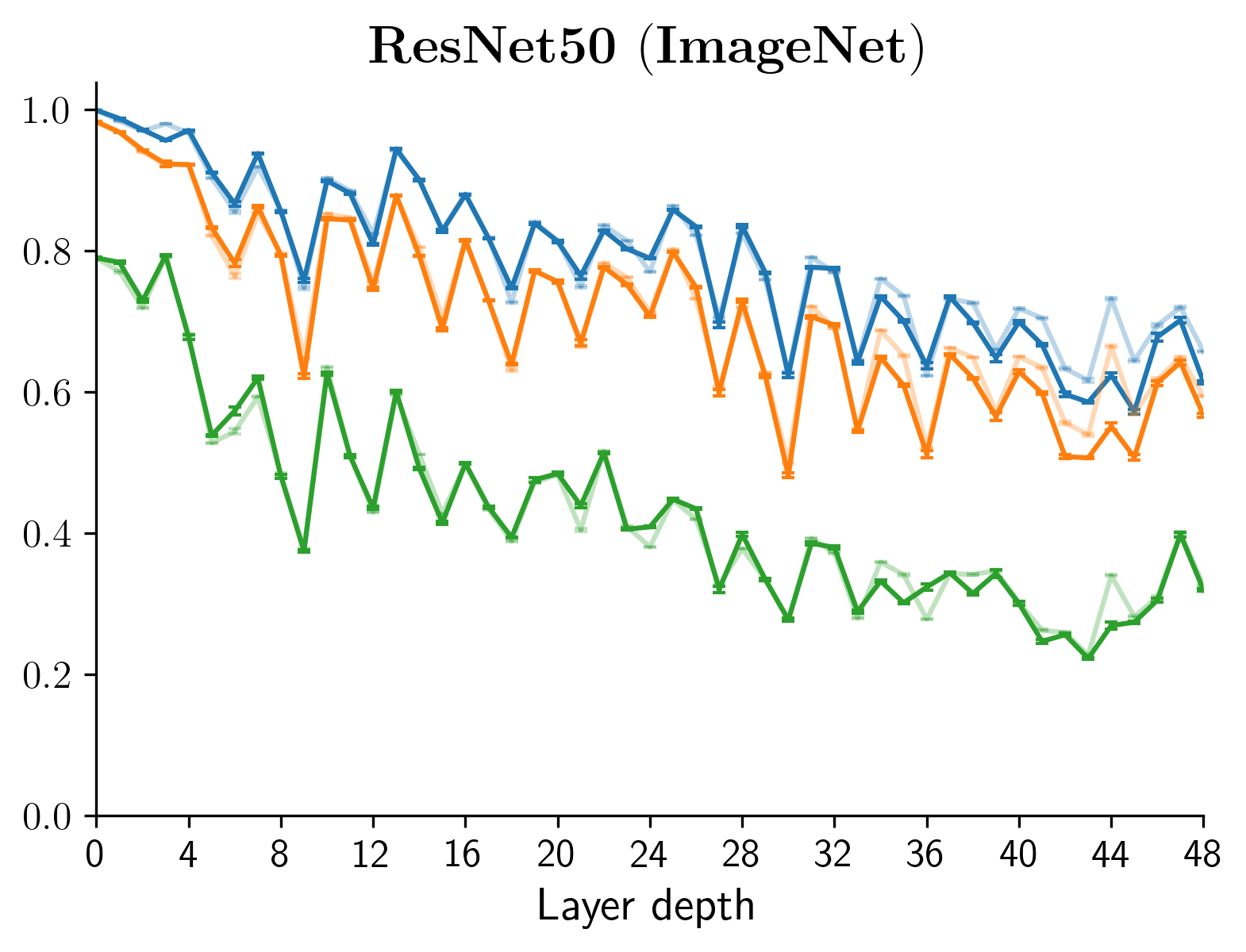}\hfill
    \includegraphics[width=.25\textwidth]{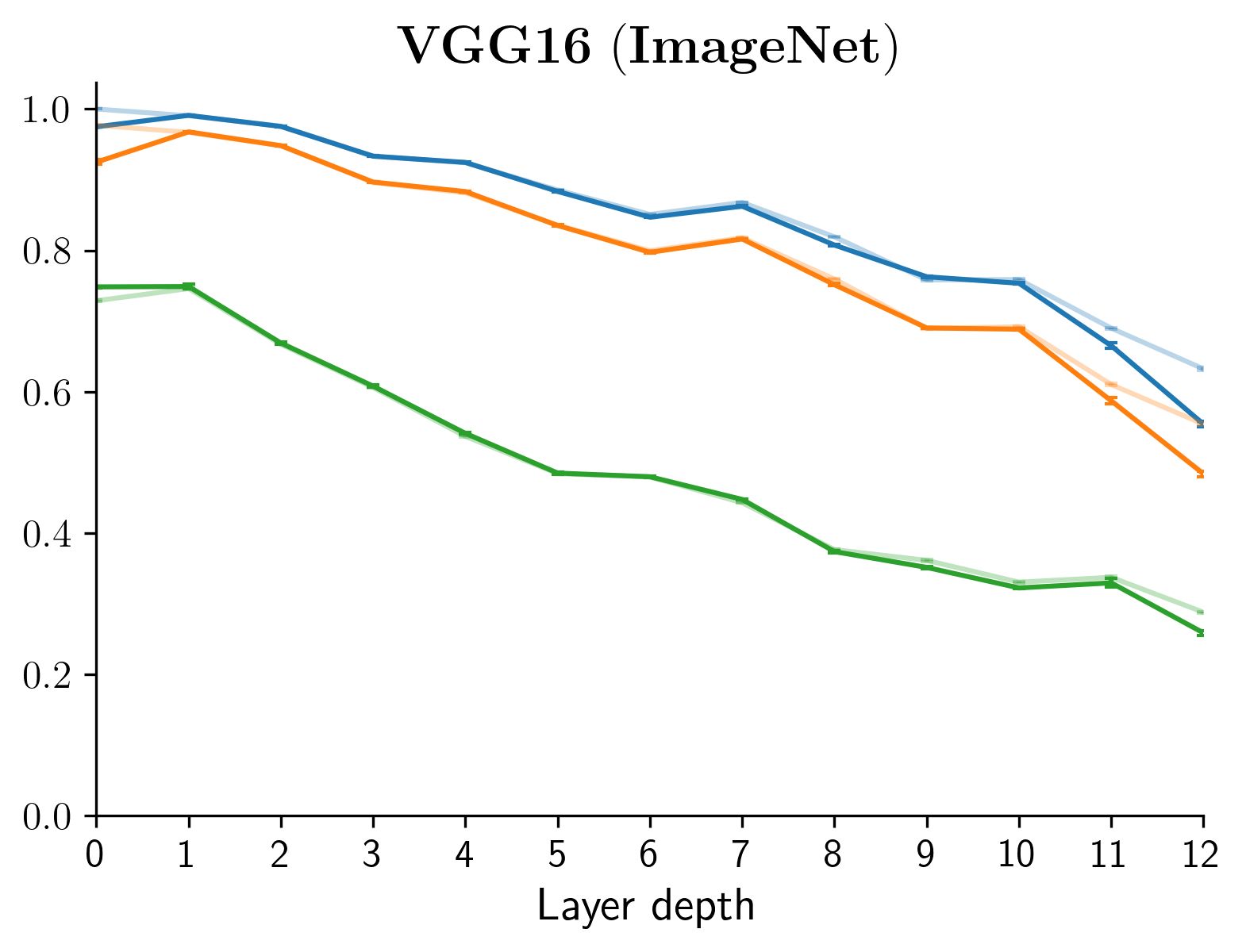}\hfill
    \includegraphics[width=.25\textwidth]{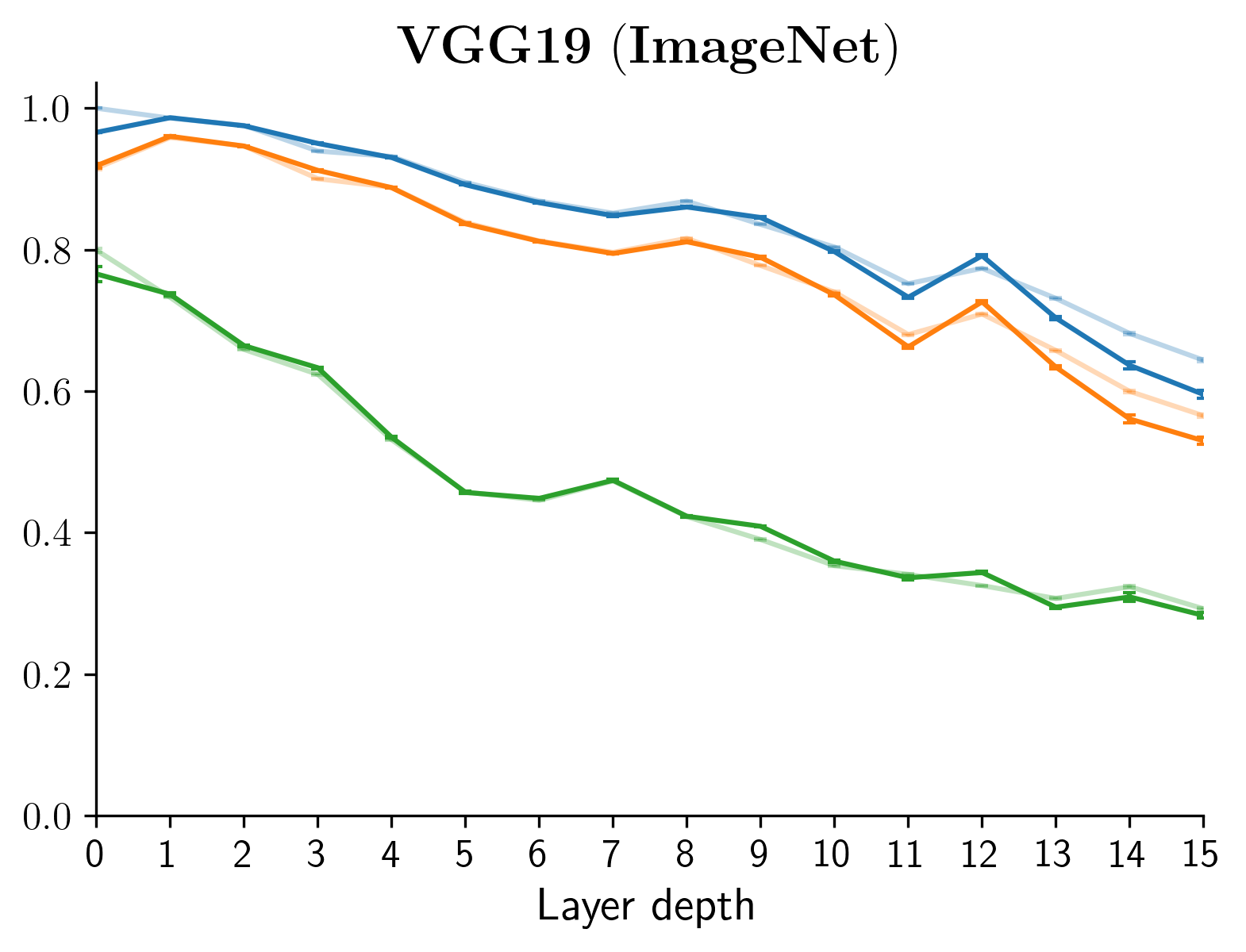}
    
    \caption{\textbf{Representational Convergence Across a Network Hierarchy Using a Random Pixel.} We plot the representational convergence across (ImageNet-trained) network hierarchies using both---the central pixel (darker shade) and a random pixel (lighter shade) for $3$ alignment metrics---\textcolor{mplblue}{Linear predictivity}, \textcolor{mplorange}{Procrustes}, \textcolor{mplgreen}{Permutation}. Across all these metrics, we observe that the spatial choice of the sample pixel leaves the alignment effectively unchanged.}
    \label{fig:network-hierarchy-random-pixel}
\end{figure*}

Through this experiment, we see that in fact choosing an arbitrary spatial location results in alignment trends across the network hierarchy remaining effectively unchanged, confirming that the choice of spatial location does not qualitatively affect our conclusions. Although using the full spatial map would be ideal, it is computationally prohibitive---scaling polynomially with dataset size---making the single-pixel approach an efficient and reliable proxy.

\end{document}